\documentclass[aps,prb,groupedaddress,twocolumn]{revtex4-1}

\usepackage{graphicx}% Include figure files
\usepackage{amsmath}
\usepackage{amssymb}
\usepackage{braket}
\usepackage{appendix}

\begin{document}

%\preprint{}

\title{Two-particle correlations in a functional renormalization group scheme using a dynamical mean-field theory approach}


\author{Michael Kinza$^{1}$}
\email[]{kinza@physik.rwth-aachen.de}
\author{Carsten Honerkamp$^{1}$}
%\thanks{}
\affiliation{
$^1$ Institute for Solid State Theory, RWTH Aachen University,
D-52056 Aachen\\ and JARA - Fundamentals of Future Information
Technology
}

\date{\today}

\begin{abstract}
We apply a recently introduced hybridization flow functional renormalization group scheme for Anderson-like impurity models as an impurity solver in a dynamical mean-field theory (DMFT) approach to lattice Hubbard models. We present how this scheme is capable of reproducing metallic and insulating solutions of the lattice model. Our setup also offers a numerically rather inexpensive method to calculate two-particle correlation functions. For the paramagnetic Hubbard model on the Bethe lattice in infinite dimensions we calculate the local two-particle vertex
for the metallic and the insulating phase. Then we go to a two-site cluster DMFT scheme for the two-dimensional
Hubbard model that includes short-range antiferromagnetic fluctuations and obtain the local and non-local two-particle vertex functions. We discuss the rich frequency structures of these vertices and compare with the vertex in the single-site solution.
\end{abstract}

%\keywords{}

\maketitle


\section{Introduction}

A prominent phenomenon in the field of strongly correlated systems is the Mott Hubbard metal-insulator transition (MIT), \cite{Mott68,Geb97} where the lattice electrons undergo a quantum phase transition from a paramagnetic metal to a paramagnetic insulator driven by the local Coulomb repulsion. For small values of the interaction the kinetic energy of the electrons dominates over their interaction energy, leading to a metallic state. For a large local repulsion doubly occupied sites become energetically costly and hence, for one electron per lattice site, the system will
minimize its energy by localizing the electrons. The system becomes an insulator. The simplest microscopic model describing the Mott MIT is the one-band Hubbard model.\cite{Hub63,Gut63,Kan63} In nature Mott MITs are found for example in transition metal oxides like chromium-doped V$_{2}$O$_{3}$, \cite{McWhan73} or in the undoped mother substances of cuprate high-temperature superconductors.\cite{Lee06} 

The qualitative features of the Mott transition regarding the ground state can be understood using the Gutzwiller approximation\cite{Gut65} and Brinkmann-Rice\cite{Bri70} theory. A controlled access in infinite dimensions and a quantitative theory of the spectral properties of models and materials near or in the Mott state are well described by the various forms of the dynamical mean-field theory (DMFT).\cite{Geo96,Pru95,Kot04} Here, a local problem for a subset of the full lattice, augmented by a dynamical Weiss field that represents the influence of the environment, is solved exactly by means of an impurity solver. Then the solution of this local problem is proliferated to the whole lattice, from which a new Weiss field is determined. Then the local problem is solved again. This procedure is iterated until the Weiss field and the local properties converge. 

The use of a local 'impurity' problem in the DMFT framework is the key approximation for any finite dimension, which makes the whole scheme applicable (and exact in infinite dimensions). There have been forceful and physically insightful attempts to include nonlocal correlations in the DMFT setup, as for example cluster extensions \cite{Mai05,Kot01,Lic00} and diagrammatic expansions around the local DMFT solution like the
dynamical vertex approximation,\cite{Tos07,Hel08,Roh11} the dual fermion method,\cite{Rub08,Rub09} the one-particle irreducible functional approach, \cite{Roh13} or multi-scale methods.\cite{Sle09} In the dual fermion strategy, for instance, the DMFT solution of a local core is used in the bare action of a non-local 'dual fermion' problem. Here one key issue is that the interaction of the dual fermions, which are obtained from the interaction vertex of the local problem, is intrinsically frequency dependent, and in order to be able to treat the dual fermion problem well, some insights about the frequency structure of these interactions will be helpful. A similar statement also holds for improvements of  strategies like the functional renormalization group (fRG, \cite{Met12}) to stronger interactions, either by starting at weak coupling and including the frequency structure and the self-energy feedback,\cite{Ueb12,Gie12} or by starting in the atomic limit by a flow in the hopping parameters, as recently shown to work for the single impurity problem \cite{Kin13}, for bosonic problems,\cite{Ran11,Ran11b} and for spin-systems.\cite{Reu13}  In all these fRG approaches, the frequency dependence of the vertex constitutes a severe complication when it has to be combined with a wavevector or space dependence. For the latter part, rather well-working approximations have been found,\cite{Met12,Hus09,Xia12,Wan13} but on the frequency part, not much is known beyond direct studies with rather large effort \cite{Ueb12} or boson-exchange parametrizations.\cite{Gie12} 
 
In this work, we follow two goals in this context. First we explore how a fRG hybridization flow method that was recently developed for the single impurity Anderson model performs as an impurity solver in DMFT cycles, i.e. in DMFT(fRG). Primarily, the fRG is still a relatively cheap impurity solver in terms of numerical effort, so studying its applicability in the DMFT framework may be useful. Furthermore, the fRG is a flexible and transparent method that nicely illustrates how nonlocal correlations emerge from local interactions, so using the fRG to build in correlations beyond the local core physics may be rewarding. If one wants to pursue this line, one should check how well the fRG works for small cores. We find that in DMFT(fRG) the hallmarks of the Mott transition can be reproduced, but also notice some technical complications that may require further improvements of the fRG scheme in order for the method to become truly competitive with other, established solvers. But as the results are qualitatively reasonable and the numerical effort is rather manageable, we can go to a second field of interest, the frequency structure of the local and nonlocal effective interaction vertices. Knowing the frequency dependence of these objects is of strong importance in the above-mentioned attempts to include nonlocal correlations beyond current DMFT schemes. We find that the effective vertices exhibit 'boson-like' frequency features, but also other 'loop coupling' features that are not easily captured by simple parametrizations of the frequency dependence in terms of frequency transfers or the total frequency. Here, our findings confirm results by the Vienna group\cite{Roh12} for the local vertex, obtained with DMFT using exact diagonalization (ED) as an impurity solver, and expand them to the non local situation.

This paper is organized as follows: In Sec. \ref{sec:method} we give a brief introduction to the single-site and the cluster DMFT framework and explain in which way the fRG hybridization flow scheme
can be used as an impurity solver in the DMFT self-consistency cycle. In the next Sec. \ref{sec:results} we present results for the spectral density, the vertex function and the spin susceptibility in the
insulating and the metallic phase, obtained with single-site DMFT (Sec. \ref{sec:resultssinglesitedmft}) and with two-site cluster DMFT (Sec. \ref{sec:resultstwositedmft}). A discussion of 
the differences in the frequency structure between single-site and two-site cluster DMFT vertices and the conclusion are given in Sec. \ref{sec:conclusions}.  

\section{Method}
\label{sec:method}
In this paper, we consider variants of the Hubbard model at half filling. The Hamiltonian is given by
\begin{equation}\label{eqhamiltonianhubbardmodel}
\hat{H}=-t\sum_{\langle i,j \rangle,\sigma}c^{\dag}_{i,\sigma}c_{j,\sigma}+U\sum_{i}\left(\hat{n}_{i,\uparrow}-1/2\right)\left(\hat{n}_{i,\downarrow}-1/2\right),
\end{equation}
where $c^{\dag}_{i,\sigma} (c_{i,\sigma})$ create (annihilate) electrons with spin $\sigma$ on site $i$ and $\hat{n}_{i,\sigma}=c^{\dag}_{i,\sigma}c_{i,\sigma}$. $t$ is the hopping amplitude between
nearest neighbours $\langle i,j \rangle$ on lattices specified below and $U>0$ is the onsite Coulomb repulsion.
If the model is defined on a bipartite lattice, the Hamiltonian (\ref{eqhamiltonianhubbardmodel}) is particle hole symmetric.

\subsection{Single-site DMFT}
In the first part of the paper we study the Hubbard model on a Bethe lattice with infinite connectivity $z\rightarrow\infty$. To make sure that this limit is physically meaningful we have to scale the
hopping parameter $t$ like $\frac{t^{\ast}}{\sqrt{z}}$ with constant $t^{\ast}$.\cite{Met89} The local density of states (DOS) is then semi-elliptic  \cite{Eco06}
\begin{equation}\label{eqlocaldos}
\text{DOS}\left(\omega\right)=\frac{1}{2\pi t^2}\sqrt{4 t^2-\omega^2}\, \Theta\left(2t-|\omega|\right)
\end{equation}
with bandwidth $W=4 t$. \footnote{Here and in the following we denote $t^{\ast}$ by $t$.} The self-energy becomes a purely local quantity i.e. $\Sigma_{ij}\left(i\omega\right)=\Sigma_{i}\left(i\omega\right)\delta_{ij}$ and
because of translational invariance it is site-independent $\Sigma_{i}\left(i\omega\right)=\Sigma\left(i\omega\right)$. This locality of the self-energy is the essential part of the DMFT, which is therefore exact in infinite dimensions.
The local lattice Green's function is then given by
\begin{equation}\label{eqlocallatticegreensfunction}
\mathcal{G}(i\omega)=\int\textrm{d}\epsilon\frac{\text{DOS}(\epsilon)}{i\omega-\Sigma(i\omega)-\epsilon}=\mathcal{G}_{0}(i\omega-\Sigma(i\omega))
\end{equation}
with the free local lattice Green's function $\mathcal{G}_0$.

The local self-energy can be written as a functional of the local lattice Green's function $\Sigma=\mathcal{S}\left[\mathcal{G}\right]$ in terms of skeleton 
diagrams.\cite{Geo92,Jar92} This can be used to map the Hubbard model to a single impurity Anderson model (SIAM)
\begin{eqnarray}
\hat{H}_{\text{And}}&=&\hat{H}_{\text{dot}}+\hat{H}_{\text{bath}}+\hat{H}_{\text{hybridization}}\\
\hat{H}_{\text{dot}}&=&\sum_{\sigma}\epsilon_{d}d^{\dag}_{\sigma}d_{\sigma}+U\hat{n}_{d,\uparrow}\hat{n}_{d,\downarrow}\\
\hat{H}_{\text{bath}}&=&\sum_{\vec{k},\sigma}\epsilon_{\vec{k}}b^{\dag}_{\vec{k},\sigma}b_{\vec{k},\sigma}\\
\hat{H}_{\text{hybridization}}&=&-\sum_{\vec{k},\sigma}\left(V_{\vec{k}}d^{\dag}_{\sigma}b_{\vec{k},\sigma}+H.c.\right),
\end{eqnarray}
that describes a dot level with onsite energy $\epsilon_d$ and local interaction $U$ that is coupled by a hybridization term $V_{\vec{k}}$ to uncorrelated bath levels
with energy $\epsilon_{\vec{k}}$. $d^{(\dag)}_{\sigma}$ create and annihilate electrons on the dot level and $b^{(\dag)}_{\vec{k},\sigma}$
on the bath levels respectively.
The local dot Green's function is given by
\begin{equation}
\mathcal{G}_{\text{dot}}(i\omega)=\frac{1}{i\omega-\epsilon_d-\Sigma_{\text{dot}}(i\omega)-\Delta(i\omega)}
\end{equation}
with the hybridization function
\begin{equation}
\Delta(i\omega)=\sum_{\vec{k}}\frac{|V_{\vec{k}}|^2}{i\omega-\epsilon_{\vec{k}}}.
\end{equation}
The self-energy is by construction local on the dot level. It has the same functional dependence on the dot Green's function as in the Hubbard model,
$\Sigma_{\text{dot}}=\mathcal{S}\left[\mathcal{G}_{\text{dot}}\right]$. If we now choose the parameters $V_{\vec{k}}$ and $\epsilon_{\vec{k}}$ such that
\begin{equation}\label{eqhybridizationfunction}
\Delta(i\omega)=i\omega-\epsilon_d-\Sigma_{\text{dot}}(i\omega)-\mathcal{G}(i\omega)^{-1}
\end{equation}
holds, we arrive at
\begin{equation}\label{eqselfenergycondition}
\Sigma_{\text{dot}}(i\omega)=\Sigma(i\omega).
\end{equation}
With Eqs. \ref{eqlocallatticegreensfunction}, \ref{eqhybridizationfunction} and \ref{eqselfenergycondition} we can express $\Delta(i\omega)$ by
the free hybridization function $\Delta_0(i\omega)=i\omega-\epsilon_d-\mathcal{G}_0(i\omega)^{-1}$ via
\begin{equation}\label{eqfreehybridizationfunction}
\Delta(i\omega)=\Delta_{0}(i\omega-\Sigma(i\omega)).
\end{equation}

The Eqs. \ref{eqlocallatticegreensfunction}, \ref{eqhybridizationfunction} and \ref{eqselfenergycondition} 
form a set of self-consistency equations for the local self-energy $\Sigma(i\omega)$. The SIAM can be solved by a large class of numerical methods 
(so-called 'impurity solvers') like for instance the numerical renormalization group,\cite{Bul98,Bul99} the quantum Monte Carlo,\cite{Jar92,Roz92} or the exact diagonalization method.\cite{Caf94,Lie12}

In the following we use a functional renormalization group scheme, introduced in Ref. \onlinecite{Kin13} and described below, as impurity solver.
In order to apply this scheme we have to map the bath of the Anderson model to a semi-infinite tight binding chain in which its first site is connected
to the impurity site,
\begin{eqnarray}\label{eqhamiltonandersonmodel}
\nonumber
\hat{H}_{\text{And}}&=&\epsilon_d\sum_{\sigma}d^{\dag}_{\sigma}d_{\sigma}+U\left(\hat{n}_{d,\uparrow}-1/2\right)\left(\hat{n}_{d,\downarrow}-1/2\right)\\
\nonumber
&-&v\sum_{\sigma}\left(d^{\dag}_{\sigma}b_{1,\sigma}+H.c.\right)\\
\nonumber
&-&\sum_{i=1}^{\infty}\sum_{\sigma}t_i\left( b^{\dag}_{i,\sigma}b_{i+1,\sigma}+H.c.\right)
+\sum_{i=1}^{\infty}\sum_{\sigma}\epsilon_{i}b^{\dag}_{i,\sigma}b_{i,\sigma}.
\end{eqnarray}
For a general bath this can be achieved by the Lanczos algorithm described for example in Ref. \onlinecite{Hew93}. For a semi-elliptic local DOS
(\ref{eqlocaldos}) we have to choose $\epsilon_i=0$ and $t_i=t$ for all $i$.
Then the free hybridization function has the form
\begin{eqnarray}
\nonumber
\Delta_{0}(i\omega)&=&v^2 g_{t}\left(i\omega\right)\\
\textrm{with}\ g_{t}\left(i\omega\right)&=&\frac{1}{2t^2}\left(i\omega-i\textrm{sgn}(\omega)\sqrt{4t^2-(i\omega)^2}\right).
\end{eqnarray}
If we now additionally choose $\epsilon_d=0$ and $v=t$ the free local dot Green's function $\mathcal{G}_0$ is given by $g_{t}$
and the local $\text{DOS}(\omega)=-\frac{1}{\pi}\text{Im}\mathcal{G}_0\left(i\omega\rightarrow\omega+i0^{+}\right)$ is semi-elliptic.

\subsection{Impurity solver: fRG hybridization flow}

In order to solve the impurity model (\ref{eqhamiltonandersonmodel}) we apply a functional renormalization group (fRG) scheme,\cite{Wet93,Sal01,Met12} introduced in Ref. \onlinecite{Kin13}. In this reference a detailed description of the formalism is given and we just repeat the main aspects.

The fRG scheme is designed to treat impurity models in the form of a semi-infinite tight binding chain, where the interaction term is located
on the first site which is the situation in the model (\ref{eqhamiltonandersonmodel}). Later we will see that this formalism can be extended to multi-impurity models in
the form of a $N$-chain ladder as shown in Figure \ref{picttwochainladder}.
As in the Figures \ref{pictandersonmodel} and \ref{pictandersonmodel2} 
we denote the interacting site by $d$ and the remaining 'bath' sites by $b_1,b_2,\text{etc}$.
First, the system is divided into two parts. One (called 'core') contains the interacting site and the first $L$ bath sites and one (called 'bath')
contains the remaining sites $b_i$ with $i>L$. We start with a situation where the core and the bath are completely decoupled. Therefore we multiply the hopping matrix element between them by a factor $\Lambda$ and set $\Lambda=0$ in the beginning. Then we solve the Hamiltonian of the isolated core exactly and calculate the one- and two-particle correlation functions. These serve
as input to fRG flow equations with $\Lambda$ as flow-parameter, that are integrated from $\Lambda=0$ to $\Lambda=1$. This means that the flow leads from the decoupled core to the fully embedded core. In Ref. \onlinecite{Kin13} we showed how the Kondo physics of a single correlated site is obtained in a qualitatively correct way for the $L=1$-core, but not for the $L=0$-core.

It turns out to be useful to implement the fRG flow in an effective theory on the bath site $b_{L+1}$, in which the interacting core and all bath sites with index $>L+1$ are integrated out. The effective action of this theory and the fRG flow equations can be found in Appendix \ref{sec:appendixfrg}. We work with two different approximation levels, called approximation 1 and approximation 2. In the first level, only the self-energy flow is considered, and the interaction vertices remain fixed to their initial values, while in approximation 2, also the vertices are allowed to change from their starting values.

As can be seen from this description, this fRG impurity solver explicitly involves the one-particle irreducible vertex function (i.e. the interaction vertex). Yet, in terms of the numerical effort, the fRG scheme is relatively inexpensive. For
example a parallelized integration of the flow equations on eight cores using 200 Matsubara frequencies in approximation 2 takes about 8 hours. For other impurity solvers, the calculation of the vertex function represents a formidable growth of the numerical effort.  Hence it appears worthwhile to explore the use of our fRG scheme as a numerically relatively inexpensive tool to study this quantity in more detail, especially in cluster DMFT calculations. Due to the truncation of the infinite set of flow equations our setup is not exact and we do not claim to obtain quantitative predictions. But as discussed below, the frequency structure of the vertex function comes out in good qualitative agreement with DMFT(ED) calculations,\cite{Roh12} and, in addition,  we can go to nonlocal correlations as well.

\subsection{Single-site DMFT(fRG)}
\subsubsection{Insulating phase}
As just mentioned, in the functional renormalization group scheme that is used to solve the SIAM the system is separated into two parts. 
One part, called 'core', contains the correlated impurity site and the first $L$ bath sites and the other part, called 'bath',  contains the remaining bath sites. In the simplest case, $L=0$, the core consists only of the impurity site. As shown in Ref. \onlinecite{Kin13} the fRG scheme with
this choice of the core fails in describing the quasi-particle properties of the SIAM, i.e., the imaginary part of the Matsubara self-energy has a finite step at $i\omega=0$, which does not become 
reduced by decreasing temperature and therefore one cannot define a finite quasiparticle weight
\begin{equation}\label{eqquasiparticleweight}
Z^{-1}=1-\left. \frac{d\text{Im}\Sigma\left(i\omega\right)}{d\omega}\right|_{\omega=0^{+}}.
\end{equation}
Due to this the $L=0$-core is not suitable to describe the metallic phase of the Hubbard model.
However one can still hope to arrive at a reasonable description of the insulating phase even with this simplest choice for the core. Below we see that this indeed works. 

\begin{figure}[htbp]
    \centering
      \includegraphics[width=0.5\textwidth]{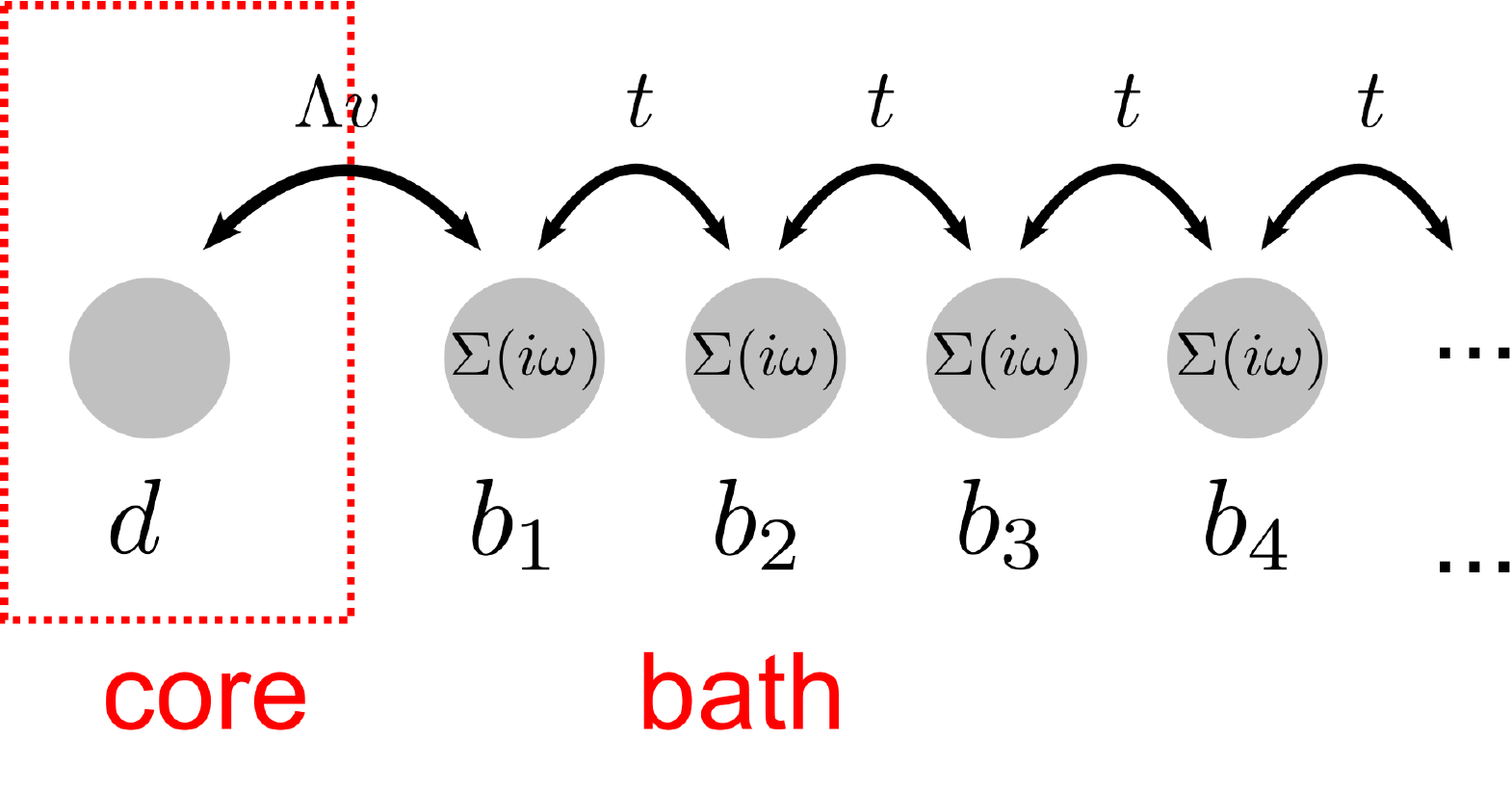}
      \caption[andersonmodel]{(color online) Anderson model (core size L=0) with semi-infinite tight binding chain as bath.}     
      \label{pictandersonmodel}
\end{figure}

The full hybridization function $\Delta(i\omega)$ is given by Eq. \ref{eqfreehybridizationfunction}. It corresponds to a semi-infinite tight binding chain
with a local term $\Sigma(i\omega)$ on each lattice site (cf. Fig.\ref{pictandersonmodel}).

\begin{figure}[htbp]
    \centering
      \includegraphics[width=0.5\textwidth]{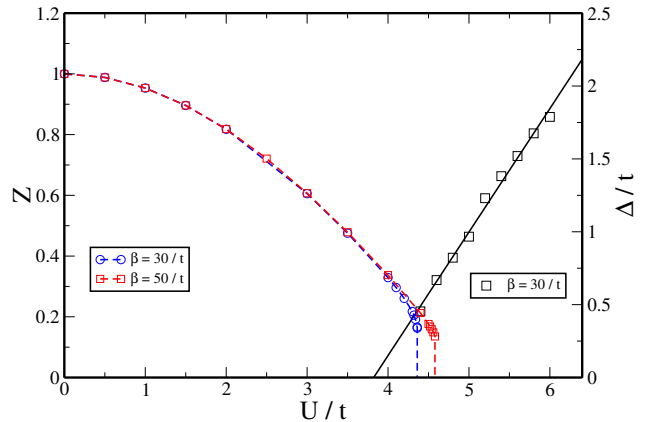}
      \caption[andersonmodel]{(color online) The plot shows the gap size $\Delta$ in the insulating phase and the quasi-particle weight $Z$ in the metallic phase as function of the interaction strength $U$ at several temperatures, calculated in single-site DMFT(fRG). The gap sizes are estimated from the spectral density calculated in approximation 1 of the fRG flow equations.
			Compared to the literature \cite{Blu03} the quasi-particle weight $Z$ goes down too slowly, which is consistent with the behavior found in the Anderson Impurity case.\cite{Kin13}}     
      \label{pictgapandzvsu}
\end{figure}

To get an estimate for which interactions $U$ this approach delivers a reasonable description of an insulating phase, we show in Fig. \ref{pictgapandzvsu} the gap $\Delta$
as function of $U$ for $\beta = 30 / t$. The gap sizes $\Delta$ are obtained from the spectral density calculated in approximation 1 of the fRG flow equations. The gap vanishes at $U_c\approx 3.8 t$.

\subsubsection{Metallic phase} \label{metphase}
In order to describe the metallic phase of the Hubbard model, the $L=1$-core, containing the correlated site and one bath site, is an appropriate starting point, as this core also successfully reproduced the Kondo central peak in the SIAM setup.\cite{Kin13} In the spectrum of the decoupled core one obtains two peaks near zero energy which lead to a continuous Matsubara self-energy at $i\omega=0$ resulting with Eq. \ref{eqquasiparticleweight} in a finite quasiparticle weight $Z$.

\begin{figure}[htbp]
    \centering
      \includegraphics[width=0.5\textwidth]{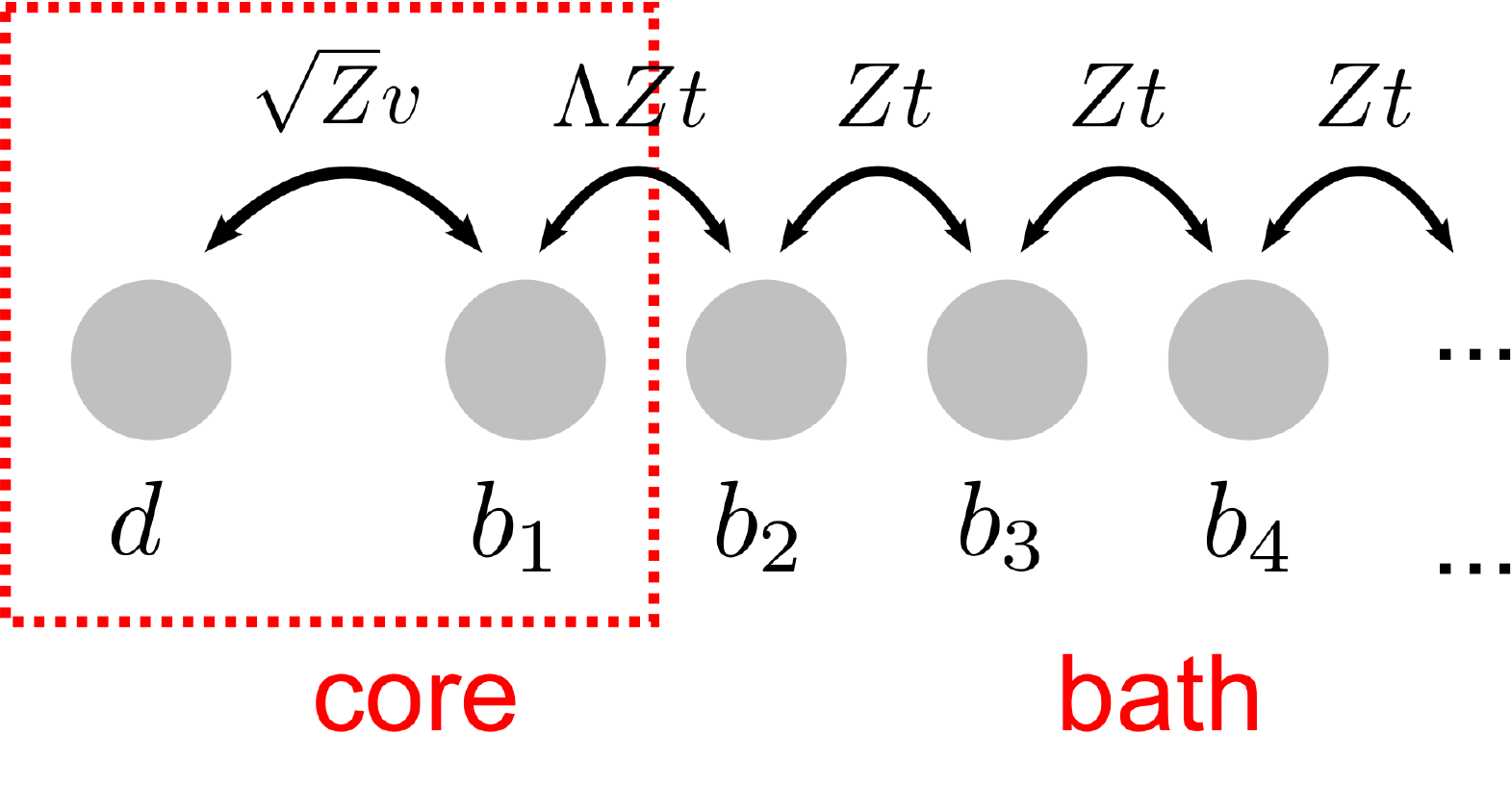}
      \caption[andersonmodel]{(color online) The Anderson model for the case $L=1$. It corresponds to a semi-infinite tight binding chain with hopping $Zt$ and impurity-bath coupling $\sqrt{Z}v$.}     
      \label{pictandersonmodel2}
\end{figure}

The full hybridization function is again given by $\Delta(i\omega)=\Delta_{0}(i\omega-\Sigma(i\omega))$, but opposite to the  $L=0$-case a local self-energy term on the first bath site that is part of the $L=1$-core, is forbidden, because the exact diagonalization
of the core requires a frequency-independent core Hamiltonian. To circumvent this we approximate $\Sigma(i\omega)$ for small 
frequencies as $\Sigma(i\omega)\approx\left(1-Z^{-1}\right)i\omega$, with the quasiparticle weight $Z$.
The full hybridization function $\Delta(i\omega)$ is then given by
\begin{eqnarray}\label{eqapproxhybridizationfunctionmetall}
\nonumber
\Delta(i\omega)&=&\Delta_{0}(i\omega-\Sigma(i\omega))\approx\Delta_{0}(i\omega/Z)\\
\nonumber
&=& \frac{v^2}{2t^2}\left(\frac{i\omega}{Z}-i\textrm{sgn}(\omega)\sqrt{4t^2-\left(\frac{i\omega}{Z}\right)^2}\right)\\
\nonumber
&=& \frac{(\sqrt{Z}v)^2}{2(Zt)^2}\left(i\omega-i\textrm{sgn}(\omega)\sqrt{4(Zt)^2-(i\omega)^2}\right)\\
&=& (\sqrt{Z}v)^2 g_{Zt}(i\omega).
\end{eqnarray}
It corresponds to a semi-infinite tight binding chain with hopping $Zt$ and impurity-bath coupling $\sqrt{Z}v$ (cf. Fig.\ref{pictandersonmodel2}).

In each selfconsistency cycle of the DMFT equations one calculates the quasi-particle weight $Z$ from the local self-energy which defines the new hopping parameters for the next cycle. For $\Lambda=0$, i.e. without solving the fRG flow equations, this scheme is equivalent
to the two-site DMFT scheme introduced in Ref. \onlinecite{Pot01}. This two-site DMFT scheme \footnote{Not to be confused with the two-site cluster DMFT scheme.} yields a satisfactory description of the Mott transition and the Fermi
liquid state in the single-band Hubbard model at $T=0$. The quasi-particle weight is predicted as $Z=1-U^2/U^2_c$ with $U_c = 1.5W$, which is very close to the result of
the numerical renormalization group. \cite{Bul99} For values of $U$ larger than $U_c$ this scheme reduces to the Hubbard-I approximation.\cite{Hub63}
Our extended scheme is implemented at finite temperatures. In Fig. \ref{pictgapandzvsu} we show the quasi-particle weight $Z$ as a function of $U$ at $\beta=30 / t$
and $\beta=50 / t$ (calculated in approximation 1). These temperatures are still
lower than the critical end point of the MIT phase diagram. The quasi-particle weight $Z$ vanishes discontinuous at certain values $U_c(T)$, which marks the
breakdown of the metallic state. For larger values of $U$ the quasi-particle weight decreases in the DMFT cycle until a linearization of the
self-energy is no more possible.
Compared to the literature,\cite{Blu03} the obtained values $U_c(T)$ come out too small. Note that the approximation for the 
hybridization function (\ref{eqapproxhybridizationfunctionmetall}) becomes very bad at large frequencies especially for small quasi-particle weights near the phase transition.
The obtained $U_c$ is larger than the interaction strength, where the gap $\Delta$ vanishes (cf. Fig. \ref{pictgapandzvsu}). Although one expects a hysteresis region at the
phase transition and the two values are indeed different, a direct comparison is of course problematic due to the distinct approaches used to describe the insulating and the
metallic phase.

\subsection{Two-site cluster DMFT(fRG)}

Although the DMFT is only exact in the limit of infinite dimensions, it turns out to be an extremely useful approximation scheme for systems with finite dimension. In these systems nonlocal correlation effects, like e.g. antiferromagnetic fluctuations or superconducting d-wave pairing, play an important role
and several extensions of the simple DMFT framework exist that capture these effects. Important examples are perturbational expansions around the local DMFT solution \cite{Stan04,Tos07,Hel08,Rub08,Rub09} or numerical cluster DMFT schemes, where short ranged correlations within a finite cluster are included.\cite{Lic00,Kot01,Mai05} 
\begin{figure}[htbp]
    \centering
      \includegraphics[width=0.4\textwidth]{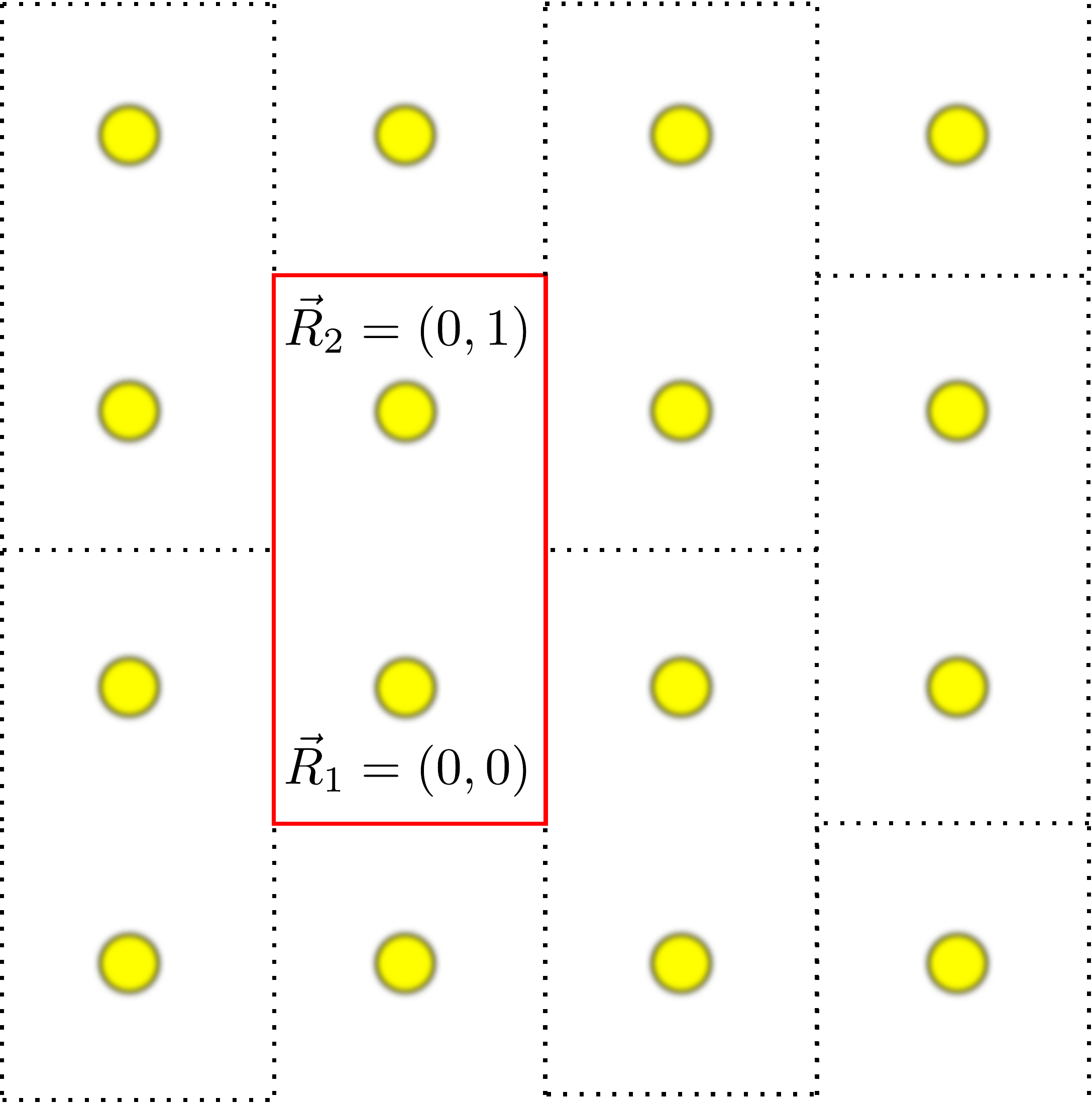}
      \caption[squarelattice]{(color online) Tiling of the square lattice with two-site clusters. Each lattice site can be uniquely described by a cluster vector and the
			site within the cluster $\vec{R}_j$. Note, that other periodic arrangements of the two-site clusters, corresponding to a different choice of the
			superlattice, would be also possible. However, this would only lead to another equivalent description of our problem and with our choice
			the quantities in the reciprocal superlattice acquire the most compact form.}     
      \label{pictsquarelattice}
\end{figure}

In the second part of the paper we extend our setup to a cluster DMFT scheme for the Hubbard model (\ref{eqhamiltonianhubbardmodel}) on a 2-dimensional square lattice
with tight binding dispersion 
\begin{equation}\label{eqtightbindingdispersion}
\epsilon_{\vec{k}}=-2t\left[\cos(k_x)+\cos(k_y)\right]
\end{equation}
and bandwidth $W=8t$. As shown in Fig. \ref{pictsquarelattice} we divide
the lattice into plaquettes with $L=2$ sites. This breaks the translational invariance of the original lattice problem and introduces a superlattice $\Gamma$ of clusters, whose sites
form a subset of the original lattice $\gamma$. Each lattice site of the original lattice $\vec{r}_{i}$ can then be uniquely described by a cluster-vector $\vec{r}_{m}$ and the site within the cluster $\vec{R}_{j}$ as $\vec{r}_{i}=\vec{r}_m+\vec{R}_j$.

The Brillouin zone of the original lattice ($BZ_{\gamma}$) contains $L$ points of the reciprocal superlattice. For the two-site clusters these are
$\vec{K}_{1}=\left(0,0\right)$ and $\vec{K}_{2}=\left(\pi,\pi\right)$. Any wavevector $\vec{k}\in BZ_{\gamma}$ can be uniquely written as $\vec{k}=\vec{K}+\vec{\tilde{k}}$, with $\vec{K}\in \{\vec{K}_{1},\vec{K}_{2}\}$ 
and $\vec{\tilde{k}}$ belonging to the Brillouin zone of the superlattice ($BZ_{\Gamma}$).\footnote{Here we follow the notation of Chapter 8 and 11 in Ref. \onlinecite{Ave12}.}

The hopping amplitude between two sites of the same cluster $\vec{R}_a$ and $\vec{R}_b$ can be obtained from the dispersion relation by the Fourier transformation
\begin{eqnarray}
\nonumber
t_{ab}&=&\frac{1}{N}\sum_{\vec{k}}\text{e}^{i\vec{k}(\vec{R}_a-\vec{R}_b)}\epsilon_{\vec{k}}\\
\nonumber
&=& \frac{1}{N}\sum_{\vec{K},\vec{\tilde{k}}}\text{e}^{i(\vec{K}+\vec{\tilde{k}})(\vec{R}_a-\vec{R}_b)}\epsilon_{\vec{K}+\vec{\tilde{k}}}\\
&=& \frac{L}{N}\sum_{\vec{\tilde{k}}}\underbrace{\text{e}^{i\vec{\tilde{k}}(\vec{R}_a-\vec{R}_b)}\frac{1}{L}\sum_{\vec{K}}\text{e}^{i\vec{K}(\vec{R}_a-\vec{R}_b)}\epsilon_{\vec{K}+\vec{\tilde{k}}}}_{\hat{t}_{ab}(\vec{\tilde{k}})}.
\end{eqnarray}
$\hat{t}(\vec{\tilde{k}})$ is the partial Fourier transformation of the band dispersion i.e. a matrix in the cluster space which
depends on the wavevector $\vec{\tilde{k}}$ of the reciprocal superlattice. For the tight binding dispersion (\ref{eqtightbindingdispersion}) it is given by
\begin{equation}
\hat{t}(\vec{\tilde{k}})=\left[\begin{array}{cc} 0 & \text{e}^{-i\tilde{k}_y}\epsilon_{\vec{\tilde{k}}}  
\\ \text{e}^{i\tilde{k}_y}\epsilon_{\vec{\tilde{k}}} & 0
\end{array}\right].
\end{equation}
If we assume that the self-energy is local on each cluster i.e. independent of $\vec{\tilde{k}}$ we obtain the local cluster Green's function as
\begin{eqnarray}
\nonumber
\hat{\mathcal{G}}(i\omega)&=&\frac{L}{N}\sum_{\vec{\tilde{k}}}\left[i\omega\textbf{1}-\hat{t}(\vec{\tilde{k}})-\hat{\Sigma}(i\omega)\right]^{-1}\\
&=&\hat{\mathcal{G}}_{0}\left(i\omega\textbf{1}-\hat{\Sigma}(i\omega)\right).
\end{eqnarray}
This can be interpreted as the local Green's function of a two-impurity Anderson model with hybridization function
\begin{eqnarray}
\nonumber
\hat{\Delta}(i\omega)&=&i\omega\textbf{1}-\hat{t}-\hat{\Sigma}(i\omega)-\hat{\mathcal{G}}(i\omega)^{-1}\\
&=&\hat{\Delta}_{0}\left(i\omega\textbf{1}-\hat{\Sigma}(i\omega)\right).
\end{eqnarray}
The free hybridization function $\hat{\Delta}_0$ is again given by
\begin{equation}
\hat{\Delta}_0(i\omega)=i\omega\textbf{1}-\hat{t}-\hat{\mathcal{G}}_0(i\omega)^{-1}.
\end{equation}
$\hat{t}$ is the cluster hopping matrix defined by
\begin{equation}
\hat{t}=\left[\begin{array}{cc} 0 & -t  
\\ -t & 0
\end{array}\right].
\end{equation}
Note that in the present work we do not allow for any symmetry-breaking, neither in the single-site DMFT, nor in the two-site cluster DMFT approach. This means that, e.g., the antiferromagnetic ground state of the square lattice Hubbard model will not be captured. In principle, the symmetry breaking could be included by considering spin- and sublattice-dependent self-energies. Here, in order to keep the first applications of the DMFT(fRG) scheme simple, we do not allow for this additional aspect and focus on the non-magnetic 'mother states' of the true ground states.

\begin{figure}[htbp]
    \centering
      \includegraphics[width=0.5\textwidth]{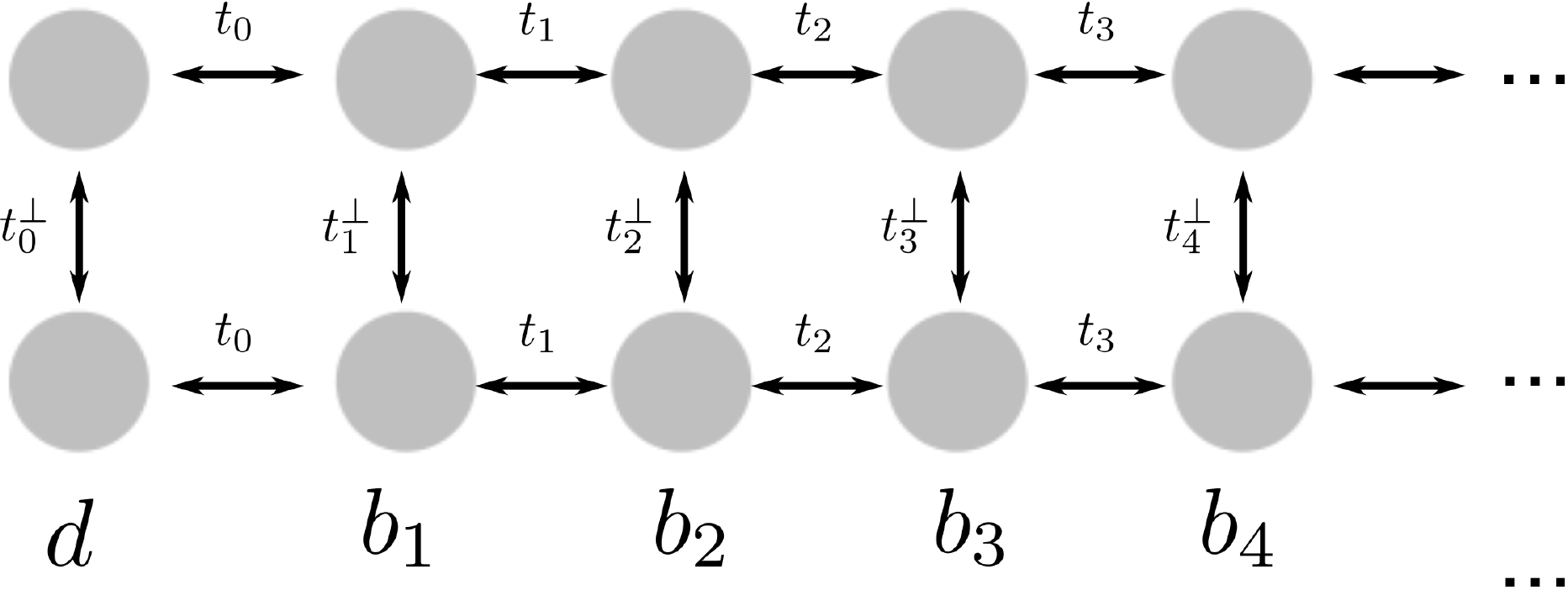}
      \caption[twochainladder]{Illustration of the two-chain ladder, corresponding to Hamiltonian (\ref{eqhamiltontwositeandersonmodel})}     
      \label{picttwochainladder}
\end{figure}

To apply our fRG scheme the two-impurity Anderson model must have the form of a two-chain ladder as shown in Fig.(\ref{picttwochainladder})
with Hamiltonian
\begin{widetext}
\begin{eqnarray}\label{eqhamiltontwositeandersonmodel}
\nonumber
\hat{H}_{\text{2-site-And}}&=&U\sum_{j=1}^{2}\hat{n}_{d,j,\uparrow}\hat{n}_{d,j,\downarrow}
-t^{\perp}_0 \sum_{\sigma}\left(d^{\dag}_{1,\sigma}d_{2,\sigma}+H.c.\right)
-t_{0}\sum_{\sigma}\sum_{j=1}^{2}\left(d^{\dag}_{j,\sigma}b_{1,j,\sigma}+H.c.\right)\\
&&-\sum_{i=1}^{\infty}\sum_{j=1}^{2}\sum_{\sigma}\left(t_i b^{\dag}_{i,j,\sigma}b_{i+1,j,\sigma}+H.c.\right)
-\sum_{i=1}^{\infty}\sum_{\sigma}t^{\perp}_{i}\left(b^{\dag}_{i,1,\sigma}b_{i,2,\sigma}+H.c.\right).
\end{eqnarray}
\end{widetext}

To determine the parameters of the two-chain ladder we fit the eigenvalues of the free hybridization function, $\Delta^{(1)}_{0}(i\omega)$
and $\Delta^{(2)}_{0}(i\omega)$, to a discretized hybridization function of the form
\begin{equation}
\Delta^{N,(i)}_{0}(i\omega)=\sum_{i=1}^{N}\frac{|v_i|^2}{i\omega-\epsilon_i},
\end{equation}
where the fit-parameters $v_i$ and $\epsilon_i$ are calculated by a conjugate gradient minimization \cite{Geo96} of the distance function
\begin{equation}
d=\frac{1}{\omega_{\text{max}}}\sum_{\omega}|\Delta^{(i)}_{0}(i\omega)-\Delta^{N,(i)}_{0}(i\omega)|^2.
\end{equation}
Note that the fit-parameters for the two eigenvalues are not independent because it is 
$\Delta^{(2)}_{0}(i\omega)=-\Delta^{(1)}_{0}(i\omega)^{\ast}$ due to particle hole symmetry. 
The finite bath can then be transformed to a tridiagonal form by the Lanczos algorithm, which determines
the hopping parameters of the two-chain ladder.

The extension of our fRG scheme to multi-impurity problems in the form of a $N$-chain ladder like in Eq. \ref{eqhamiltontwositeandersonmodel} can be found in Appendix \ref{sec:appendixfrg}.

\section{Results}
\label{sec:results}
\subsection{Single-site DMFT}
\label{sec:resultssinglesitedmft}
First let us discuss the results for using the hybridization flow as DMFT solver for the case of single-site DMFT, embedded in a Bethe lattice.  We show that the approach can reasonably describe both the insulating as well as the metallic phase, and give results for the effective interaction vertices in these phases.

\subsubsection{Insulating phase}
\begin{figure}[htbp]
    \centering
      \includegraphics[width=0.5\textwidth]{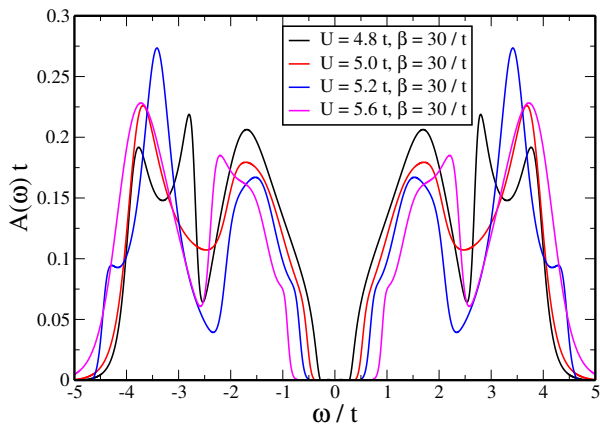}
      \caption[spectraldensityinsulatingphase]{(color online) spectral density for $U=4.8 t, 5 t, 5.2 t, 5.6 t$ at $\beta=30 / t$.}     
      \label{pictspectraldensityinsulatingphase}
\end{figure}

From our numerical data for the self-energy on the Matsubara frequency axis we obtain the spectral density $A(\omega)=-\frac{1}{\pi}\text{Im}\mathcal{G}(\omega+i0^{+})$ by an analytical
continuation using a Pad$\acute{\text{e}}$-algorithm described in Ref. \onlinecite{Vid77}.
The spectral density  for several values of $U / t$ is shown in Fig. \ref{pictspectraldensityinsulatingphase}. 
One obtains an opening of a Mott gap around $\omega=0$ with an average center-to-center separation of the two Hubbard bands of $\sim U$. The width of the Hubbard bands for these moderate $U$-values is only a little smaller than the band width of the noninteracting problem, $W=4t$.
The rich multi-peak structure of the Hubbard bands (with a variable number of maxima) is probably an artifact of our approximation, most likely due to the
discrete core used in the initial condition of the flow equation.

\begin{figure*}[htbp]
    \centering
      \includegraphics[width=0.45\textwidth]{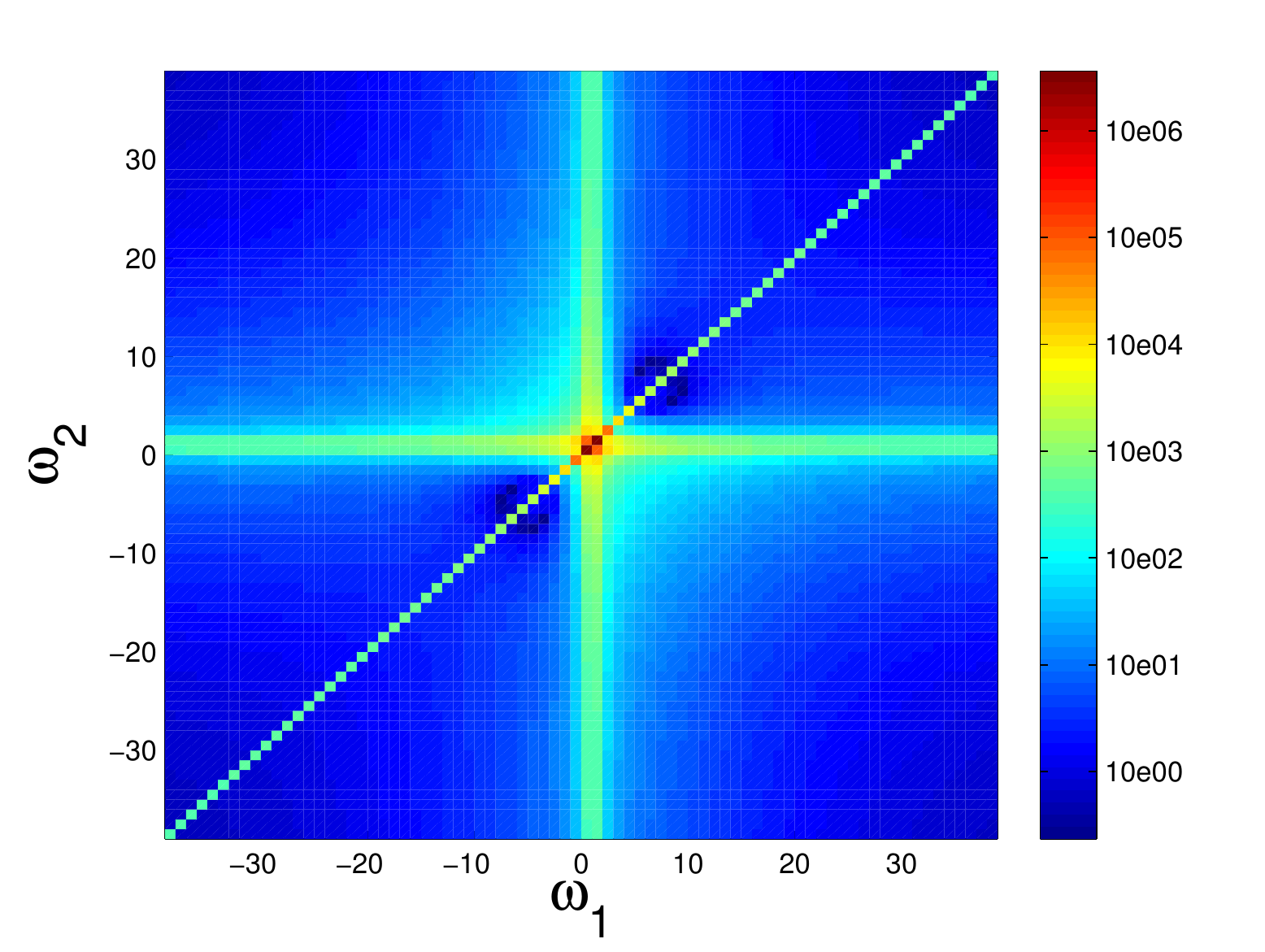}
      \includegraphics[width=0.45\textwidth]{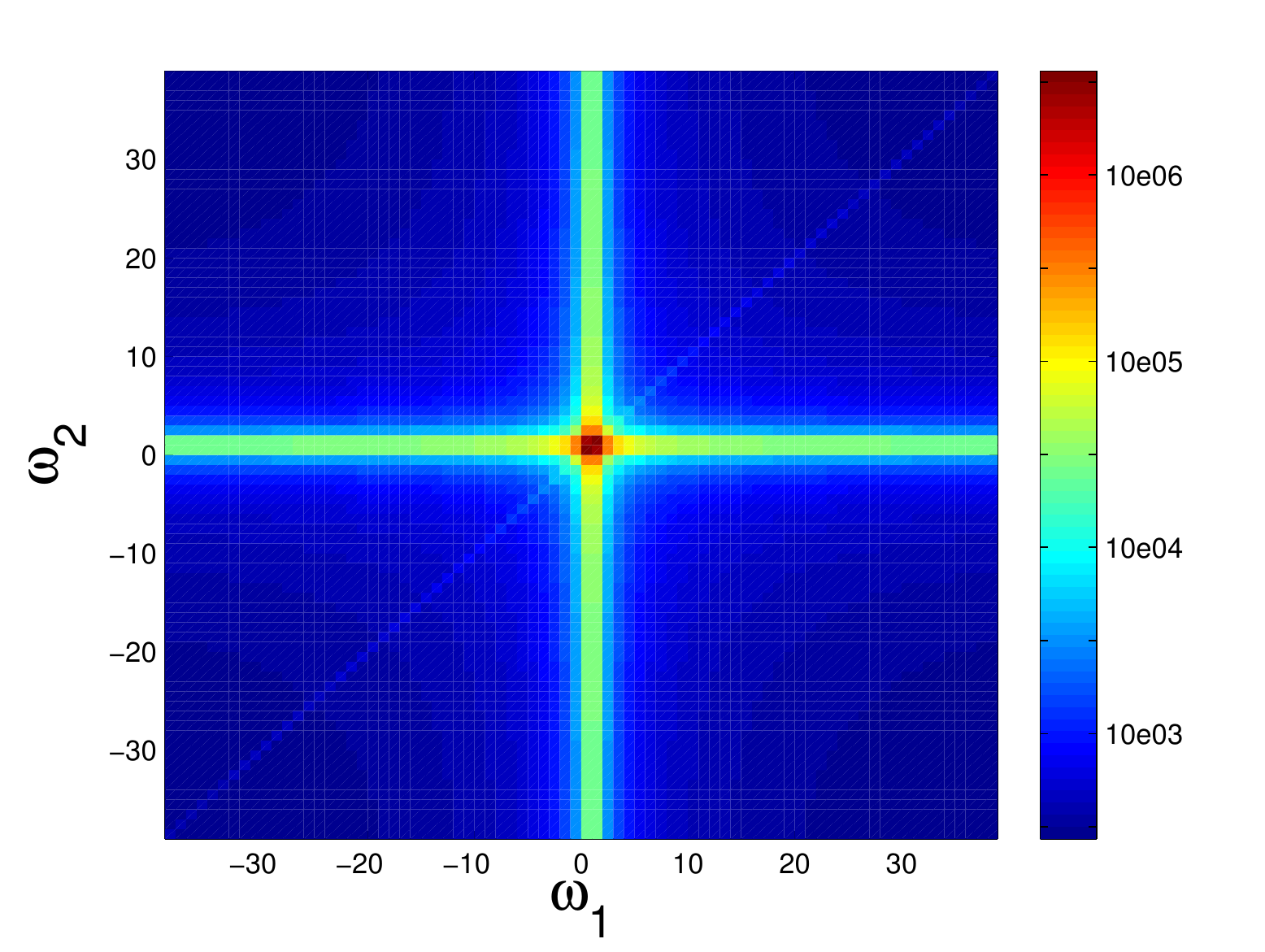}
      \includegraphics[width=0.45\textwidth]{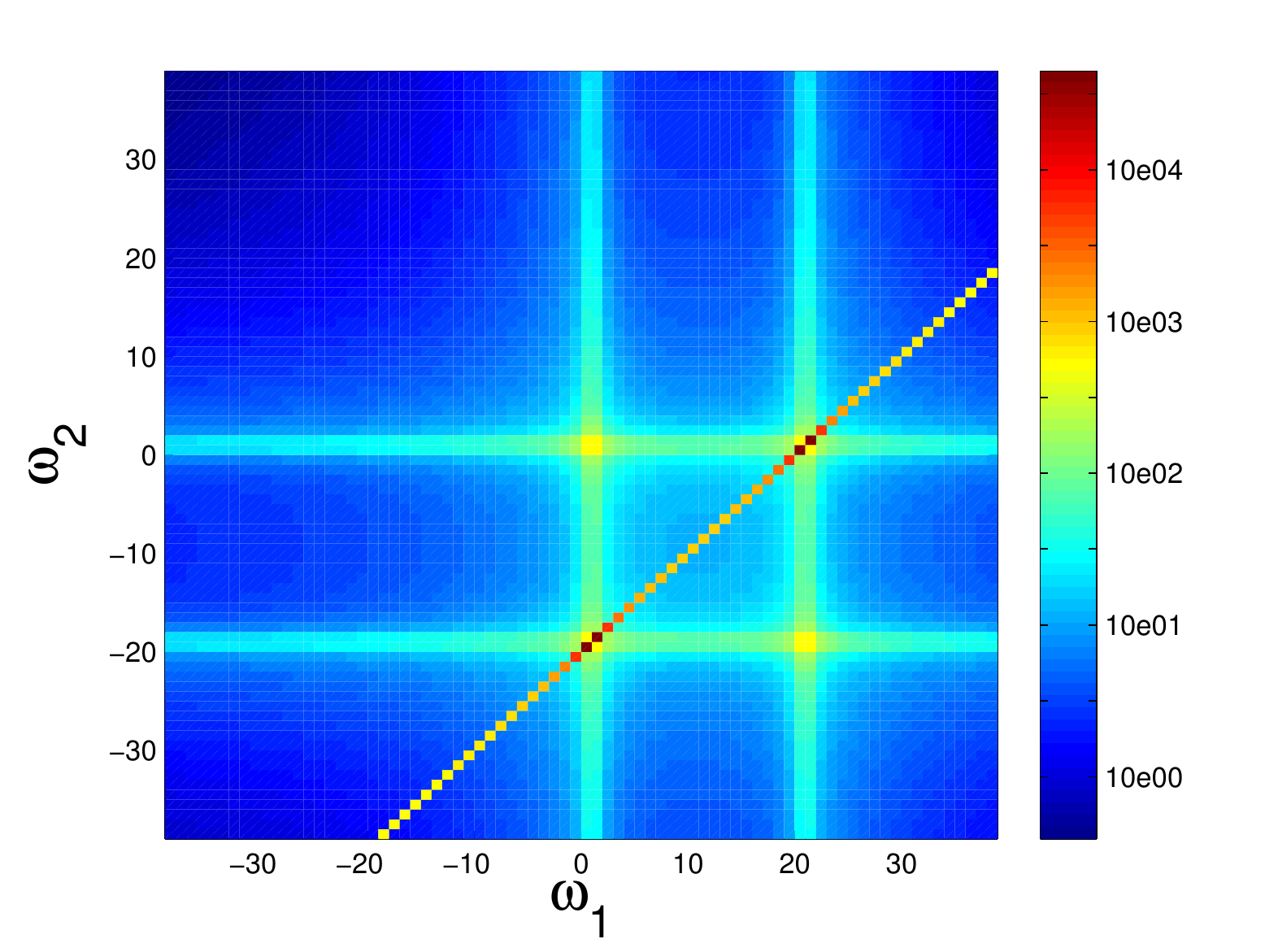}
      \includegraphics[width=0.45\textwidth]{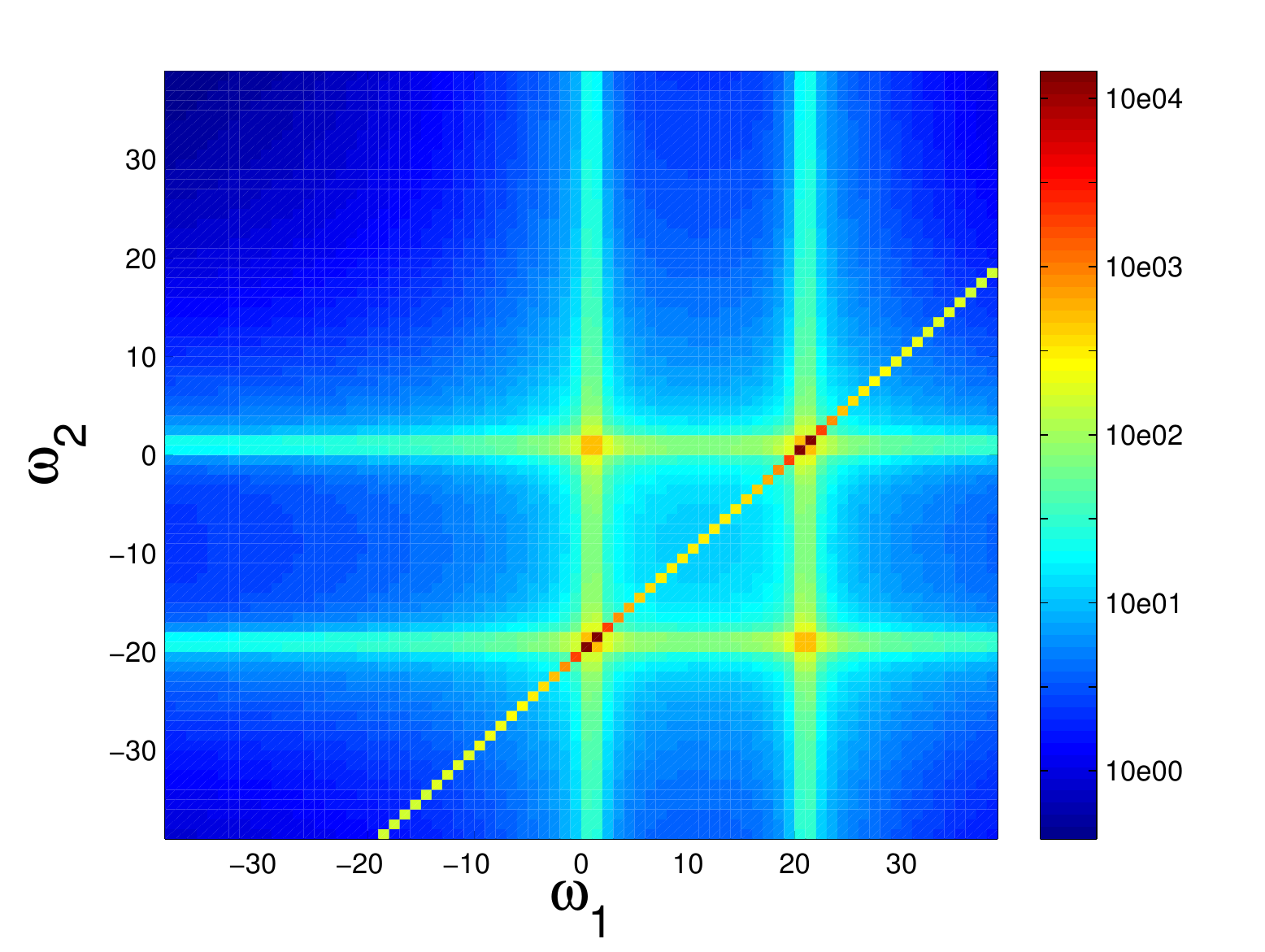}
      \caption[vertexplot1]{(color online) Absolute values of the 1PI local vertex functions for $U = 5 t$, $\beta = 30 / t$, obtained with single-site DMFT(fRG). The left plots show the density part $|\Gamma_{d}(i\omega_1,i\omega_2|i\omega_1-i\nu)-U/2|$, and the right plots the magnetic part 
      $|\Gamma_{m}(i\omega_1,i\omega_2|i\omega_1-i\nu)+U/2|$. Upper panel: Transfer frequency $\nu=0$, Lower panel: $\nu=40\frac{\pi}{\beta}$. The frequencies are signed by their Matsubara index.}     
      \label{vertexplot1}
\end{figure*}

Next let us discuss the frequency structure of the local, one-particle irreducible (1PI) interaction vertex (defined in Appendix \ref{sec:appendixa}, Eq. (\ref{eqrelationcongreensfunctionvertex})) at the converged DMFT solution as it comes out of the fRG flow that embeds the core into the lattice. In Fig. \ref{vertexplot1} we show the density part of this 1PI local vertex, $|\Gamma_{d}(i\omega_1,i\omega_2|i\omega_1-i\nu,i\omega_2+i\nu)-U/2|$, and the
magnetic part, $|\Gamma_{m}(i\omega_1,i\omega_2|i\omega_1-i\nu,i\omega_2+i\nu)+U/2|$,  for $U = 5 t$ and $\beta = 30 / t$
as functions of the incoming frequencies $\omega_1$ (x-axis) and $\omega_2$ (y-axis). The decomposition of the general vertex into the density and magnetic part is described in Appendix B, in Eqs. \ref{eqdensitypart} and \ref{eqmagneticpart}.
The outgoing frequencies are parametrized
by the bosonic Matsubara frequency $\nu$. We show the two cases $\nu=0$ and $\nu=40\frac{\pi}{\beta}$. To visualize the frequency structure better we subtracted the frequency independent  term $U/2$ ($-U/2$) from the density (magnetic) part. For particle hole symmetry, these vertices are purely real.

Note that the connected part of the dynamic charge and spin susceptibilities $\chi^{\text{charge/spin},c}(i\nu)$ is obtained from the connected two-particle Green's function $\mathcal{G}^{c,(2)}_{d/m}(i\omega_1,i\omega_2|i\omega_1-i\nu,i\omega_2+i\nu)$
by summations with respect to $\omega_1$ and $\omega_2$ (cf. Eq. \ref{eqdynamicchargesusceptibilitydef} and \ref{eqdynamicspinsusceptibilitydef}).
 $\mathcal{G}^{c,(2)}_{d/m}$ and $\Gamma_{d/m}$ are related by Eq. \ref{eqrelationcongreensfunctionvertex}, from which follows that the frequency structure of $\Gamma_{d/m}$ determines the local charge and spin response of the system.

The main features of the obtained frequency structure correspond to those that are already visible in the single-site Hubbard vertex (\ref{eqlocalvertexdensity}) and (\ref{eqlocalvertexmagnetic}) at half filling that describes the response of a free spin 1/2. This is of course expected, because the insulating phase in the single-site DMFT is given by a paramagnetic insulator with local uncoupled spin degrees of freedom.

In all vertices of Fig. \ref{vertexplot1} one recognizes a sharply peaked diagonal structure for $\omega_2 = \omega_1 - \nu$. 
In the single-site Hubbard vertex (\ref{eqlocalvertexdensity}) and (\ref{eqlocalvertexmagnetic}) this corresponds to the term proportional to $\delta_{\omega_2,\omega_{1'}}$.
In the DMFT vertex for the embedded site, it remains very sharp and no broadening is observed. As discussed in Ref. \onlinecite{Roh12}
it diverges in the Mott phase for $T \rightarrow 0$, which explains the strong enhancement of this structure at these low temperatures.

The first $\delta$-term in the single-site Hubbard vertex (\ref{eqlocalvertexdensity}) and (\ref{eqlocalvertexmagnetic}), that is proportional to $\delta_{\omega_1,-\omega_2}$, would lead to an additional peak
structure on the secondary diagonal in the $\omega_1,\omega_2$-plane. But for repulsive interactions $U > 0$ it is exponentially suppressed already for the isolated single site. Hence, also in the DMFT vertex no such structure is obtained.

The last term in the single-site Hubbard vertex (\ref{eqlocalvertexdensity}) and (\ref{eqlocalvertexmagnetic}) proportional to $\delta_{\omega_1,\omega_{1'}}$ only gives a contribution for $\nu=0$. In the density part this contribution is not visible, because this term is again exponentially suppressed. In the magnetic part it is finite and occurs in the
DMFT vertex as large difference in the offset between $\nu=0$ and $\nu\neq0$ (right column of Figure \ref{vertexplot1}). This difference leads to a term $\propto\delta_{\nu,0}$ in the spin susceptibility,
which will be discussed further below. 

Furthermore there is a '+'-shaped cross structure in the DMFT vertex, that is centered at $(\omega_1=0,\omega_2=0)$ (for $\nu=0$). At nonzero $\nu=40\frac{\pi}{\beta}$ four of those structures can be found, centered at 
$(0,0)$, $(-\nu,0)$, $(-\nu,\nu)$ and $(0,\nu)$. In the local Hubbard vertex these correspond to the terms proportional to $U^3$ and $U^5$.

Summarizing these observations we can state that the 1PI interaction vertex is by no means a structureless object. At least for this insulating regime it appears difficult to parametrize the vertex in a simple way. In particular, the cross structures indicate that a parametrization in terms of bosonic transfer frequencies does not capture the vertex in all aspects.

\begin{figure}[htbp]
    \centering
      \includegraphics[width=0.5\textwidth]{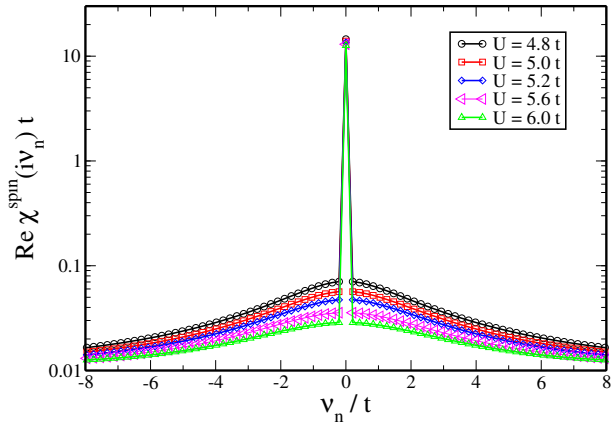}
      \caption[spinsusceptibilitymatsubarainsulatingphase]{(color online) Real part of the Matsubara local spin susceptibility in the insulating regime of single-site DMFT(fRG) for $U = 4.8 t, 5 t, 5.2 t, 5.6 t, 6 t$ at $\beta=30 / t$.}     
      \label{pictspinsusceptibilitymatsubarainsulatingphase}
\end{figure}

\begin{figure}[htbp]
    \centering
      \includegraphics[width=0.5\textwidth]{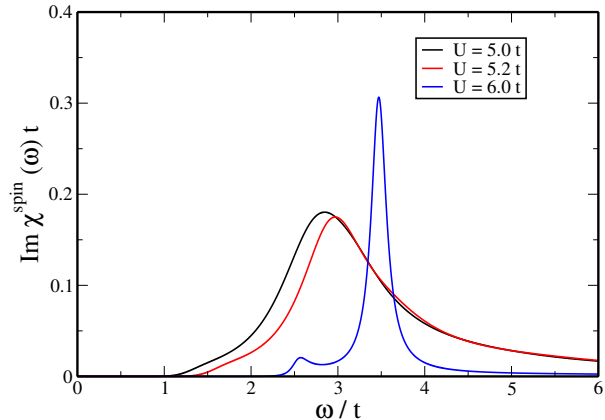}
      \caption[spinsusceptibilityinsulatingphase]{(color online) Imaginary part of the spin susceptibility on the real frequency axis in the insulating regime of single-site DMFT(fRG) for $U=5 t, 5.2 t, 6 t$ at $\beta=30 / t$.}     
      \label{pictspinsusceptibilityinsulatingphase}
\end{figure}

In order to see that these vertices make physical sense, we now compute the local dynamical spin susceptibility from the 1PI vertex, by Eq. \ref{eqdynamicspinsusceptibilitydef}. Due to our finite frequency patching (we included $200$ Matsubara frequencies at $\beta=30 / t$) our results become inaccurate especially for large frequencies because of the different speed of convergence of the connected and the disconnected part of the susceptibility. Nevertheless we obtain reasonable results by an analytical continuation of our data at least at low frequencies. In Fig.\ref{pictspinsusceptibilitymatsubarainsulatingphase} we show the real part of the spin susceptibility on the Matsubara axis (the imaginary part vanishes due to particle hole symmetry). Beside a continuous frequency dependence at nonzero frequencies we obtain an additional term proportional to $\delta_{\nu,0}$, which is characteristic for a free spin degree of freedom. This feature is already visible in the spin susceptibility of the local Hubbard model (\ref{eqspinsusceptibilitylocalhubbardsite}).
It does not occur in the imaginary part of the spin susceptibility on the real frequency axis, shown in Fig. \ref{pictspinsusceptibilityinsulatingphase}, because this vanishes at $\omega=0$ due to $\text{Im}\chi^{\text{spin}}(\omega)=-\text{Im}\chi^{\text{spin}}(-\omega)$. In this quantity we obtain a broad spectrum of spin excitations with an onset of twice the single-particle gap in agreement with Ref. \onlinecite{Raa09} or the data shown in Fig. \ref{pictspectraldensityinsulatingphase}. The two peak structure for $U = 6t$ could be an artifact of our approximation. Note that in this single-site DMFT approach nonlocal collective spin excitations that should appear below the particle hole continuum are not included.


\subsubsection{Metallic phase}
\begin{figure}[htbp]
    \centering
    \includegraphics[width=0.5\textwidth]{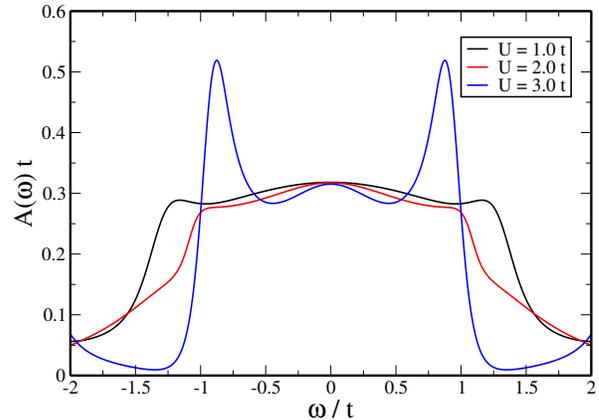}
    \caption[spectraldensitymetallicphase]{(color online) Single-particle spectral density for  $U = 1t, 2t, 3t$ at $\beta = 30 / t$ in the single-site DMFT(fRG)-solution of the Bethe lattice.}     
    \label{pictspectraldensitymetallicphase}
\end{figure}

Next let us explore the results of single-site DMFT(fRG) for the metallic regime of the Bethe lattice Hubbard model, using the scheme presented in Sec. \ref{metphase}. 
In Figure \ref{pictspectraldensitymetallicphase} we show the spectral density for $U = 1t, 2t, 3t$ at $\beta = 30 / t$. In all cases
we get only stable Pad$\acute{\text{e}}$-results for frequencies  $|\omega|<2t$. The spectral weight at $\omega=0$ is pinned to the
noninteracting value $A(\omega=0)=\text{DOS}(\omega=0)=\frac{1}{\pi t}$, which is for $T=0$ expected from Luttinger's theorem.\cite{Mue89,Mue89b}
Here we find it also for nonzero temperature values.
The shoulders at the sides of the quasi-particle are located near the position of the low energy peaks at energies $\pm\frac{1}{4}\left(\sqrt{U^2+64 z(U) v^2}-\sqrt{U^2+16 z(U) v^2}\right)$ in
the spectrum of the $L=1$-core and remain as artifacts in the DMFT spectra (cf. the discussion in Sec. V.B. in Ref. \onlinecite{Kin13}).

\begin{figure*}[htbp]
    \centering
      \includegraphics[width=0.45\textwidth]{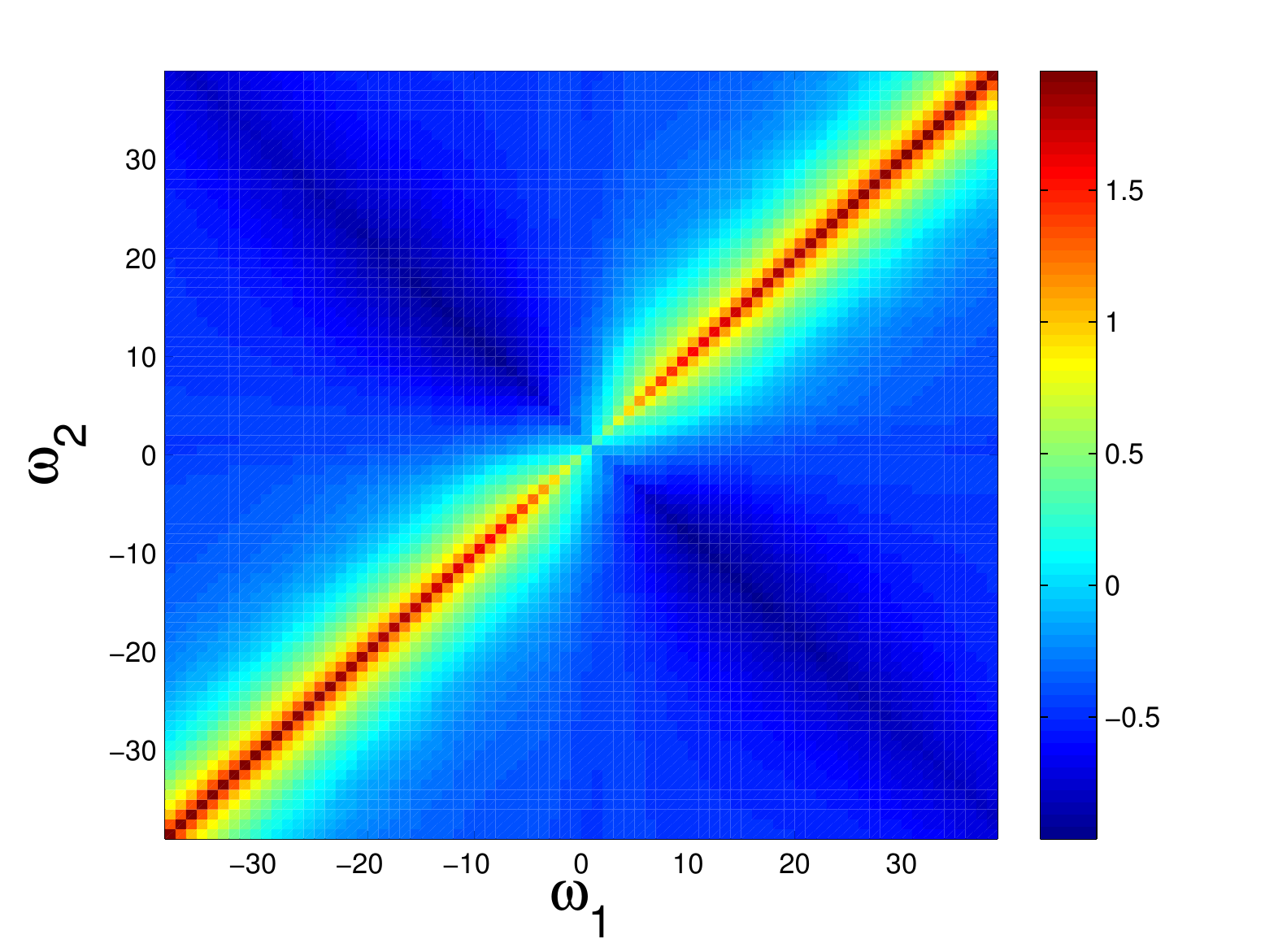}
      \includegraphics[width=0.45\textwidth]{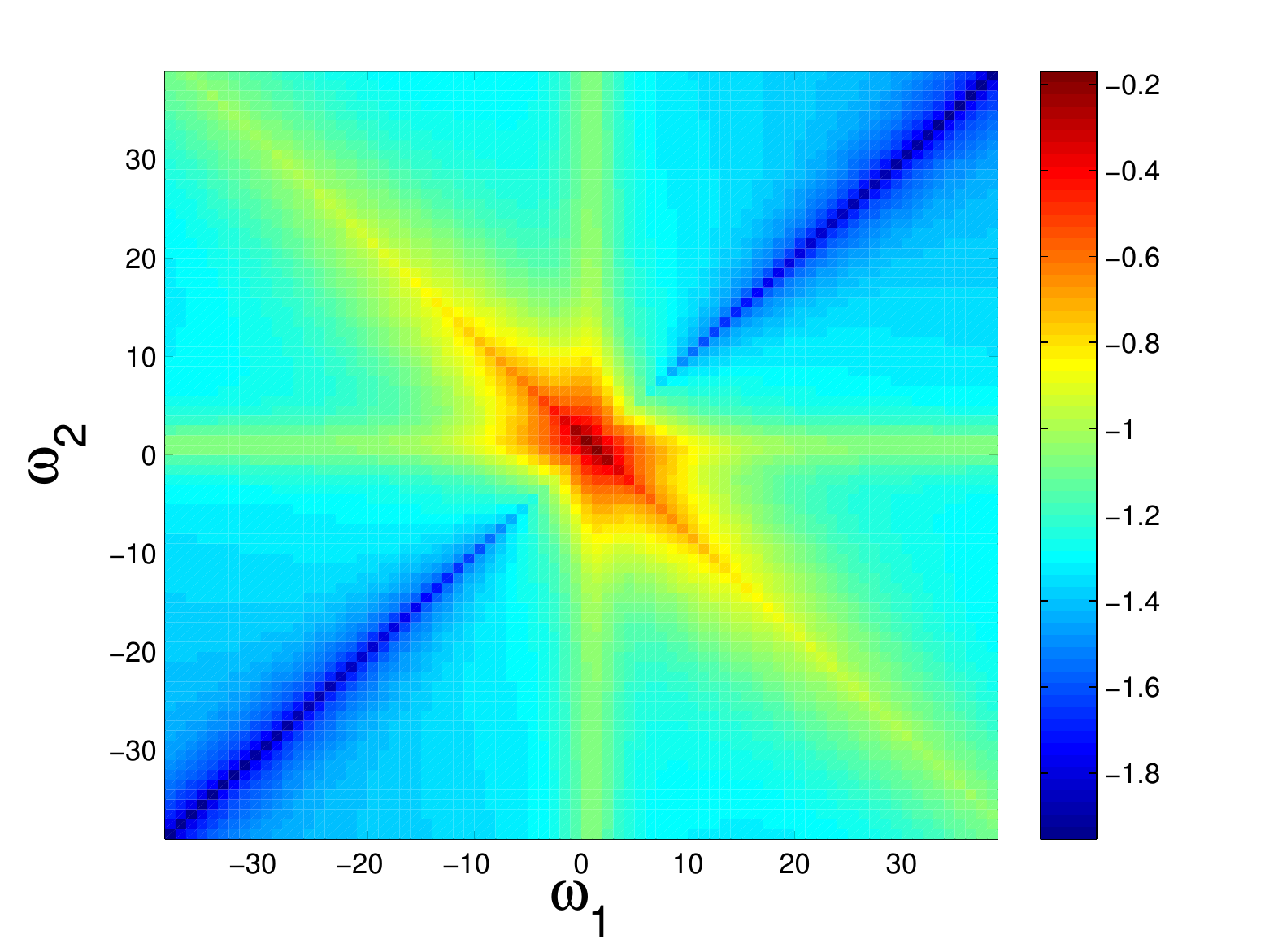}
      \includegraphics[width=0.45\textwidth]{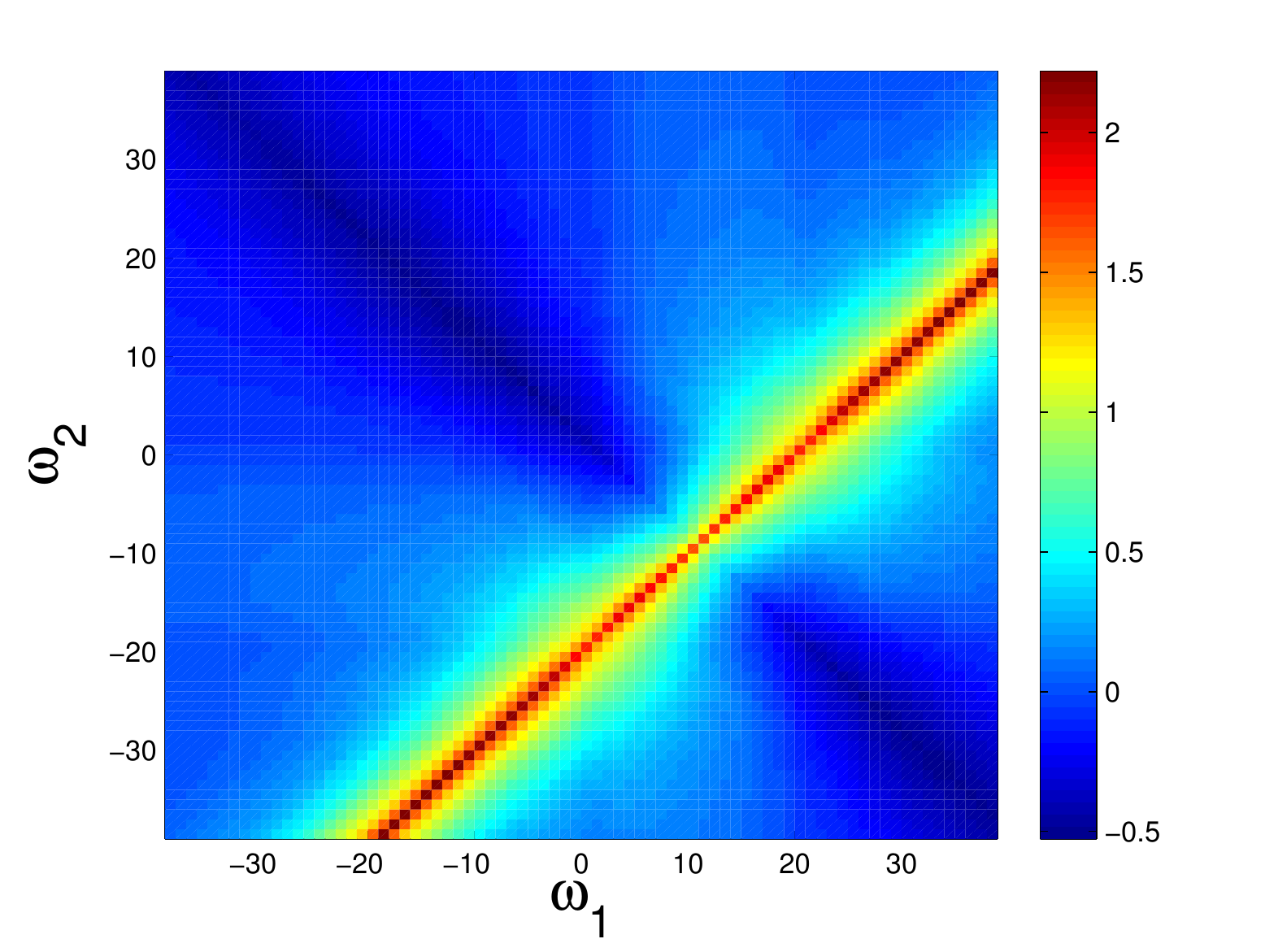}
      \includegraphics[width=0.45\textwidth]{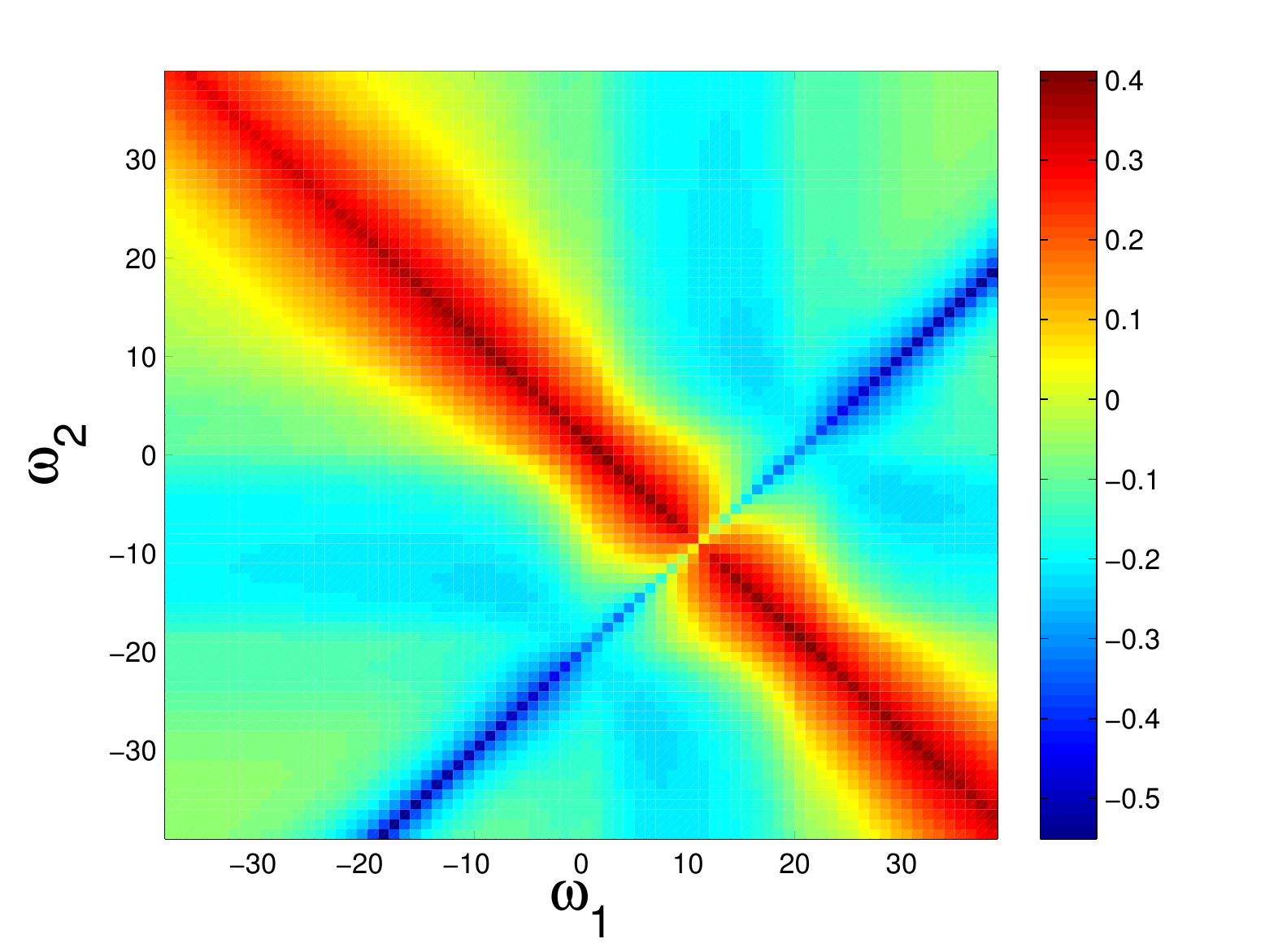}
      \caption[vertexplot2]{(color online) 1PI vertex functions in the metallic solution of single-site DMFT(fRG) for $U = 2 t$, $\beta = 30 / t$. The density part $\Gamma_{d}(i\omega_1,i\omega_2|i\omega_1-i\nu)-U/2$ is shown on the left and magnetic part 
      $\Gamma_{m}(i\omega_1,i\omega_2|i\omega_1-i\nu)+U/2$ on the right. Upper panel: $\nu=0$, Lower panel: $\nu=40\frac{\pi}{\beta}$. The frequencies are signed by their Matsubara index.}     
      \label{vertexplot2}
\end{figure*}

In Fig. \ref{vertexplot2} the density part $\Gamma_{d}(i\omega_1,i\omega_2|i\omega_1-i\nu,i\omega_2+i\nu)-U/2$ and the
magnetic part $\Gamma_{m}(i\omega_1,i\omega_2|i\omega_1-i\nu,i\omega_2+i\nu)+U/2$ of the 1PI vertex function for $U = 2 t$ and $\beta = 30 / t$
are shown. Again, the vertices are purely real due to particle hole symmetry.

The main features of the frequency structure described above for the insulating phase are also visible in the metallic phase, but there
are also certain differences. It can be clearly seen that now the vertices are continuous in the whole frequency plane and no sharp $\delta$-like features or singularities, as in the insulating phase, occur.

On the main diagonal at $\omega_2=\omega_1 - \nu$ one observes again a pronounced structure, which is much more broadened compared to the
insulating phase. In addition there is a similar structure on the secondary diagonal at $\omega_1=-\omega_2$, which was absent in
the insulating phase. As discussed in Ref. \onlinecite{Roh12} these features stem diagrammatically from particle hole and particle-particle
scattering processes respectively. Both are already visible in the vertices of the $L=1$-core and become only more pronounced in the fRG flow.

There is also a '+'-shaped structure at the same position as in the insulating phase. As seen in the lower panel of Fig. \ref{vertexplot2} 
this structure evolves into a band with width $|\nu|$ for $\nu=40\frac{\pi}{\beta}$. In perturbation theory these structures correspond to third-order diagrams \cite{Roh12} which involve mixing of particle-particle and particle-hole bubbles. No such structures occur in the vertices of the $L=1$-core. This means that they are generated entirely in the fRG flow that accomplishes the embedding into the lattice.

We compared our vertex data with DMFT(ED)-vertices, calculated by the Vienna Group\cite{Roh12}  for the same set of parameters. All described features are also visible in the frequency structure of the DMFT(ED)-vertices and even their relative size and sign are qualitatively reproduced in our scheme. Quantitatively there are differences.
For example the vertical structure at $\omega_2=\omega_1-\nu$ is broadened and its absolute size comes out smaller in our scheme. 

Summarizing the description of the single-site vertices, we can state that both in insulating as well as in the metallic state, the interaction vertices exhibit a lot of structure. The bosonic (diagonal) features could be captured by simpler parametrizations using functions depending on certain transfer frequencies only,\cite{Kar08} but other features like the '+'-structures would not be captured by that. 
In Ref. \onlinecite{Roh12} the decomposition of the 1PI vertex into  two-particle irreducible (2PI) vertices and the fully irreducible vertex is discussed. We have reproduced this reasoning for some examples. In the 2PI vertices, certain bosonic features are removed, but other bosonic features due to the channel coupling remain, e.g. in the particle-particle 2PI vertices one still sees sharp features for specific frequency transfers that originate from particle-hole insertions. The fully irreducible vertex has a nontrivial frequency structure as well.\cite{Roh12,Schae13}

\subsection{Two-site cluster DMFT}
\label{sec:resultstwositedmft}
\begin{figure*}[htbp]
    \centering
      \includegraphics[width=0.5\textwidth]{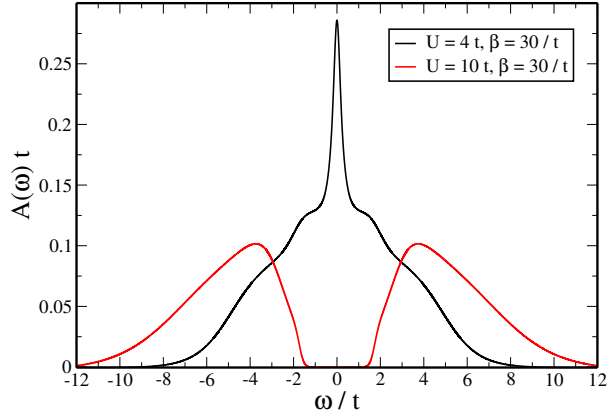}
      \caption[spectraldensitytwositecluster]{(color online) Local single-particle spectral density $A(\omega)=-\frac{1}{\pi}\text{Im}\ \mathcal{G}_{ii}(\omega+i0^{+})$ for $U = 4 t$ and $U = 10 t$ at $\beta=30/t$, obtained by two-site DMFT(fRG).}     
      \label{pictspectraldensitytwositecluster}
\end{figure*}

In Fig. \ref{pictspectraldensitytwositecluster} we show the local spectral density $A(\omega)=-\frac{1}{\pi}\text{Im}\ \mathcal{G}_{ii}(\omega+i0^{+})$ for $U = 4 t$ and $U = 10 t$ at $\beta=30/t$. 
Unlike for the single-site DMFT(fRG) scheme, using the two-site cluster as core, we can describe metallic and insulating behavior with the same fRG-scheme, without having to parametrize the self-energy by a $Z$-factor. For $U = 10 t$ we find an insulating spectrum with two Hubbard bands at $\omega=\pm 5 t$ separated by a gap. In the metallic spectrum for $U = 4 t$ these Hubbard bands are still visible as weakly pronounced shoulders at $\omega = \pm 2 t$. The sharp peak at $\omega=0$ is due to the Van Hove singularity in the  free density of states of the two-dimensional square lattice. Hence the single-particle spectra are qualitatively correct and show the expected energy scales. This gives us a robust starting point for studying the 1PI interaction vertex for the two-site core, now including its nonlocal part.

As for the single-site DMFT, we discuss the frequency structure of the 1PI vertex functions for the insulating and the metallic phase in terms of the density  and magnetic parts. Note that in units of the bandwidth $W$, the onsite interaction $U$ is in both cases the same as in the data shown for the single-site DMFT. Therefore, the vertices can be directly compared to each other on the energy axis. \footnote{To be more precise, one should compare the ration $\frac{U}{\sigma}$,
where $\sigma$ is the standard deviation of the noninteracting density of states. Anyhow, one has $\sigma=t$ for the Bethe lattice and $\sigma=2t$ for the two-dimensional
square lattice, so that both criteria are equivalent in our case.} Opposite to the single-site DMFT, the two-site cluster DMFT includes antiferromagnetic fluctuations between neighbored sites. These should be characterized by the energy scale $J$ that is for large $U$ given by $J\sim 4\frac{t^2}{U}$.

By the Fourier transformation $U_{\vec{K}_{i},\vec{R}_{j}}=\frac{1}{\sqrt{2}}\exp(i\vec{K}_{i}\vec{R}_{j})$ we transform the vertices to cluster momentum space  with the cluster momenta $\vec{K}_{1}=(0,0)$ and $\vec{K}_{2}=(\pi,\pi)$.
$\vec{R}_{1}=(0,1)$ and $\vec{R_2}=(0,0)$ are shown in Fig. \ref{pictsquarelattice}.
Due to momentum conservation the only non-negative contributions are given by 
\begin{widetext}
\begin{eqnarray}
\nonumber
\Gamma_{d/m}(\vec{K}_{1},i\omega_{1};\vec{K}_{1},i\omega_{2}|\vec{K}_{1},i\omega_{1'};\vec{K}_{1},i\omega_{2'})&\equiv&\Gamma^{1111}_{d/m}(i\omega_{1};i\omega_{2}|i\omega_{1'};i\omega_{2'}),\\
\nonumber
\Gamma_{d/m}(\vec{K}_{1},i\omega_{1};\vec{K}_{2},i\omega_{2}|\vec{K}_{1},i\omega_{1'};\vec{K}_{2},i\omega_{2'})&\equiv&\Gamma^{1212}_{d/m}(i\omega_{1};i\omega_{2}|i\omega_{1'};i\omega_{2'}),\\
\nonumber
\Gamma_{d/m}(\vec{K}_{1},i\omega_{1};\vec{K}_{2},i\omega_{2}|\vec{K}_{2},i\omega_{1'};\vec{K}_{1},i\omega_{2'})&\equiv&\Gamma^{1221}_{d/m}(i\omega_{1};i\omega_{2}|i\omega_{1'};i\omega_{2'}),\\
\nonumber
\Gamma_{d/m}(\vec{K}_{1},i\omega_{1};\vec{K}_{1},i\omega_{2}|\vec{K}_{2},i\omega_{1'};\vec{K}_{2},i\omega_{2'})&\equiv&\Gamma^{1122}_{d/m}(i\omega_{1};i\omega_{2}|i\omega_{1'};i\omega_{2'})
\end{eqnarray}
\end{widetext}
and the same quantities with $\vec{K}_{1}\leftrightarrow\vec{K}_{2}$ respectively.
Due to particle hole symmetry one has $\Gamma^{2222}_{d/m}=\left(\Gamma^{1111}_{d/m}\right)^{\ast}$,
$\Gamma^{2121}_{d/m}=\left(\Gamma^{1212}_{d/m}\right)^{\ast}$, $\Gamma^{2112}_{d/m}=\left(\Gamma^{1221}_{d/m}\right)^{\ast}$ and
$\Gamma^{2211}_{d/m}=\left(\Gamma^{1122}_{d/m}\right)^{\ast}$. Hence we can restrict the discussion to the former vertices.

If we plot $\Gamma^{\#_1\#_2\#_3\#_4}_{d/m}(i\omega_1;i\omega_2|i\omega_1-i\nu;i\omega_2+i\nu)$ in the $\omega_1-\omega_2$-plane we have the symmetry axes (A) at $\omega_2=\omega_1-\nu$
and (B) at $\omega_1=-\omega_2$. $M_{\text{A}}$ and $M_{\text{B}}$ are mirror operators at axis (A) and (B) respectively. In Table \ref{tabellespiegelung}
we show the transformation behavior of $\Gamma^{1111}_{d/m}$, $\Gamma^{1212}_{d/m}$, $\Gamma^{1221}_{d/m}$ and $\Gamma^{1122}_{d/m}$ under $M_{\text{A}}$
and $M_{\text{B}}$ which follows from time reversal symmetry and particle hole symmetry.
\begin{table}
\caption[tabellespiegelung]{In the $\omega_1-\omega_2$-plane there are two symmetry axes: (A) at $\omega_2=\omega_1-\nu$ and (B) at $\omega_1=-\omega_2$
with the corresponding mirror operators $M_{\text{A}}$ and $M_{\text{B}}$ respectively. In the table we show the transformation behavior
of $\Gamma^{\#_1\#_2\#_3\#_4}_{d/m}(i\omega_1;i\omega_2|i\omega_1-i\nu;i\omega_2+i\nu)$ under $M_{\text{A}}$ and $M_{\text{B}}$.}
\begin{tabular}{|p{1.5cm}|p{1.5cm}|p{1.5cm}|}
\hline
   & $M_{\text{A}}$ & $M_{\text{B}}$ \\
\hline\hline
  $\Gamma^{1111}_{d/m}$ & $\Gamma^{1111}_{d/m}$ & $\left(\Gamma^{1111}_{d/m}\right)^{\ast}$ \\
\hline
  $\Gamma^{1212}_{d/m}$ & $\left(\Gamma^{1212}_{d/m}\right)^{\ast}$ & $\Gamma^{1212}_{d/m}$  \\
\hline
  $\Gamma^{1221}_{d/m}$ & $\Gamma^{1221}_{d/m}$ & $\Gamma^{1221}_{d/m}$  \\
\hline
  $\Gamma^{1122}_{d/m}$ & $\left(\Gamma^{1122}_{d/m}\right)^{\ast}$ & $\left(\Gamma^{1122}_{d/m}\right)^{\ast}$ \\
\hline
\end{tabular}
\label{tabellespiegelung}
\end{table}
For $\nu=0$ one can furthermore show that $\Gamma^{1221}_{d/m}\in\mathbb{R}$ and $\Gamma^{1122}_{d/m}\in\mathbb{R}$.
In presenting the data, we will restrict the discussion to the case of zero transfer frequencies $\nu$, either for the charge or the magnetic channel. Based on the experience from the single-site vertex, this data contains the main features, which would get shifted or split, but not changed drastically in the case of finite frequency transfer.
\subsubsection{Insulating phase}

\begin{figure*}[htbp]
    \centering
      \includegraphics[width=0.4\textwidth]{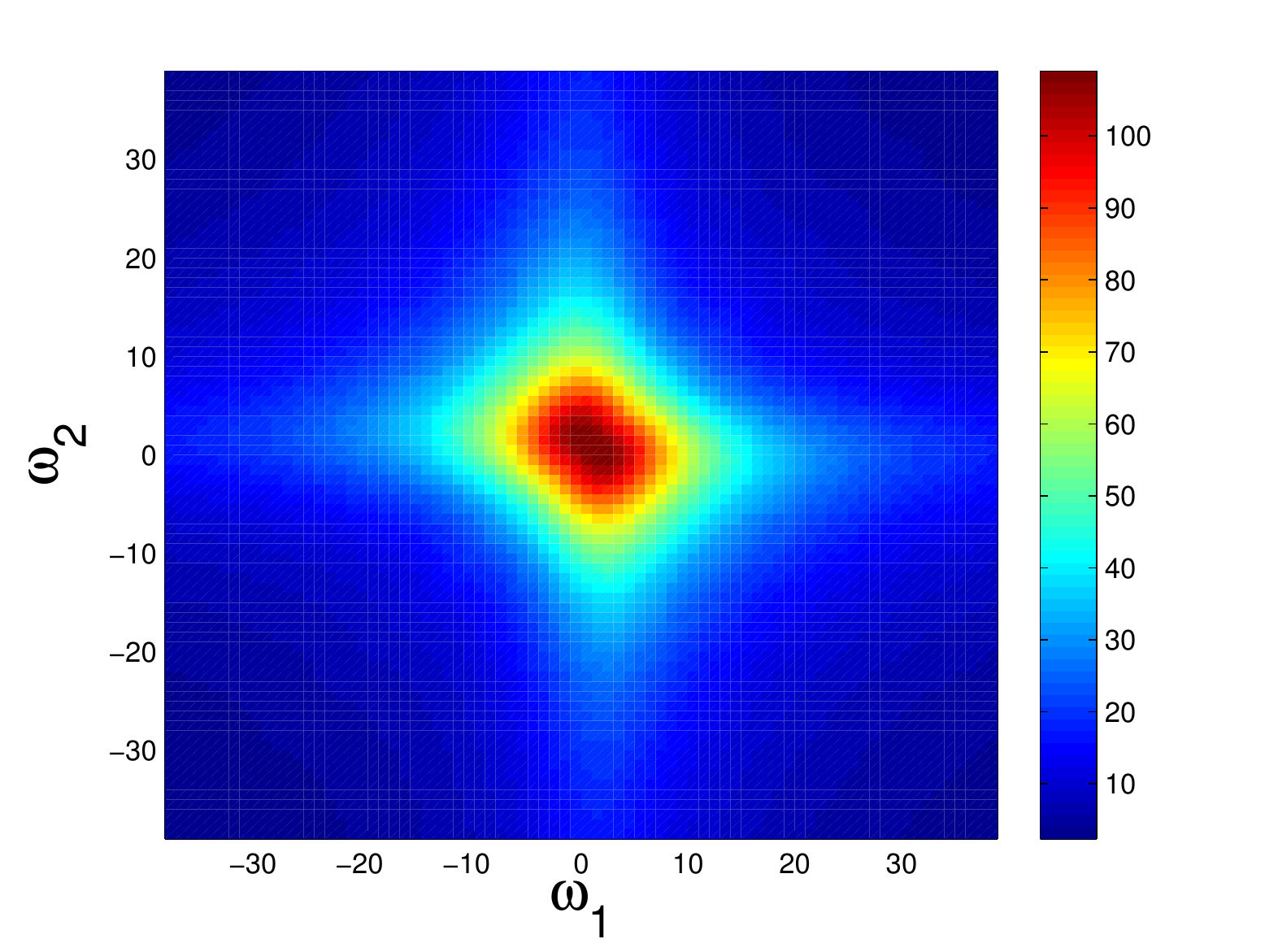}
      \includegraphics[width=0.4\textwidth]{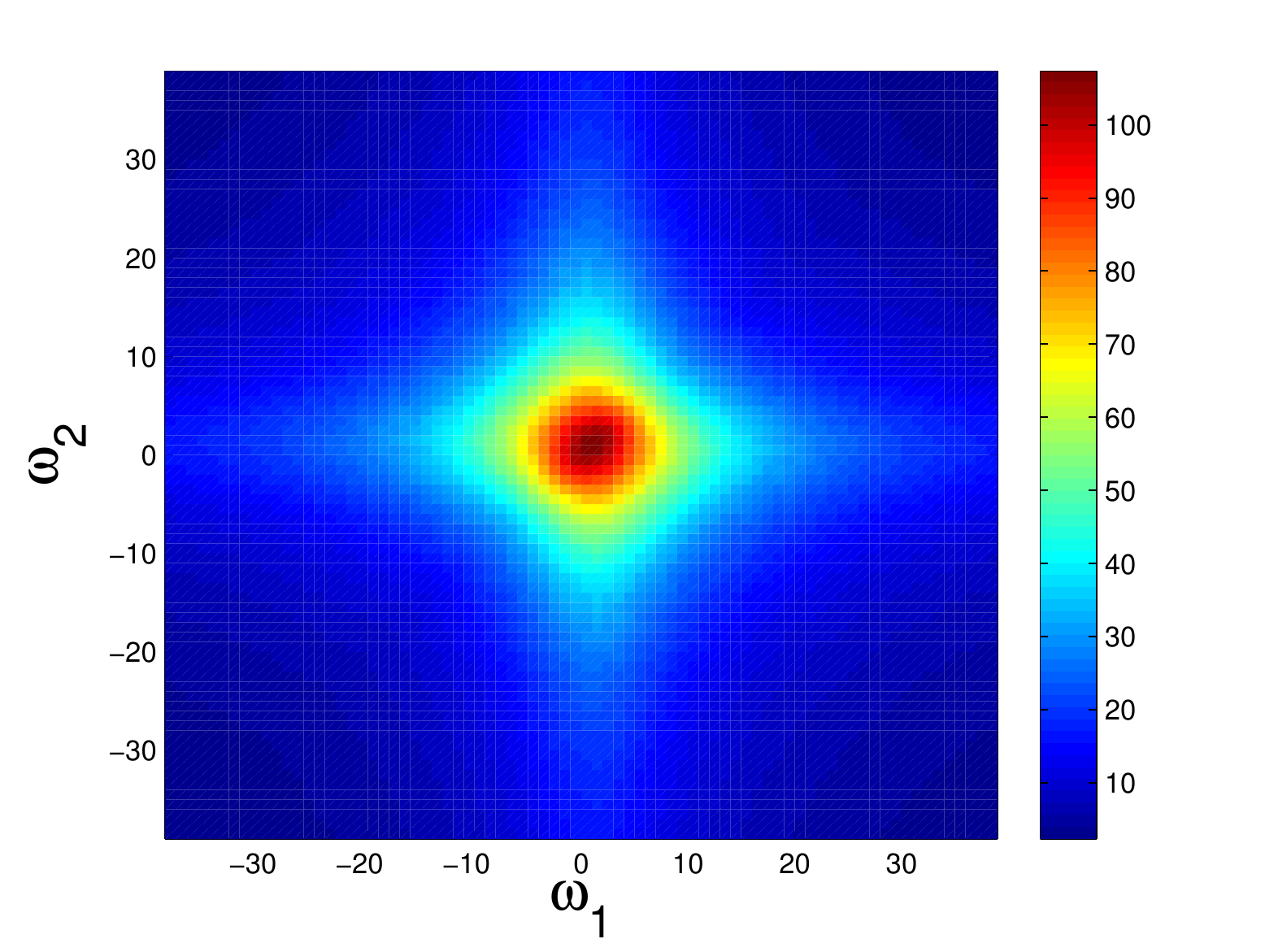}
      \includegraphics[width=0.4\textwidth]{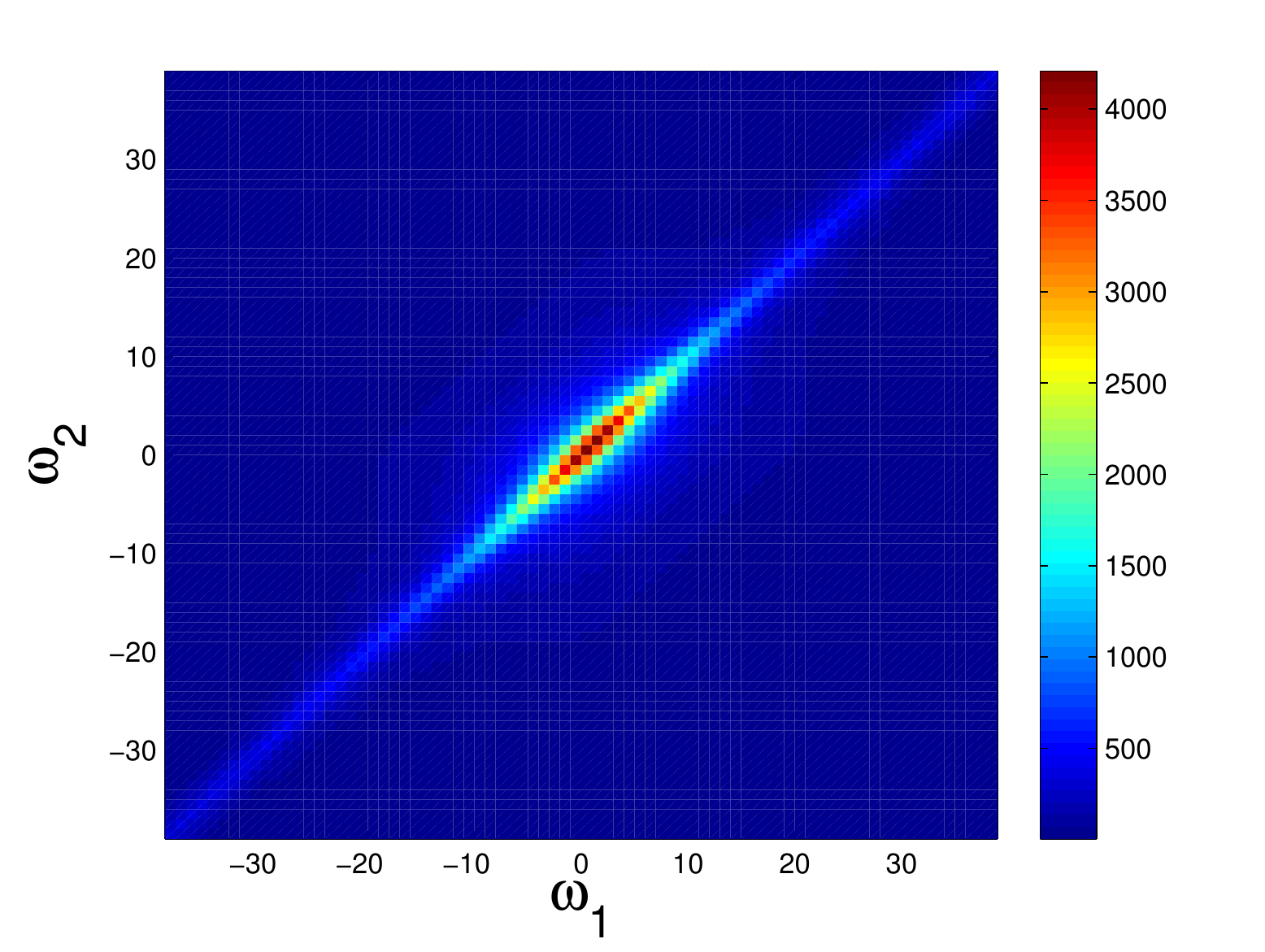}
      \includegraphics[width=0.4\textwidth]{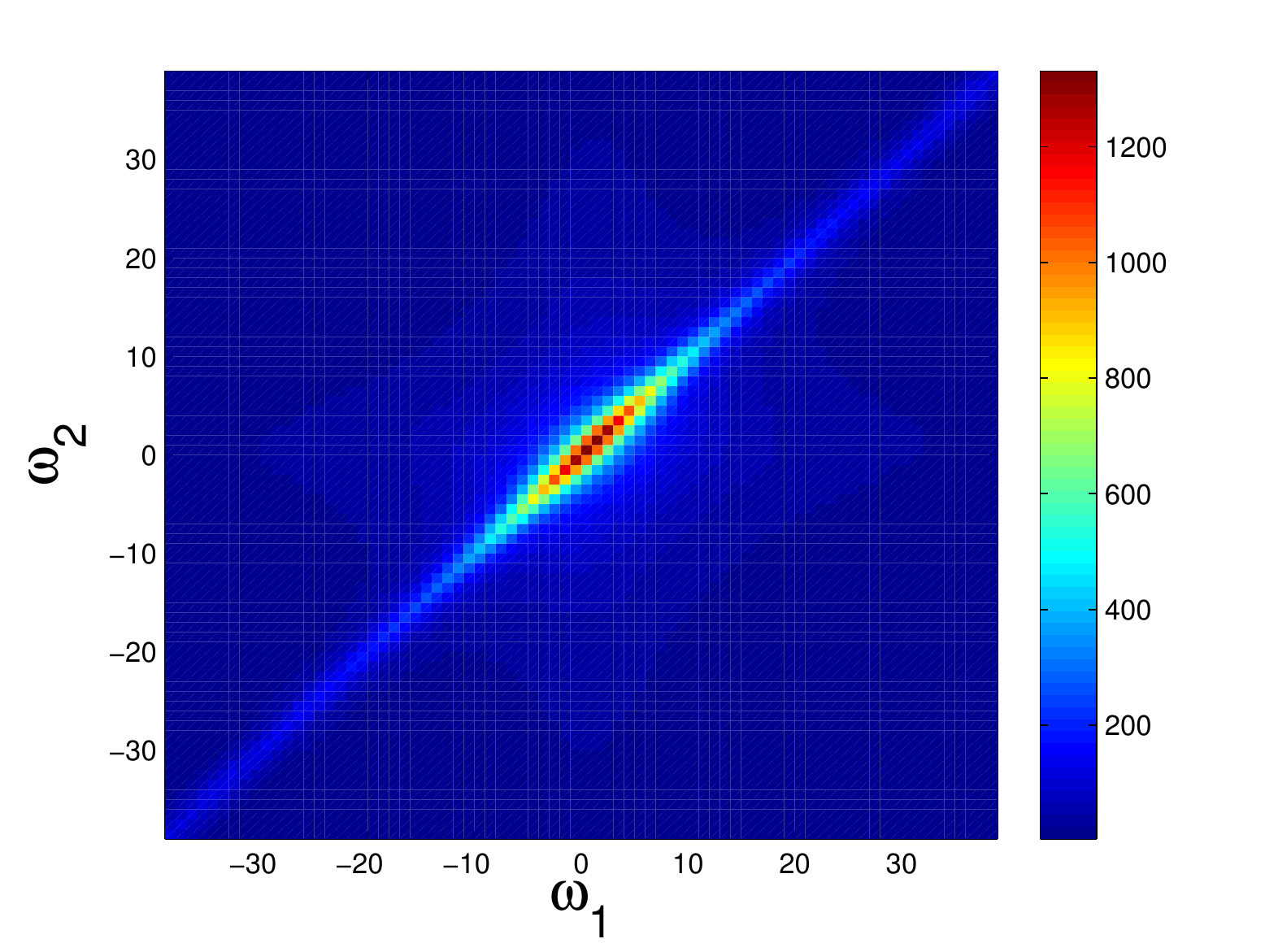}
      \includegraphics[width=0.4\textwidth]{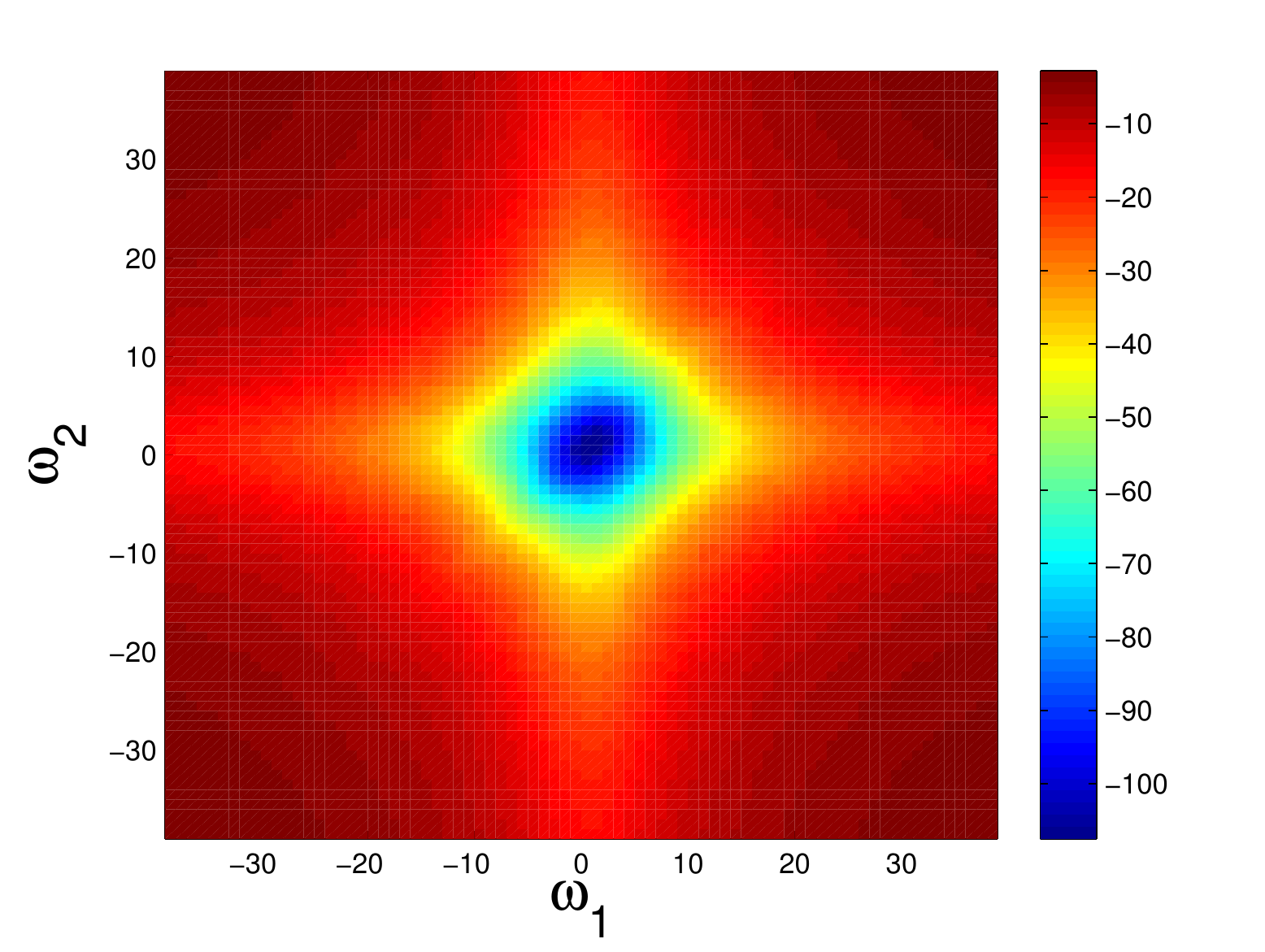}
      \includegraphics[width=0.4\textwidth]{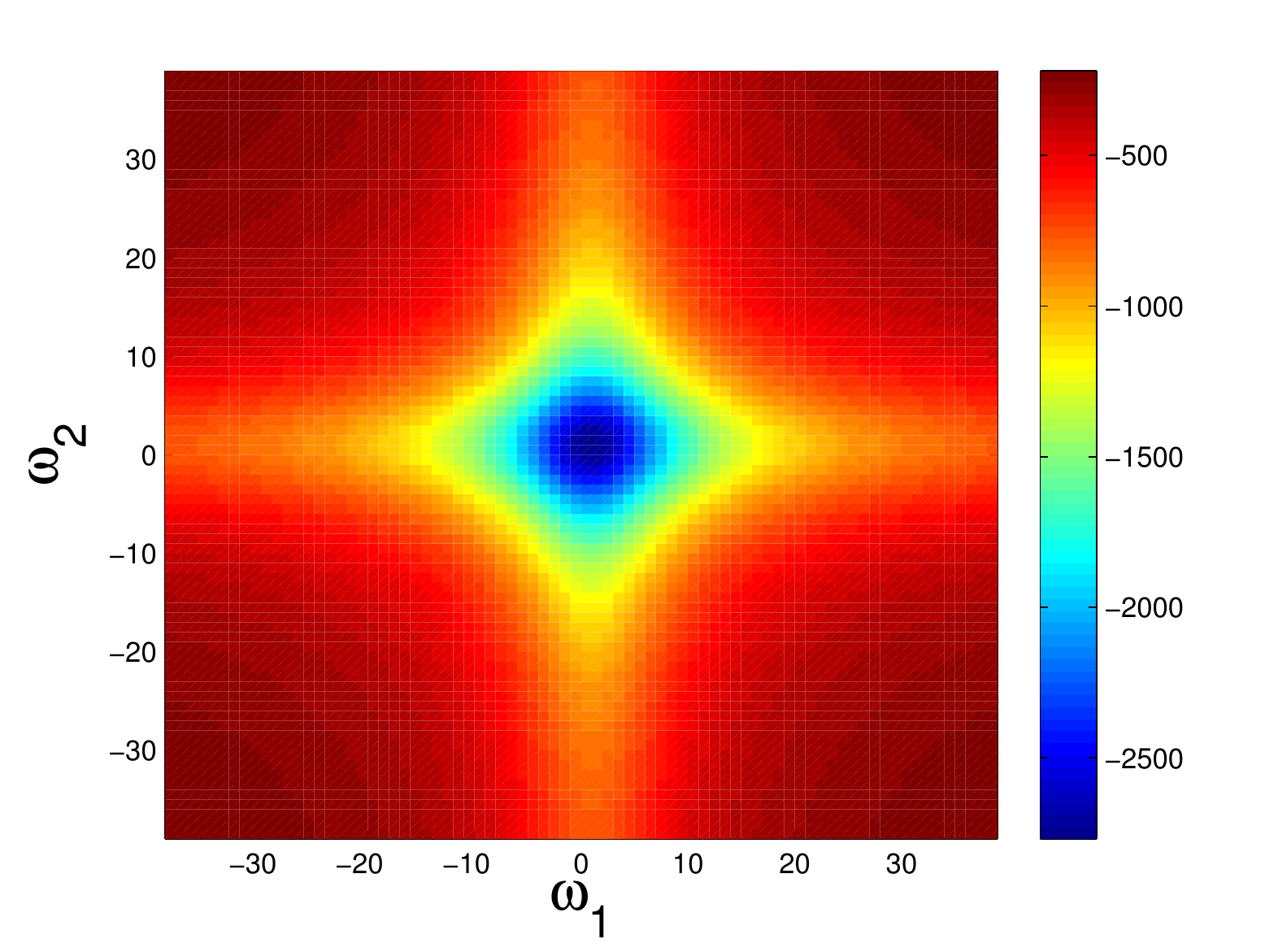}
      \includegraphics[width=0.4\textwidth]{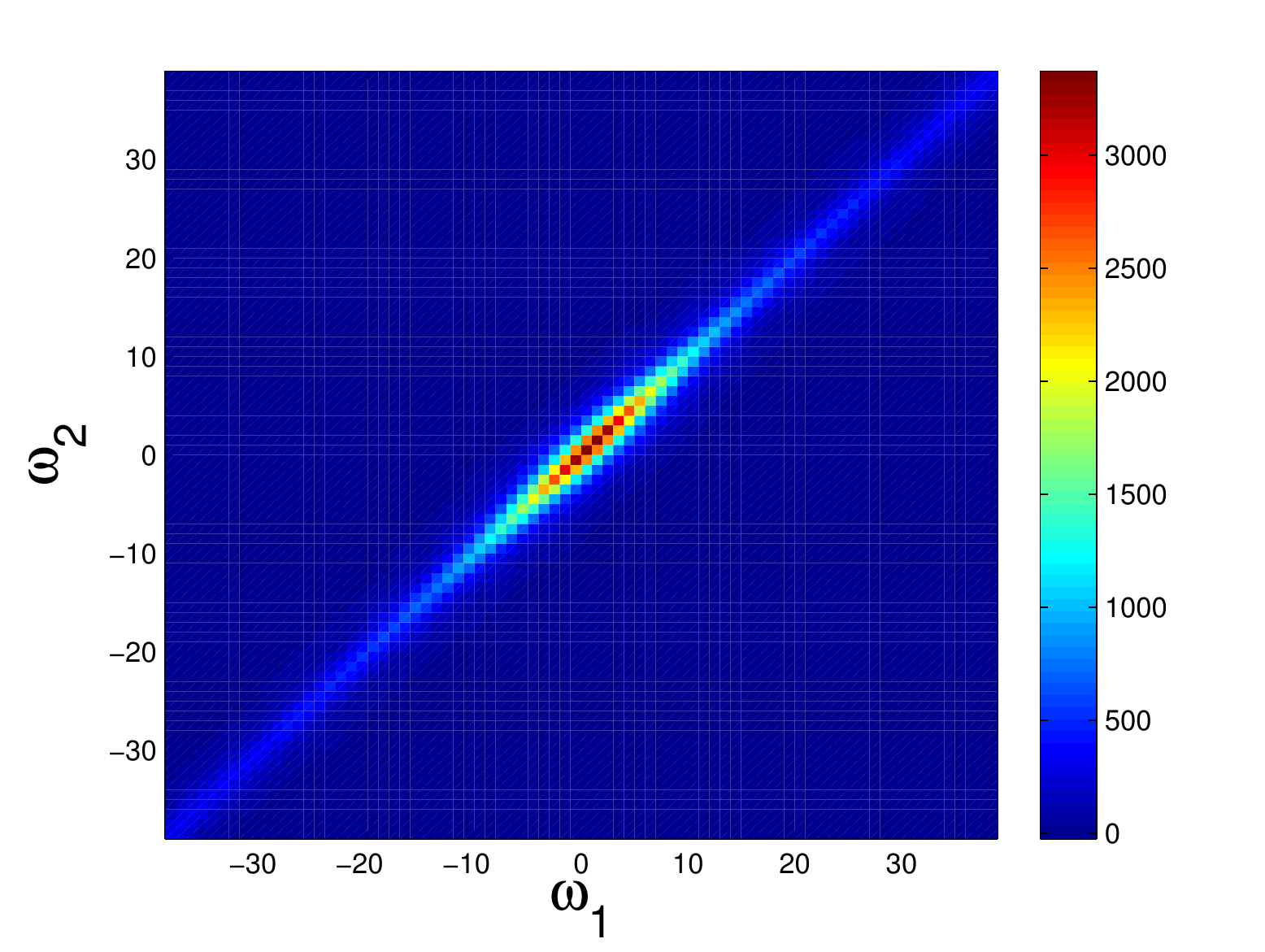}
      \includegraphics[width=0.4\textwidth]{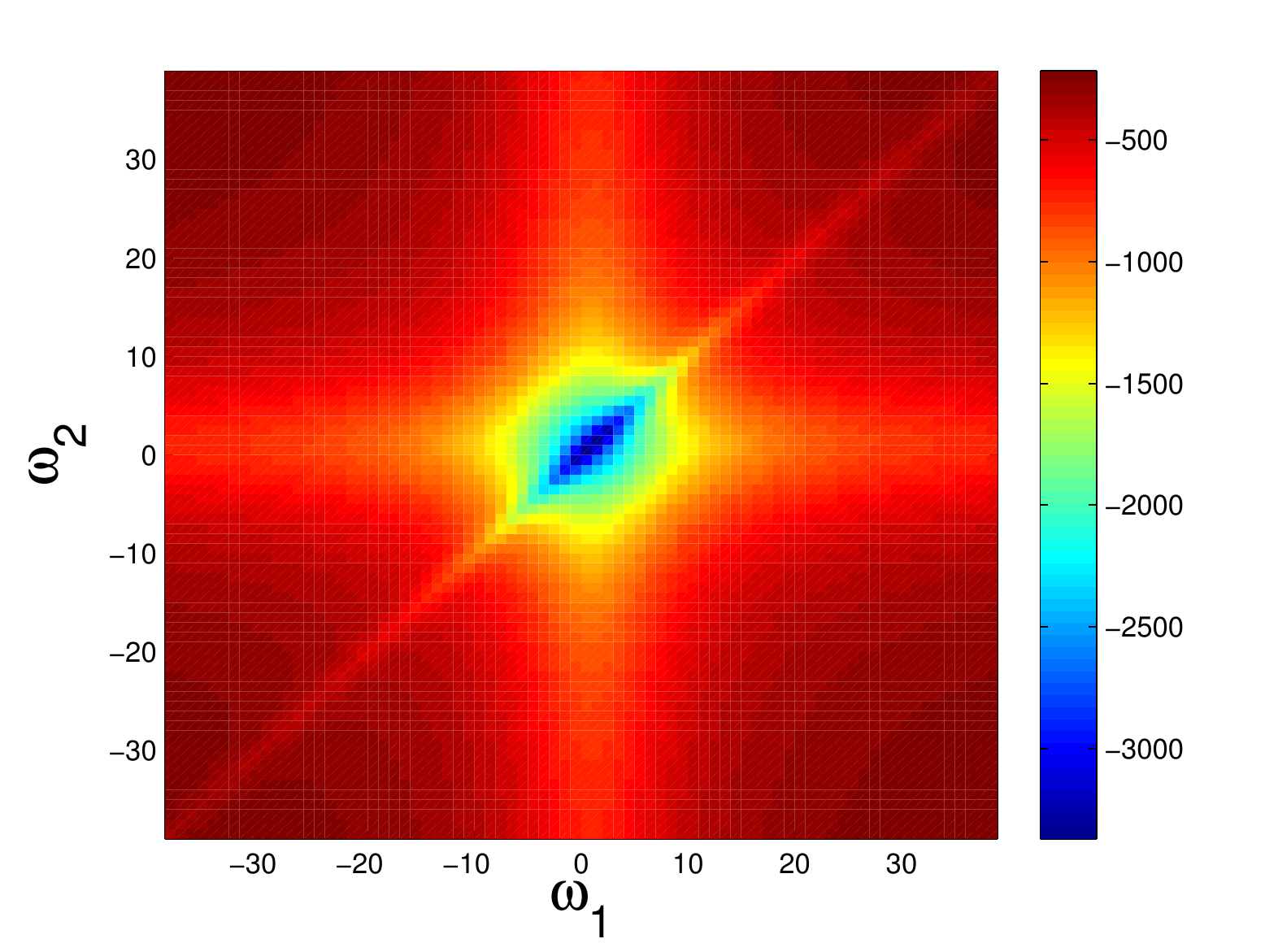}
      \caption[vertexplot3]{(color online) Vertex functions $|\Gamma^{1111}_{d/m}(i\omega_1;i\omega_2|i\omega_1;i\omega_2)\mp U/4|$, $|\Gamma^{1212}_{d/m}(i\omega_1;i\omega_2|i\omega_1;i\omega_2)\mp U/4|$,
      $\Gamma^{1221}_{d/m}(i\omega_1;i\omega_2|i\omega_1;i\omega_2)\mp U/4$ and $\Gamma^{1122}_{d/m}(i\omega_1;i\omega_2|i\omega_1;i\omega_2)\mp U/4$ for $U = 10 t$, $\beta = 30 / t$, obtained in two-site DMFT(fRG). Left column: density part. Right column: magnetic part. The frequencies are signed by their Matsubara index.}     
      \label{vertexplot3}
\end{figure*}

In Fig. \ref{vertexplot3} we show the vertices $|\Gamma^{1111}_{d/m}(i\omega_1;i\omega_2|i\omega_1;i\omega_2)\mp U/4|$, $|\Gamma^{1212}_{d/m}(i\omega_1;i\omega_2|i\omega_1;i\omega_2)\mp U/4|$,
$\Gamma^{1221}_{d/m}(i\omega_1;i\omega_2|i\omega_1;i\omega_2)\mp U/4$ and $\Gamma^{1122}_{d/m}(i\omega_1;i\omega_2|i\omega_1;i\omega_2)\mp U/4$ for $U = 10 t$ and $\beta = 30 / t$.
Since $\Gamma^{1111}$ and $\Gamma^{1212}$ are complex-valued we plot their absolute values.

In the density and magnetic part of $\Gamma^{1111}$ and $\Gamma^{1221}$, the only apparent feature is  a '+'-shaped structure, which reaches its maximum in
the center at $(\omega_1=0,\omega_2=0)$. It is much more broadened compared to the single-site DMFT (Fig. \ref{vertexplot1}) and its width increases with the interaction $U$.

The density and magnetic part of $\Gamma^{1212}$ and $\Gamma^{1122}$ are dominated by a peaked diagonal frequency structure at $\omega_1=\omega_2$, which reaches its maximum at $(\omega_1=0,\omega_2=0)$.  Except for the magnetic part of $\Gamma^{1122}$,  an additional '+'-shaped structure is  only very weakly pronounced.
Snapshots of the peaked structure at $\omega_1=\omega_2$ along or parallel to the main diagonal can be described by a Lorentzian with width $\approx J$.
This should be compared to the local vertex of the single-site DMFT (cf. Fig. \ref{vertexplot1}). Here the antiferromagnetic coupling $J$ is absent and also the peaked structure at $\omega_2=\omega_1$
is $\delta$-shaped, i.e. its width is equal to zero. This difference is mainly caused by the fact that in the two-site core, the localized spins couple antiferomagnetically and form a singlet.  The embedding of this core in the gapped bath only leads to quantitative changes, but without allowing for longer-ranged spin correlations in this cluster DMFT framework, the  singlet character does not change.
Therefore, qualitatively, the important features in the frequency structure of the embedded vertex are already visible in the vertex of the isolated two-site Hubbard model, which serves as 'core' in our cluster DMFT scheme. Hence, if one tries to describe a short-range correlated system, using a finite-site vertex of a core with qualitatively similar properties may be a good approximation or guide to look for viable parametrizations. Near phase transitions the picture may become more complicated \cite{Roh11}.

\begin{figure*}[htbp]
    \centering
      \includegraphics[width=0.45\textwidth]{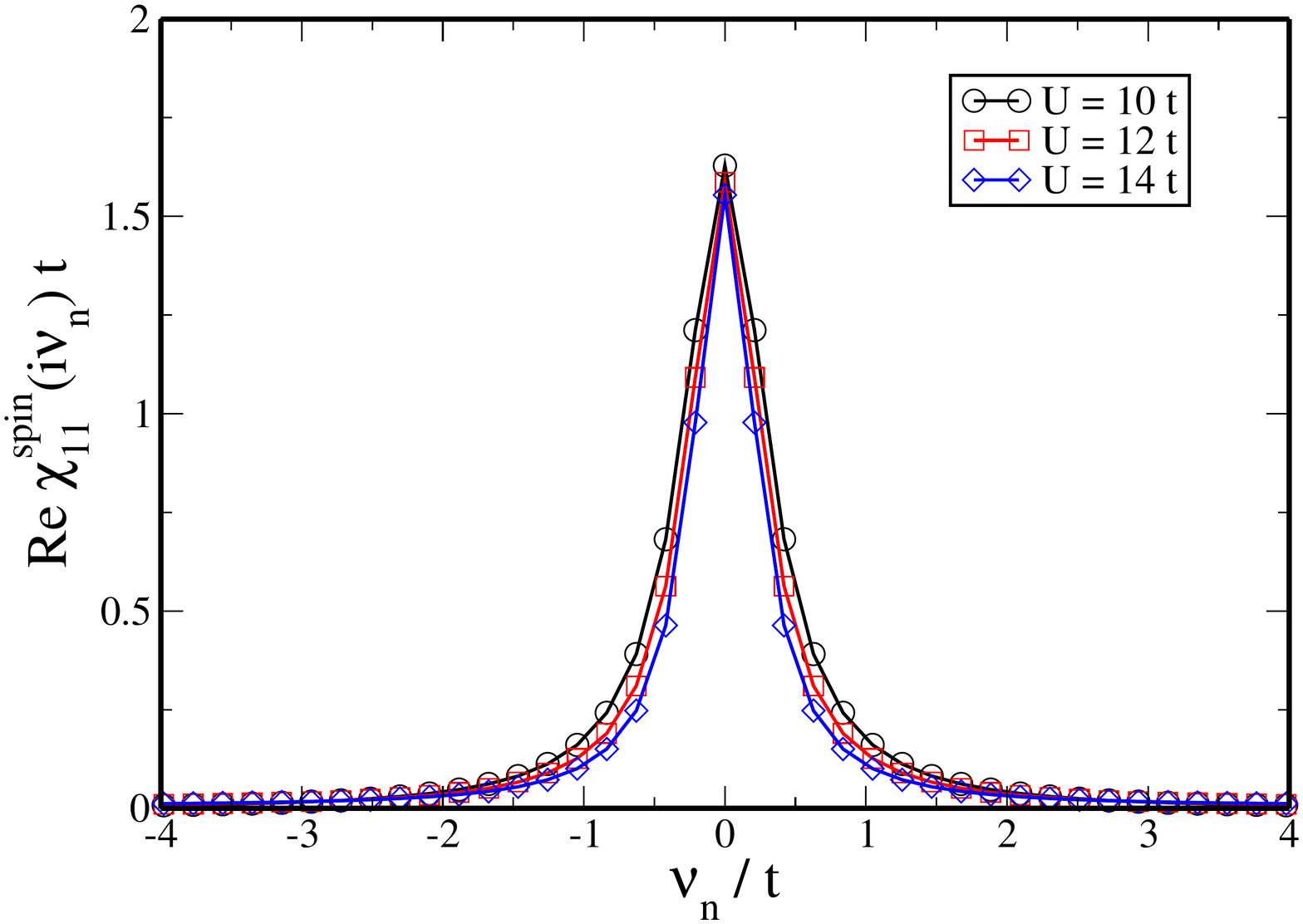}
      \includegraphics[width=0.45\textwidth]{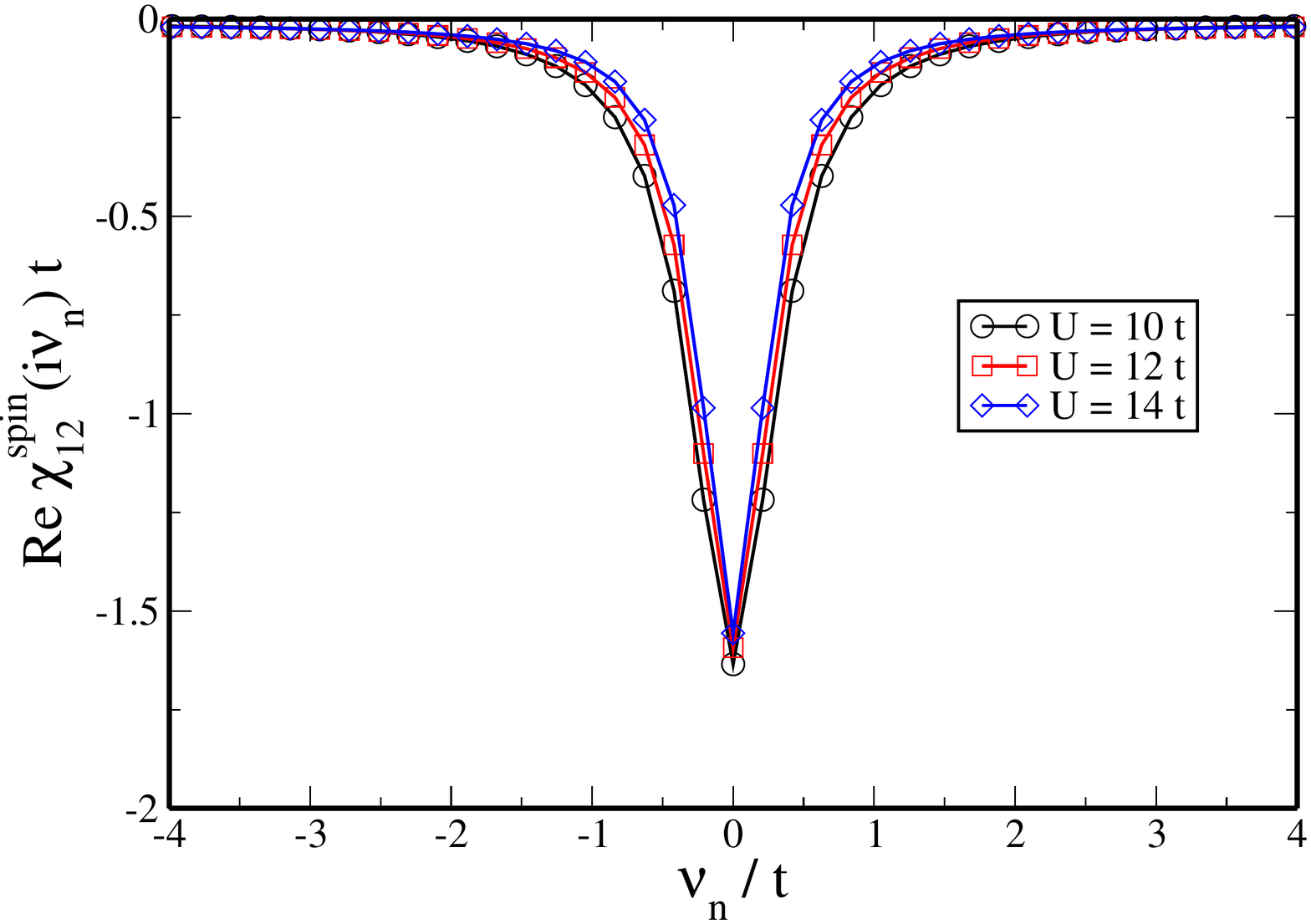}
       \caption[spinsusceptibility2sitecluster]{(color online) Spin susceptibility $\chi^{\text{spin}}_{ij}(i\nu)$ on the Matsubara-axis, obtained in the insulating regime of the two-site DMFT(fRG) for the Hubbard model on the square lattice. Left plot: local spin susceptibility. Right plot: next-neighbor spin susceptibility.}     
      \label{pictspinsusceptibility2sitecluster}
\end{figure*}

In Fig.  \ref{pictspinsusceptibility2sitecluster} we show the local and next-neighbor spin susceptibilities on the Matsubara-axis. In contrast to the single-site DMFT (cf. Figure \ref{pictspinsusceptibilitymatsubarainsulatingphase}) no term $\propto \delta_{\nu,0}$ occurs in the local spin susceptibility,
which was characteristic for a free spin degree of freedom. Now, the spin moments are screened by an antiferromagnetic exchange interaction. The Pad$\acute{\text{e}}$-spectra show sharp spin excitations at certain values $\pm \Delta E^{\text{spin}}_{ij}$ and the Matsubara data are consistent with a functional dependence of the form $\chi^{\text{spin}}_{ij}(i\nu)\sim (-1)^{(i-j)} \frac{\Delta E^{\text{spin}}_{ij}}{\nu^2+(\Delta E^{\text{spin}}_{ij})^2}$. The spin excitation energy in the two-site Hubbard model is given by  $\Delta E^{\text{spin}}_{\text{2-site},11}=\Delta E^{\text{spin}}_{\text{2-site},12}=\Delta E^{\text{spin}}_{\text{2-site}}=\left(\sqrt{U^2+16 t^2}-U\right)/2$. It is equal to the antiferromagnetic exchange energy $J_\text{2-site}$ in the
corresponding two-site Heisenberg model.  In Table \ref{tabellespinspincoupling} we present the fitted values $\Delta E^{\text{spin}}_{11}$, $\Delta E^{\text{spin}}_{12}$ and $\Delta E^{\text{spin}}_{\text{2-site}}$ for the data in Fig. \ref{pictspinsusceptibility2sitecluster}. Not unexpectedly, the trend shows that for increasing insulating character, i.e. larger $U$, the excitation energies come closer to  the value of the isolated two-site cluster. 
\begin{table}
\caption[tabellespinspincoupling]{Spin excitation energies $\Delta E^{\text{spin}}_{11}$ and $\Delta E^{\text{spin}}_{12}$ obtained from the data in Figure (\ref{pictspinsusceptibility2sitecluster}) 
in comparison with the two-site Hubbard model $\Delta E^{\text{spin}}_{\text{2-site}}=J_{\text{2-site}}=\left(\sqrt{U^2+16 t^2}-U\right)/2$.}
\begin{tabular}{|p{1.5cm}|p{1.5cm}|p{1.5cm}|p{1.5cm}|}
\hline
  $U / t$ & $\Delta E^{\text{spin}}_{11} / t$ & $\Delta E^{\text{spin}}_{12} / t$ & $\Delta E^{\text{spin}}_{\text{2-site}}/t$\\
\hline\hline
  10 & 0.351 & 0.357 & 0.385 \\
\hline
  12 & 0.310 & 0.316 & 0.325 \\
\hline
  14 & 0.274 & 0.280 & 0.280 \\
\hline
\end{tabular}
\label{tabellespinspincoupling}
\end{table}

\subsubsection{Metallic phase}
\begin{figure*}[htbp]
    \centering
      \includegraphics[width=0.4\textwidth]{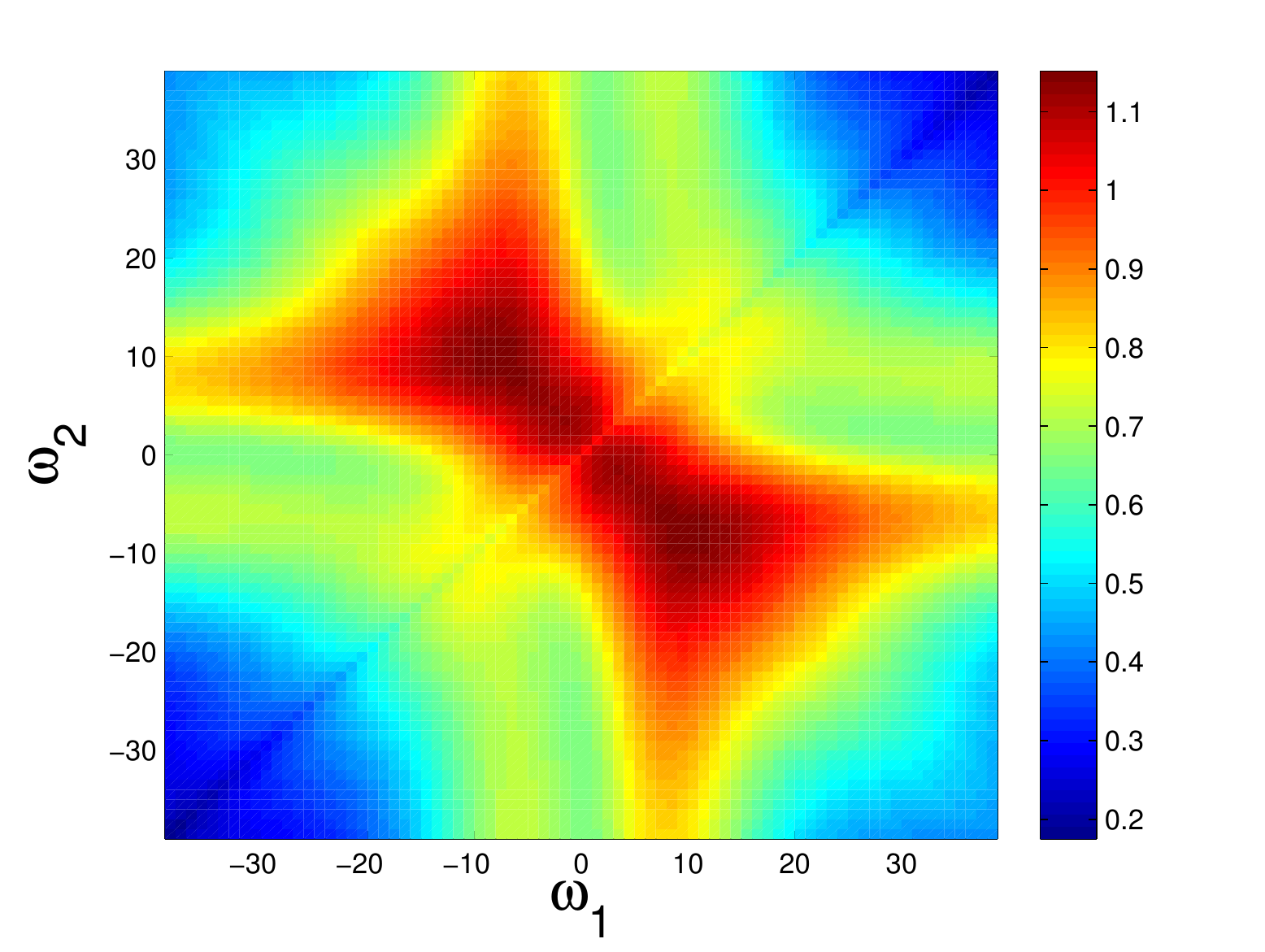}
      \includegraphics[width=0.4\textwidth]{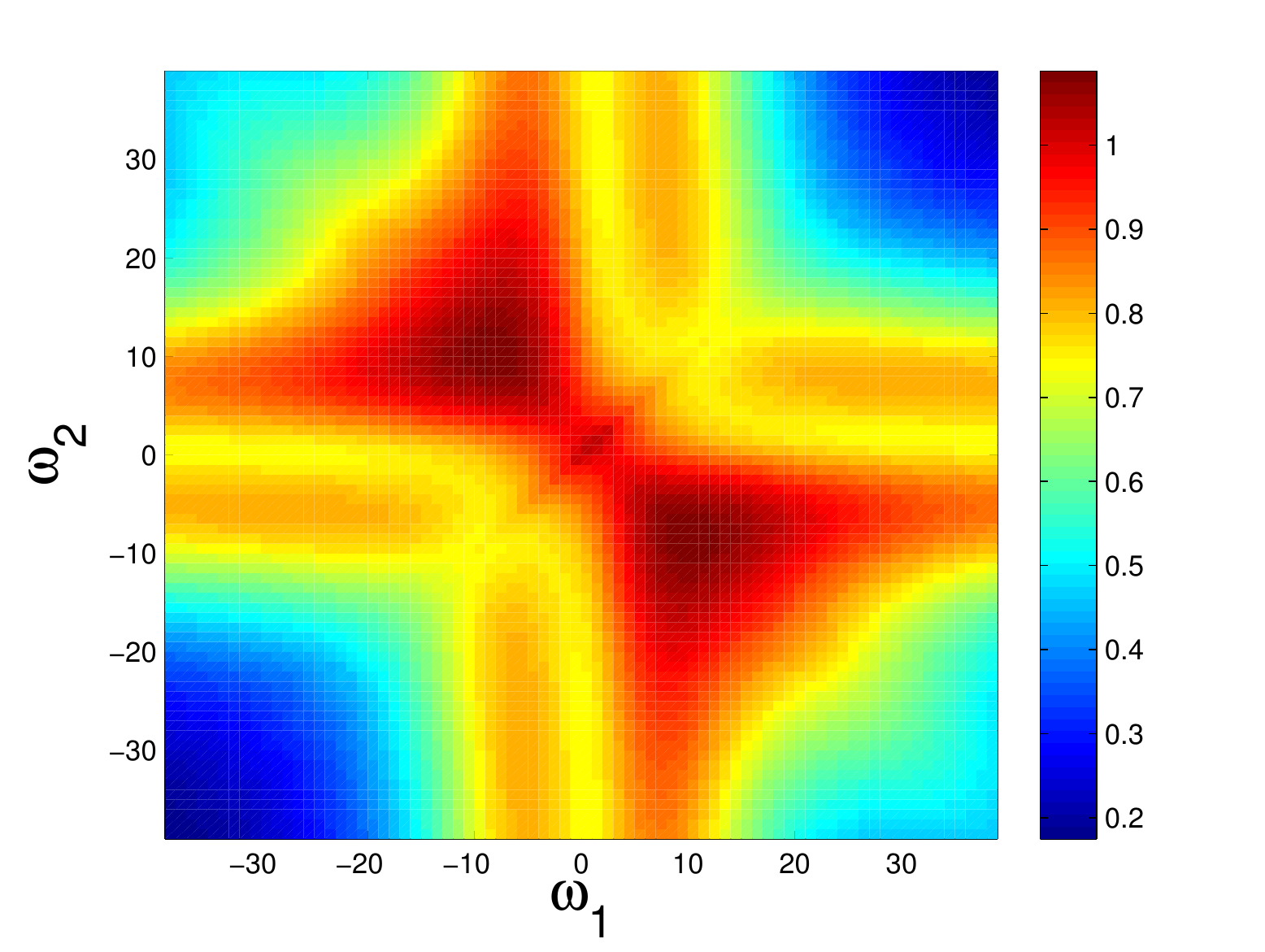}
      \includegraphics[width=0.4\textwidth]{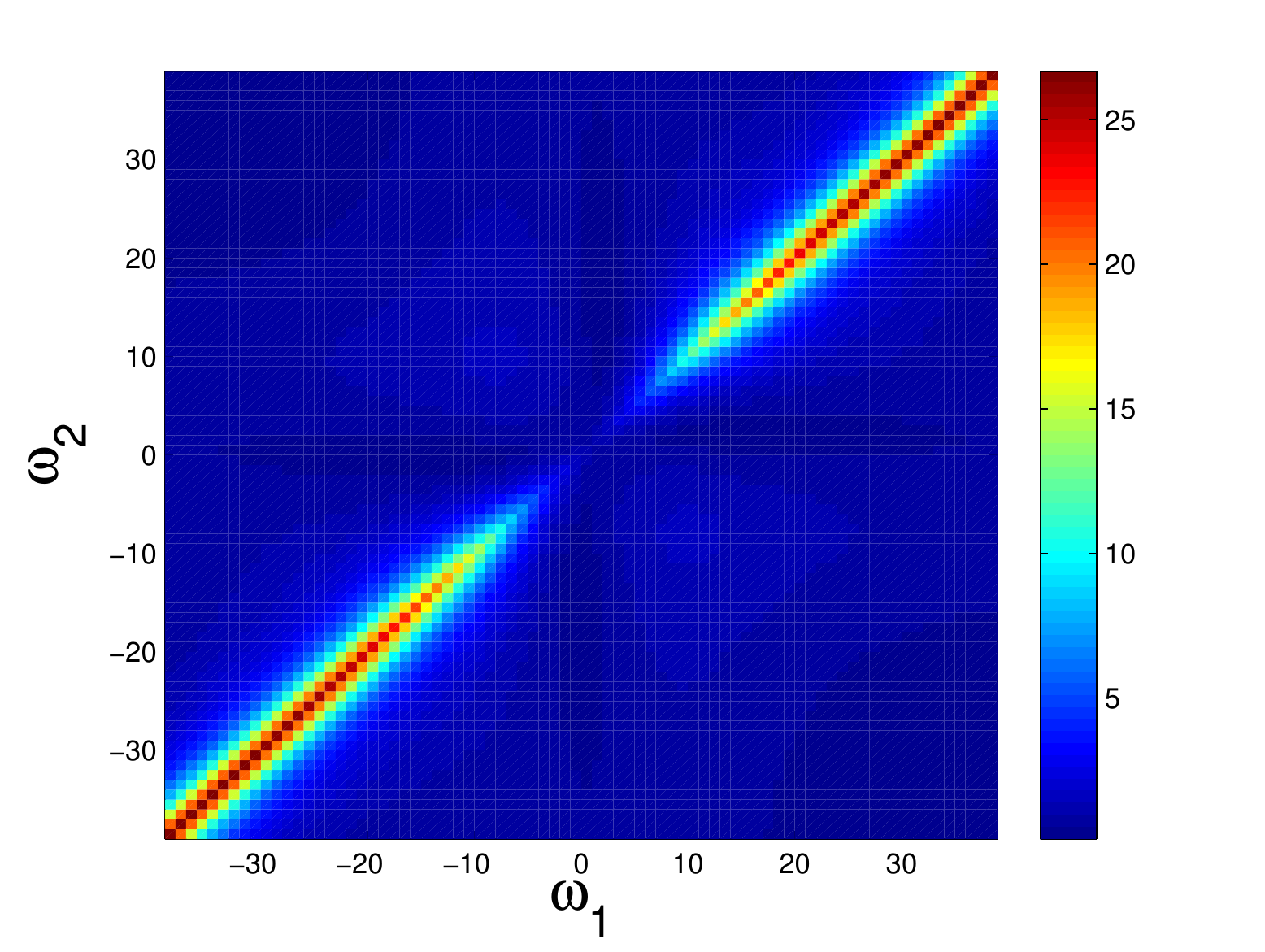}
      \includegraphics[width=0.4\textwidth]{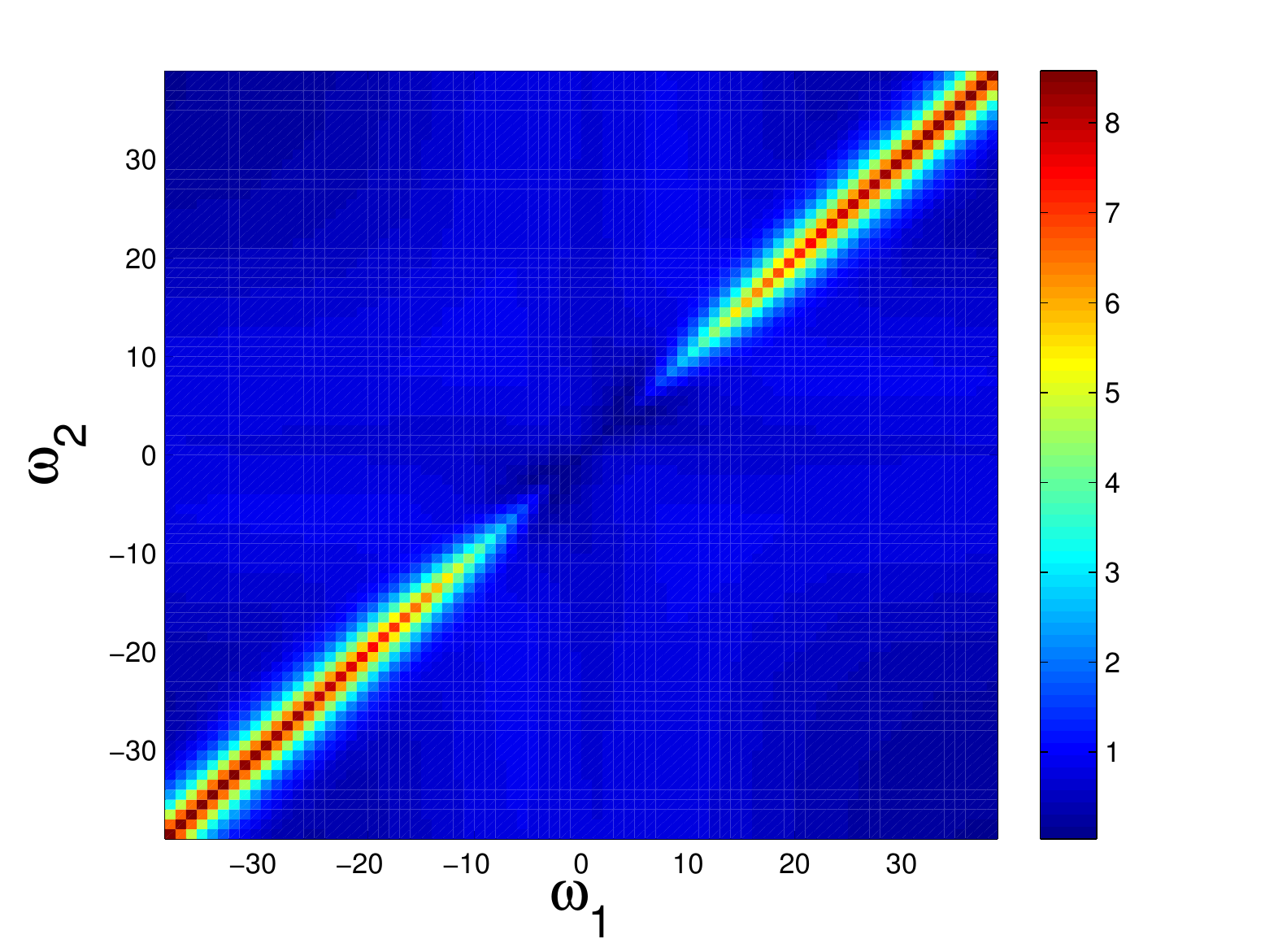}
      \includegraphics[width=0.4\textwidth]{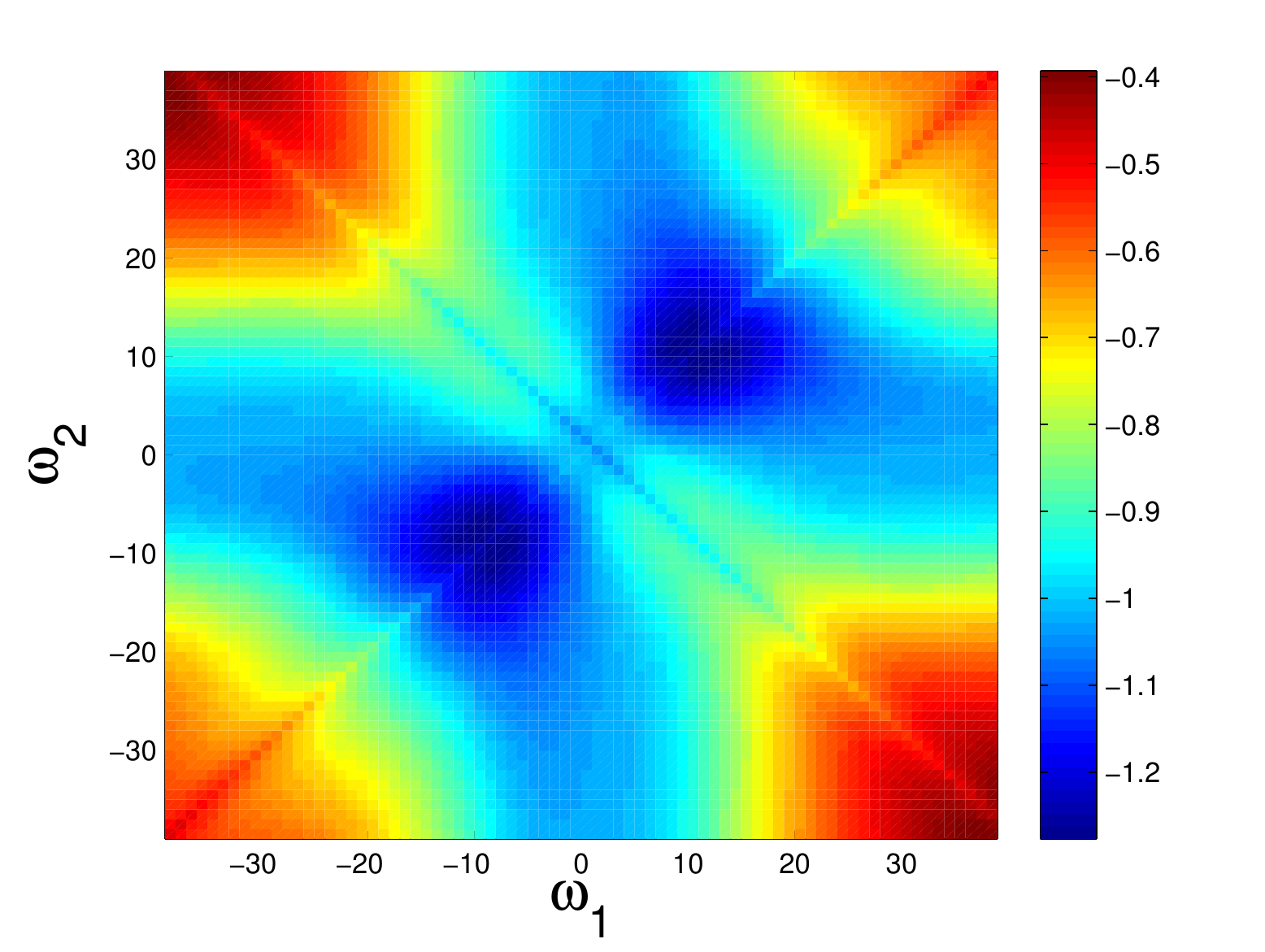}
      \includegraphics[width=0.4\textwidth]{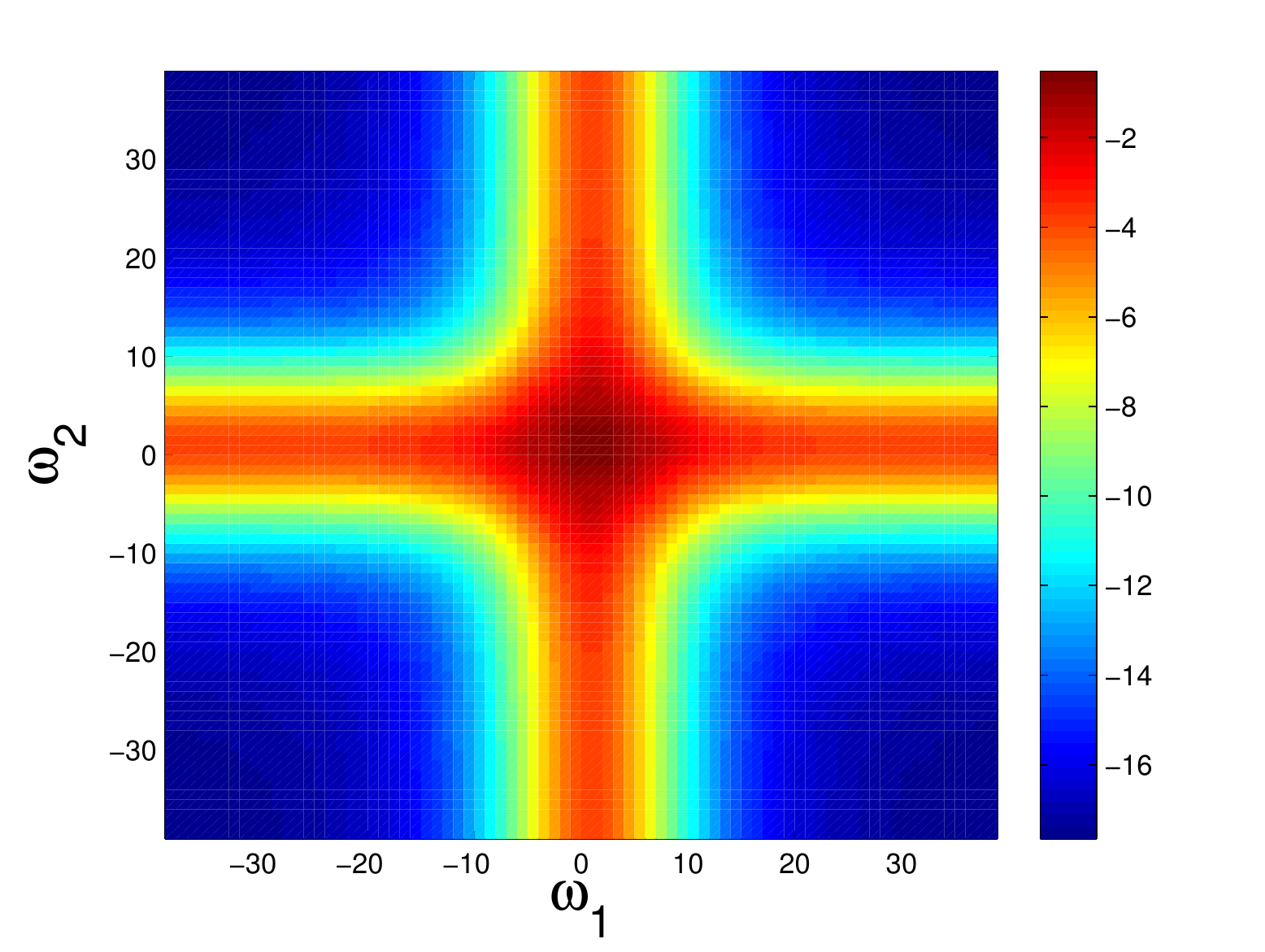}
      \includegraphics[width=0.4\textwidth]{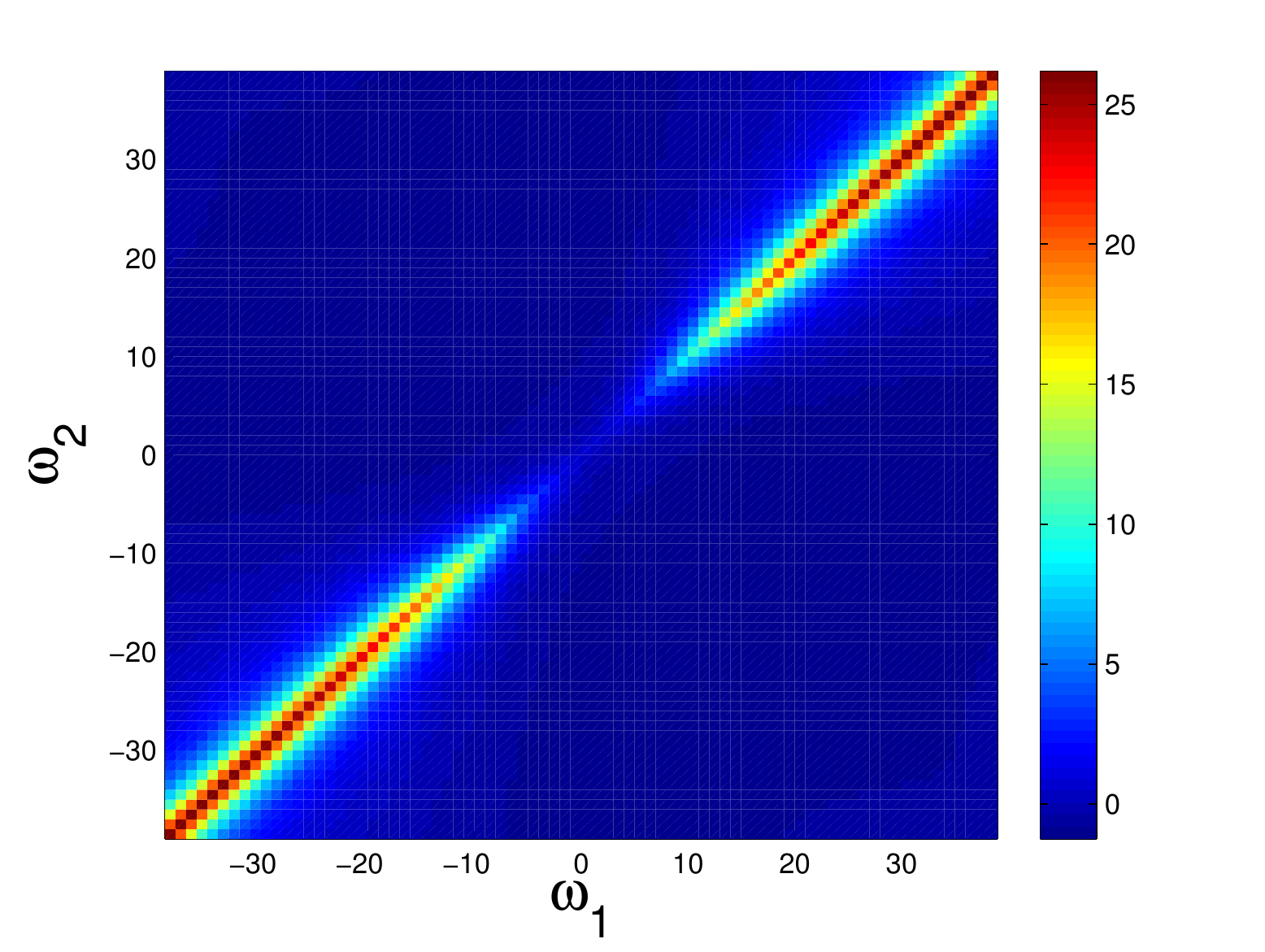}
      \includegraphics[width=0.4\textwidth]{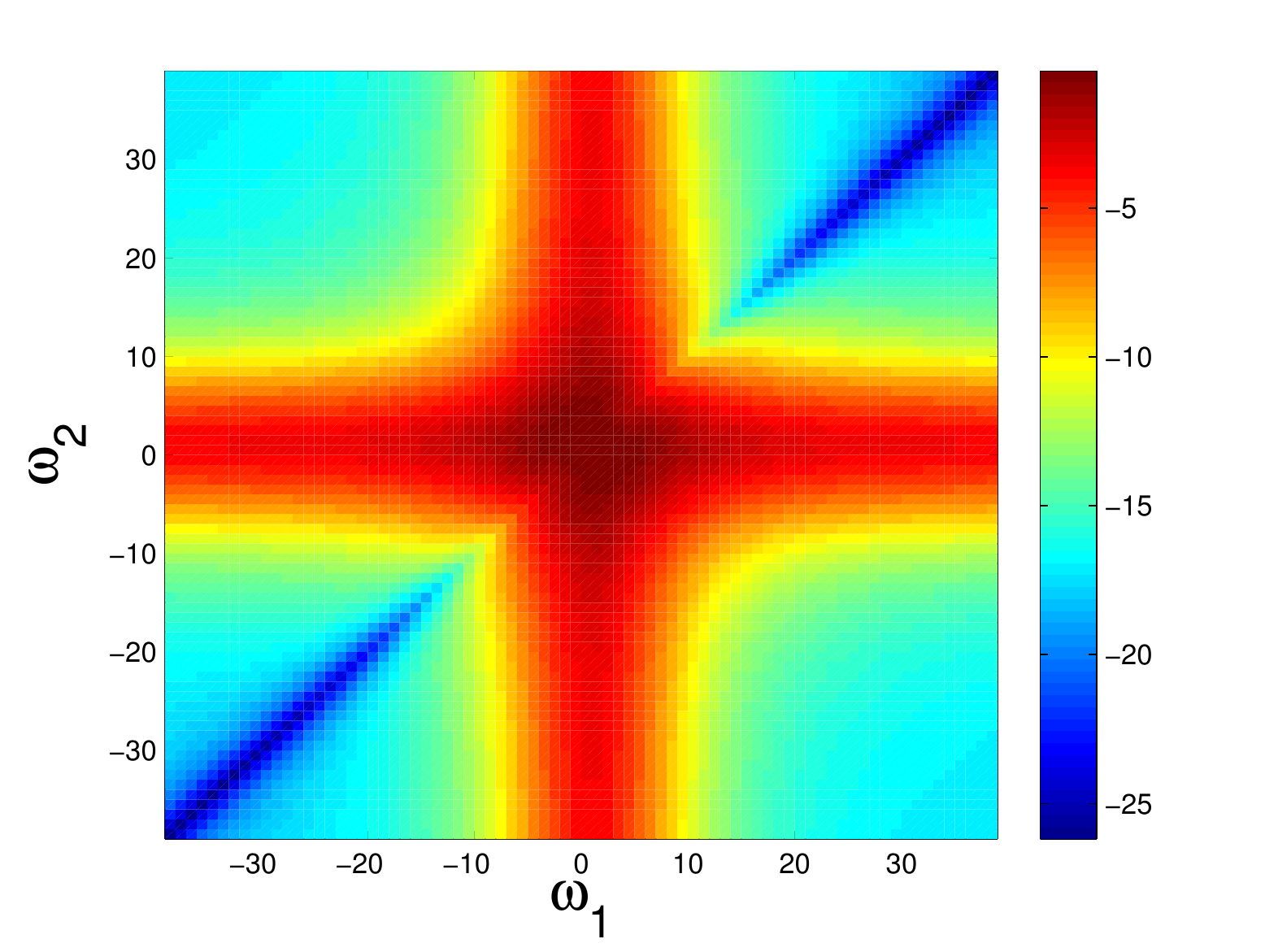}
      \caption[vertexplot4]{(color online)Vertex functions $|\Gamma^{1111}_{d/m}(i\omega_1;i\omega_2|i\omega_1;i\omega_2)\mp U/4|$, $|\Gamma^{1212}_{d/m}(i\omega_1;i\omega_2|i\omega_1;i\omega_2)\mp U/4|$,
      $\Gamma^{1221}_{d/m}(i\omega_1;i\omega_2|i\omega_1;i\omega_2)\mp U/4$ and $\Gamma^{1122}_{d/m}(i\omega_1;i\omega_2|i\omega_1;i\omega_2)\mp U/4$ for $U = 4 t$, $\beta = 30 / t$. Left column: density part. Right column: magnetic part. The frequencies are signed by their Matsubara index.}     
      \label{vertexplot4}
\end{figure*}

In Fig. \ref{vertexplot4} we show the vertices $|\Gamma^{1111}_{d/m}(i\omega_1;i\omega_2|i\omega_1;i\omega_2)\mp U/4|$, $|\Gamma^{1212}_{d/m}(i\omega_1;i\omega_2|i\omega_1;i\omega_2)\mp U/4|$,
$\Gamma^{1221}_{d/m}(i\omega_1;i\omega_2|i\omega_1;i\omega_2)\mp U/4$ and $\Gamma^{1122}_{d/m}(i\omega_1;i\omega_2|i\omega_1;i\omega_2)\mp U/4$ for $U = 4 t$ and $\beta = 30 / t$, i.e. in the metallic phase.

Compared to the insulating phase, the obtained frequency structures are now even richer. The density and magnetic parts of $\Gamma^{1111}$
and the density part of $\Gamma^{1221}$ are beyond a simple description and posses rather detailed structures along the $\omega_1=\omega_2$ and $\omega_1=-\omega_2$
lines, overlaid by an additional '+'-shaped structure. Opposite to the insulating case, the vertices become minimal in their absolute values at this '+'-shaped structure, especially
at the point $(\omega_{1}=0,\omega_{2}=0)$, rather than reaching a maximum. This is best visible in the magnetic part of $\Gamma^{1221}$, which is determined solely by this structure.
Except to this different behavior at the '+'-shaped structure, the vertices $\Gamma^{1212}$ and $\Gamma^{1122}$ are similar to the corresponding vertices in the insulating phase. 

\section{Conclusions}
\label{sec:conclusions}
In this paper we showed that a recently introduced\cite{Kin13} fRG scheme for Anderson impurity problems can serve as an efficient and flexible impurity solver for the dynamical mean-field theory. Using this new impurity solver, we studied in the first part of the paper the half-filled Hubbard model on a Bethe lattice
in infinite dimension. This showed that the hallmarks of metallic and insulating phases can be reproduced, although the transition region could not be resolved very clearly, at least with the current implementation. 

While we think that it is interesting and important to explore new impurity solvers, we certainly do not claim that the current version of this fRG impurity solver is superior to the established techniques with respect to single-particle properties. However, a quantity that has not been investigated thoroughly in the past but that is quite easily accessible  in the fRG impurity solver is the local one-particle irreducible vertex function. It explicitly appears in the fRG solution of the impurity problem and is hence obtained at no additional cost.
We obtained its density and magnetic part for the insulating and the metallic phase in good qualitative agreement with recent calculations using DMFT with exact diagonalization as impurity solver.\cite{Roh12} Understanding the frequency
structure of this vertex function in DMFT is important for several reasons. On the one hand it is an important ingredient of perturbative DMFT extensions that include
nonlocal degrees of freedom,\cite{Stan04,Tos07,Hel08,Rub08,Rub09,Roh13,Tar13} but also in the single-site DMFT two-particle correlation functions can be used to identify nonperturbative precursors of the
Mott physics inside the metallic phase of the MIT.\cite{Schae13} Note that there are additional ways to separate the 1PI vertex into other parts, like the fully irreducible vertex and 2PI vertices, see Ref. \onlinecite{Roh12}. As shown in this reference, these other vertices show slightly reduced complexity in their frequency structures, but also remain a nontrivial function of the frequencies. In order to keep the discussion manageable, we have not taken this road and only present data for the 1PI vertex. 

In the second part of the paper we studied the Hubbard model on a square lattice in two dimensions within a two-site cluster DMFT approach. In this scheme, antiferromagnetic
fluctuations between nearest neighbored sites are included. We obtained the density and magnetic part of the cluster vertex functions for the insulating and the metallic case. From the local and next-neighbor spin susceptibility we obtained the spin-spin coupling $J$ as function of $U$ in the insulating phase. 

Quite generally, our data support the findings of Ref. \onlinecite{Roh12} that the vertices show rich structure, including '+'-structures that cannot be parametrized in terms of the 'bosonic' transfer frequencies. While the physical meaning of these structures beyond a connection to higher-order diagrams is not obvious, they represent a formidable challenges for above-mentioned approaches that want to use the DMFT vertices as input in order to explore correlations on longer scales, in particular if wave vector dependencies are supposed to be added. Beyond this principal statement, we can use our data to make two valuable comparisons. First we can study {\em a)} the difference between the vertices in the metallic and  the insulating phase. Second we can {\em b)} scrutinize what changes occur when nonlocal correlations are included.

Regarding comparison {\em a)}, we find much milder frequency dependences in the metallic case. In particular, the sharp bosonic features of the single-site solution are smeared out, and the '+'-structures are broadened as well. Furthermore, many (but not all) cross sections of the vertices in the metallic phase show a reduction at low frequencies compared to  high frequencies which points to a screening effect. In the insulating state, the opposite is found. Here, the low frequency vertices are mostly enhanced. Finally, in the metallic phase of the single-site solution one can also
find enhancement features at zero incoming frequency, pointing to the role of pairing fluctuations. These cannot be seen in the insulating state, and these features are also much weaker in the two-site solution, possibly due to the spin gap. Note that the possible soft collective fluctuations that are not captured by the present cluster schemes could lead to additional frequency structures. Their systematics should however correspond to what is known from random-phase approximation or related approaches.

Comparison {\em b)} between single-site and two-site DMFT vertices shows on one hand that new features and energy scales can come in. Our data describe how the sharp delta-like diagonal features for fixed frequency transfer get broadened. These structures were indicative of a free moment in the insulating single-site solution and caused a peak in the spin susceptibility at zero Matsubara frequency. Now, the insulating two-site solution displays the exchange energy scale $J$, both in the vertex and the spin susceptibility. The embedding of the two-site core in the DMFT causes only slight additional broadenings. Beyond these expected changes,  the frequency structures are definitely dispersive, as can be seen from the vertices for different wavevector combinations. From our work one can only see that nonlocal correlations have a definite effect on the vertices. However, we are far away from understanding how far one should go in the cluster size to obtain convergence, e.g., of the local vertex. Yet, at least for larger $U$, the behavior on a nearest neighbor bond captured in our results should contain the dominant strong coupling physics, unless phase transitions with diverging length scales or geometric frustration comes into play.

To obtain the local dynamic charge- and spin susceptibilites from our vertex data is more challenging due to the finite frequency patching and the different speed of convergence of the connected and disconnected parts of the susceptibilities, but we still managed to estimate effective exchange coupling from the data.  Yet, the analytical continuation by a Pad$\acute{\text{e}}$-algorithm  does not deliver meaningful results for all sets of parameters. One might try to achieve better results by an appropriate parametrization of the vertex function in the lines of Ref. \onlinecite{Ort}.
Note, that frequency dependent vertex corrections are found to be essential for understanding experimentally observed dynamic susceptibilites in realistic
material calculations, as for example in the case of iron-based superconductors.\cite{Par11,Tos12,Liu12} Therefore, there is a great need for developing new flexible solvers, that
facilitate the heavy calculation of these quantities in realistic multi-orbital cases.

We thank S. Andergassen, A. Liebsch, C. Taranto and A. Toschi for helpful discussions and for providing data for comparison. 
This work was supported by the DFG research units FOR 732 and FOR 912.

\appendix
\begin{widetext}
\section{Green's and Vertex functions}
\label{sec:appendixa}
The n-particle Green's functions are defined as the time ordered expectation values by \cite{Neg88}
\begin{eqnarray}
\nonumber
&&\mathcal{G}^{(n)}(\alpha_1\tau_1,...,\alpha_n\tau_n|\alpha_{1'}\tau_{1'},...,\alpha_{n'}\tau_{n'})\\
\nonumber
&=&(-1)^n\langle\langle T_{\tau}a_{\alpha_1}(\tau_1)...a_{\alpha_n}(\tau_n)a^{\dag}_{\alpha_{n'}}(\tau_{n'})...a^{\dag}_{\alpha_{1'}}(\tau_{1'})\rangle\rangle_{Z}\\
&=& \frac{(-1)^n}{Z}\textrm{Tr}\left[\exp\left(-\beta\hat{H}\right)T_{\tau}a_{\alpha_1}(\tau_1)...a_{\alpha_n}(\tau_n)a^{\dag}_{\alpha_{n'}}(\tau_{n'})...a^{\dag}_{\alpha_{1'}}(\tau_{1'})\right]
\end{eqnarray} 
with the time dependent Heisenberg operators
\begin{equation}
a_{i}^{(\dag)}(\tau)=\exp\left(\tau\hat{H}\right)a_{i}^{(\dag)}\exp\left(-\tau\hat{H}\right).
\end{equation}
The Fourier transform of the Green's functions is given by
\begin{eqnarray}
\nonumber
\mathcal{G}^{(n)}(\alpha_1\tau_1,...,\alpha_n\tau_n|\alpha_{1'}\tau_{1'},...,\alpha_{n'}\tau_{n'})
&=& \frac{1}{\beta^{2n}}\sum_{i\omega_{1},...,i\omega_{n}}\sum_{i\omega_{1'},...,i\omega_{n'}}
\textrm{e}^{-\sum_{j}i\omega_{j}\tau_{j}}\textrm{e}^{\sum_{j}i\omega_{j'}\tau_{j'}}\\
&&\times\mathcal{G}^{(n)}(\alpha_1\omega_1,...,\alpha_n\omega_n|\alpha_{1'}\omega_{1'},...,\alpha_{n'}\omega_{n'}),\\
\nonumber
\mathcal{G}^{(n)}(\alpha_1\omega_1,...,\alpha_n\omega_n|\alpha_{1'}\omega_{1'},...,\alpha_{n'}\omega_{n'}))
&=& \int_{0}^{\beta}\textrm{d}\tau_{1}...\int_{0}^{\beta}\textrm{d}\tau_{n}\int_{0}^{\beta}\textrm{d}\tau_{1'}...\int_{0}^{\beta}\textrm{d}\tau_{n'}
\textrm{e}^{\sum_{j}i\omega_{j}\tau_{j}}\textrm{e}^{-\sum_{j}i\omega_{j'}\tau_{j'}}\\
&&\times\mathcal{G}^{(n)}(\alpha_1\tau_1,...,\alpha_n\tau_n|\alpha_{1'}\tau_{1'},...,\alpha_{n'}\tau_{n'}).
\end{eqnarray}
In the following, if not stated otherwise, the multi index $\alpha_i$ stands for either $(\alpha_i,\tau_i)$ or $(\alpha_i,\omega_i)$.
By subtracting the disconnected parts of $\mathcal{G}^{(2)}$ one gets the connected two-particle Green's function
\begin{eqnarray}
\mathcal{G}^{c,(2)}(\alpha_1,\alpha_2|\alpha_{1'},\alpha_{2'})&=&\mathcal{G}^{(2)}(\alpha_1,\alpha_2|\alpha_{1'},\alpha_{2'})
-\mathcal{G}^{(1)}(\alpha_1|\alpha_{1'})\mathcal{G}^{(1)}(\alpha_2|\alpha_{2'})
+\mathcal{G}^{(1)}(\alpha_1|\alpha_{2'})\mathcal{G}^{(1)}(\alpha_2|\alpha_{1'}).
\end{eqnarray}
From $\mathcal{G}^{c,(2)}$ one obtains the 1PI-Vertex-function by amputing the full one-particle Green's functions at the outer legs
\begin{eqnarray}\label{eqrelationcongreensfunctionvertex}
\Gamma(\alpha_1,\alpha_2|\alpha_{1'},\alpha_{2'})
=-\sum_{\beta_1,\beta_2\atop\beta_{1'}\beta_{2'}}\left[\mathcal{G}^{(1)}\right]^{-1}_{\alpha_1,\beta_1}
\left[\mathcal{G}^{(1)}\right]^{-1}_{\alpha_2,\beta_2}\mathcal{G}^{c,(2)}(\beta_{1},\beta_{2}|\beta_{1'},\beta_{2'})\left[\mathcal{G}^{(1)}\right]^{-1}_{\beta_{1'},\alpha_{1'}}
\left[\mathcal{G}^{(1)}\right]^{-1}_{\beta_{2'},\alpha_{2'}}.
\end{eqnarray}

\section{Dynamic susceptibilities}

If we assume spin rotation invariance, a general two-particle Green's function can be parameterized in the 
following way:
\begin{equation}\label{eqzerlegunggreensfunktionspinchannel}
\mathcal{G}^{(2)}\left(\alpha_1,\sigma_1;\alpha_2,\sigma_2|\alpha_{1'},\sigma_{1'};\alpha_{2'},\sigma_{2'}\right)
=\mathcal{A}\left(\alpha_1,\alpha_2|\alpha_{1'},\alpha_{2'}\right)\delta_{\sigma_1,\sigma_{1'}}\delta_{\sigma_2,\sigma_{2'}}
+\mathcal{B}\left(\alpha_1,\alpha_2|\alpha_{1'},\alpha_{2'}\right)\delta_{\sigma_1,\sigma_{2'}}\delta_{\sigma_2,\sigma_{1'}}
\end{equation}
Because $\mathcal{G}^{(2)}(1,2|1',2')$ is antisymmetric under the permutations $1\leftrightarrow 2$ and $1'\leftrightarrow 2'$ 
the functions $\mathcal{A}$ and $\mathcal{B}$ obey the relation
\begin{equation}
\mathcal{A}\left(\alpha_1,\alpha_2|\alpha_{1'},\alpha_{2'}\right)=-\mathcal{B}\left(\alpha_1,\alpha_2|\alpha_{2'},\alpha_{1'}\right)=
-\mathcal{B}\left(\alpha_2,\alpha_1|\alpha_{1'},\alpha_{2'}\right).
\end{equation}
Using the identity
\begin{equation}
2\delta_{\sigma_1,\sigma_{2'}}\delta_{\sigma_2,\sigma_{1'}}=\delta_{\sigma_1,\sigma_{1'}}\delta_{\sigma_2,\sigma_{2'}}
+\vec{\sigma}_{\sigma_1,\sigma_{1'}}\vec{\sigma}_{\sigma_2,\sigma_{2'}},
\end{equation}
we write the two-particle Green's function as
\begin{equation}
\mathcal{G}^{(2)}\left(\alpha_1,\sigma_1;\alpha_2,\sigma_2|\alpha_{1'},\sigma_{1'};\alpha_{2'},\sigma_{2'}\right)=
\mathcal{G}^{(2)}_{d}\left(\alpha_1,\alpha_2|\alpha_{1'},\alpha_{2'}\right)\delta_{\sigma_1,\sigma_{1'}}\delta_{\sigma_2,\sigma_{2'}}
+\mathcal{G}^{(2)}_{m}\left(\alpha_1,\alpha_2|\alpha_{1'},\alpha_{2'}\right)\vec{\sigma}_{\sigma_1,\sigma_{1'}}\vec{\sigma}_{\sigma_2,\sigma_{2'}}
\end{equation}
with the density part $\mathcal{G}^{(2)}_{d}$ and the magnetic part $\mathcal{G}^{(2)}_{m}$ given by
\begin{eqnarray}
\label{eqdensitypart}
\mathcal{G}^{(2)}_{d}\left(\alpha_1,\alpha_2|\alpha_{1'},\alpha_{2'}\right)&=&
\mathcal{A}\left(\alpha_1,\alpha_2|\alpha_{1'},\alpha_{2'}\right)+\frac{1}{2}\mathcal{B}\left(\alpha_1,\alpha_2|\alpha_{1'},\alpha_{2'}\right),\\
\label{eqmagneticpart}
\mathcal{G}^{(2)}_{m}\left(\alpha_1,\alpha_2|\alpha_{1'},\alpha_{2'}\right)&=&
\frac{1}{2}\mathcal{B}\left(\alpha_1,\alpha_2|\alpha_{1'},\alpha_{2'}\right).
\end{eqnarray}

In an analogue way one can define a density and magnetic part of the connected Green's function $\mathcal{G}^{c,(2)}_{d/m}$ and
of the 1PI vertex function $\Gamma_{d/m}$.

\subsection{Dynamic charge susceptibility}
The dynamic charge susceptibility is defined as
\begin{equation}
\chi^{\text{charge}}_{ij}\left(i\nu\right)=\int_{0}^{\beta}\textrm{d}\tau \textrm{e}^{i\nu\tau}\left[\left\langle\left\langle T_{\tau}\left(\hat{\rho}_{i}(\tau)\hat{\rho}_{j}(0)\right)\right\rangle\right\rangle
-\langle\langle\hat{\rho}_i\rangle\rangle \langle\langle\hat{\rho}_j\rangle\rangle\right]
\end{equation}
with the density-operator
\begin{equation}
\hat{\rho}_{i}=\sum_{\sigma}c_{i\sigma}^{\dag}c_{i\sigma}.
\end{equation}
The expectation value is given by a two-particle Green's function
\begin{eqnarray}\label{eqdynamicchargesusceptibilitydef}
\nonumber
\chi^{\text{charge}}_{ij}\left(i\nu\right)&=&\int_{0}^{\beta}\textrm{d}\tau \textrm{e}^{i\nu\tau}
\sum_{\sigma,\sigma'}\mathcal{G}^{(2)}\left(i,\sigma,\tau;j,\sigma',0^{-}|i,\sigma,\tau;j,\sigma'\right)
-\beta\delta_{\nu,0}\langle\langle\hat{\rho}_i\rangle\rangle \langle\langle\hat{\rho}_j\rangle\rangle\\
\nonumber
&=& \frac{1}{\beta^2}\sum_{i\omega_1,i\omega_2}\sum_{\sigma,\sigma'}\mathcal{G}^{(2)}\left(i,\sigma,i\omega_1;j,\sigma',i\omega_2|
i,\sigma,i\omega_1-i\nu;j,\sigma'\right)
-\beta\delta_{\nu,0}\langle\langle\hat{\rho}_i\rangle\rangle \langle\langle\hat{\rho}_j\rangle\rangle\\
\nonumber
&=& \frac{4}{\beta^2}\sum_{i\omega_1,i\omega_2}\mathcal{G}^{(2)}_{d}\left(i,i\omega_1;j,i\omega_2|i,i\omega_1-i\nu;j\right)
-\beta\delta_{\nu,0}\langle\langle\hat{\rho}_i\rangle\rangle \langle\langle\hat{\rho}_j\rangle\rangle\\
\nonumber
&=& \frac{4}{\beta^2}\sum_{i\omega_1,i\omega_2}\mathcal{G}^{c,(2)}_{d}\left(i,i\omega_1;j,i\omega_2|i,i\omega_1-i\nu;j\right)
 -\frac{2}{\beta}\sum_{i\omega}\mathcal{G}\left(i\omega,i,j\right)\mathcal{G}\left(i\omega-i\nu,j,i\right)\\
&=& \chi^{\text{charge},c}_{ij}\left(i\nu\right)+\chi^{\text{charge},dc}_{ij}\left(i\nu\right).
\end{eqnarray}
Here we used that the density part $\mathcal{G}^{(2)}_d$ is given by
\begin{equation}
\mathcal{G}^{(2)}_{d}\left(i,i\omega_1;j,i\omega_2|i,i\omega_1-i\nu;j\right)=
\frac{1}{2}\left(\mathcal{G}^{(2)}\left(i,\uparrow,i\omega_1;j,\uparrow,i\omega_2|i,\uparrow,i\omega_1-i\nu;j,\uparrow\right)
+\mathcal{G}^{(2)}\left(i,\uparrow,i\omega_1;j,\downarrow,i\omega_2|i,\uparrow,i\omega_1-i\nu;j,\downarrow\right)\right)
\end{equation}
and
\begin{equation}
\langle\langle\hat{\rho}_i\rangle\rangle=\frac{2}{\beta}\sum_{i\omega}\textrm{e}^{i\omega0^{+}}\mathcal{G}\left(i\omega,i,i\right).
\end{equation}

\subsection{Dynamic spin susceptibility}
The dynamic spin susceptibility is defined as
\begin{equation}
\chi^{\text{spin}}_{ij}\left(i\nu\right)=\int_{0}^{\beta}\textrm{d}\tau \textrm{e}^{i\nu\tau}\left\langle\left\langle T_{\tau}\left(\hat{S}_{i}^{+}(\tau)\hat{S}_{j}^{-}(0)\right)\right\rangle\right\rangle
\end{equation}
with the spin operators
\begin{eqnarray}
\hat{S}_{i}^{+}(\tau)=c_{i\uparrow}^{\dag}(\tau)c_{i\downarrow}(\tau),\\
\hat{S}_{i}^{-}(\tau)=c_{i\downarrow}^{\dag}(\tau)c_{i\uparrow}(\tau).
\end{eqnarray}
The expectation value is given by a two-particle Green's function
\begin{eqnarray}\label{eqdynamicspinsusceptibilitydef}
\nonumber
\chi^{\text{spin}}_{ij}\left(i\nu\right)&=&-\int_{0}^{\beta}\textrm{d}\tau \textrm{e}^{i\nu\tau}
\mathcal{G}^{(2)}\left(j,\uparrow,0^{-};i,\downarrow,\tau|i,\uparrow,\tau;j,\downarrow\right)\\
\nonumber
&=& -\frac{1}{\beta^2}\sum_{i\omega_{1},i\omega_{2}}\mathcal{G}^{(2)}\left(j,\uparrow,i\omega_{1};i,\downarrow,i\omega_{2}|i,\uparrow,i\omega_{2}-i\nu;j,\downarrow\right)\\
\nonumber
&=& \frac{2}{\beta^2}\sum_{i\omega_{1},i\omega_{2}}\mathcal{G}^{(2)}_{m}\left(i,i\omega_{2};j,i\omega_{1}|i,i\omega_{2}-i\nu;j\right)\\
\nonumber
&=& \frac{2}{\beta^2}\sum_{i\omega_{1},i\omega_{2}}\mathcal{G}^{c,(2)}_{m}\left(i,i\omega_{2};j,i\omega_{1}|i,i\omega_{2}-i\nu;j\right)-\frac{1}{\beta}\sum_{i\omega}\mathcal{G}\left(i\omega,i,j\right)\mathcal{G}\left(i\omega-i\nu,j,i\right)\\
&=& \chi_{ij}^{\text{spin},c}\left(i\nu\right)+\chi_{ij}^{\text{spin},dc}\left(i\nu\right).
\end{eqnarray}

\subsection{Local Hubbard model}
The Hamiltonian of the one-site Hubbard model (for $\mu=-U/2$ i.e. particle-hole symmetry), which serves as 'core' in our single-site DMFT(fRG) scheme 
for the insulating phase, is given by
\begin{equation}
\hat{H}=-\frac{U}{2}\sum_{\sigma}n_{\sigma}+U n_{\uparrow} n_{\downarrow}.
\end{equation}
From the one-particle Green's function
\begin{eqnarray}
\nonumber
\mathcal{G}(i\omega)&=&\frac{i\omega}{(i\omega)^2-\frac{U^2}{4}}\\
&=& \frac{1}{i\omega+U/2-\Sigma(i\omega)}
\end{eqnarray}
follows the self-energy as
\begin{equation}\label{eqlocalhubbardmodelselfenergy}
\Sigma(i\omega)=\frac{U}{2}+\frac{U^2}{4i\omega}.
\end{equation}
The two-particle 1PI vertex function \cite{Haf09} is given by
\begin{eqnarray}\label{eqvertexsinglehubbardsite}
\Gamma\left(\uparrow,i\omega_1;\uparrow,i\omega_2|\uparrow,i\omega_1';\uparrow\right)
&=&-\beta \frac{U^2}{4}\frac{\delta_{\omega_1,\omega_1'}-\delta_{\omega_2,\omega_1'}}{(i\omega_1)^2 (i\omega_2)^2}\left[(i\omega_1)^2-\frac{U^2}{4}\right]\left[(i\omega_2)^2-\frac{U^2}{4}\right],\\
\nonumber
\Gamma\left(\uparrow,i\omega_1;\downarrow,i\omega_2|\uparrow,i\omega_1';\downarrow\right)
&=& U + \left(\frac{U}{2}\right)^3\frac{\sum_{i=1,2,1',2'}(i\omega_i)^2}{\prod_{i=1,2,1',2'}(i\omega_i)}
-6\left(\frac{U}{2}\right)^5\prod_{i=1,2,1',2'}\frac{1}{(i\omega_i)}\\
\nonumber
&&-\beta \delta_{\omega_1,-\omega_2} \frac{U^2}{2} n_{F}\left(\frac{U}{2}\right)\frac{\left[(i\omega_1)^2-\frac{U^2}{4}\right]\left[(i\omega_{1'})^2-\frac{U^2}{4}\right]}{(i\omega_1)^2 (i\omega_{1'})^2}\\
\nonumber
&&+\beta \delta_{\omega_2,\omega_1'} \frac{U^2}{2} n_{F}\left(-\frac{U}{2}\right)\frac{\left[(i\omega_1)^2-\frac{U^2}{4}\right]\left[(i\omega_{1'})^2-\frac{U^2}{4}\right]}{(i\omega_1)^2 (i\omega_{1'})^2}\\
&&-\beta \delta_{\omega_1,\omega_1'} \frac{U^2}{4} \left(n_{F}\left(\frac{U}{2}\right)-n_{F}\left(-\frac{U}{2}\right)\right)
\frac{\left[(i\omega_1)^2-\frac{U^2}{4}\right]\left[(i\omega_2)^2-\frac{U^2}{4}\right]}{(i\omega_1)^2 (i\omega_2)^2}
\end{eqnarray}
with $\omega_2'=\omega_1+\omega_2-\omega_1'$ and the Fermi function
\begin{equation}
n_{F}(x)=\frac{1}{1+\exp\left(\beta x\right)}.
\end{equation}
The density and magnetic part of $\Gamma$ follow with Eqs. (\ref{eqdensitypart}) and (\ref{eqmagneticpart}) as
\begin{eqnarray}
\nonumber
\Gamma_{d}\left(i\omega_1,i\omega_2|i\omega_{1'}\right)&=&\frac{1}{2}\left[\Gamma\left(\uparrow,i\omega_1;\uparrow,i\omega_2|\uparrow,i\omega_1';\uparrow\right)
+\Gamma\left(\uparrow,i\omega_1;\downarrow,i\omega_2|\uparrow,i\omega_1';\downarrow\right)\right]\\
\nonumber
&=&\frac{U}{2} + \frac{1}{2}\left(\frac{U}{2}\right)^3\frac{\sum_{i=1,2,1',2'}(i\omega_i)^2}{\prod_{i=1,2,1',2'}(i\omega_i)}
-3\left(\frac{U}{2}\right)^5\prod_{i=1,2,1',2'}\frac{1}{(i\omega_i)}\\
\nonumber
&&-\beta\delta_{\omega_1,-\omega_2}\frac{U^2}{4}n_{F}\left(\frac{U}{2}\right)
\frac{\left[(i\omega_1)^2-\frac{U^2}{4}\right]\left[(i\omega_{1'})^2-\frac{U^2}{4}\right]}{(i\omega_1)^2 (i\omega_{1'})^2}\\
\nonumber
&&+\beta\delta_{\omega_2,\omega_{1'}}\frac{U^2}{4}\left[\frac{1}{2}+n_{F}\left(-\frac{U}{2}\right)\right]
\frac{\left[(i\omega_1)^2-\frac{U^2}{4}\right]\left[(i\omega_{2})^2-\frac{U^2}{4}\right]}{(i\omega_1)^2 (i\omega_{2})^2}\\
\label{eqlocalvertexdensity}
&&-\beta\delta_{\omega_1,\omega_{1'}}\frac{U^2}{4}n_{F}\left(\frac{U}{2}\right)
\frac{\left[(i\omega_1)^2-\frac{U^2}{4}\right]\left[(i\omega_{2})^2-\frac{U^2}{4}\right]}{(i\omega_1)^2 (i\omega_{2})^2},
\end{eqnarray}
\begin{eqnarray}
\nonumber
\Gamma_{m}\left(i\omega_1,i\omega_2|i\omega_{1'}\right)
&=&\frac{1}{2}\left[\Gamma\left(\uparrow,i\omega_1;\uparrow,i\omega_2|\uparrow,i\omega_1';\uparrow\right)
-\Gamma\left(\uparrow,i\omega_1;\downarrow,i\omega_2|\uparrow,i\omega_1';\downarrow\right)\right]\\
\nonumber
&=&-\frac{U}{2} - \frac{1}{2}\left(\frac{U}{2}\right)^3\frac{\sum_{i=1,2,1',2'}(i\omega_i)^2}{\prod_{i=1,2,1',2'}(i\omega_i)}
+3\left(\frac{U}{2}\right)^5\prod_{i=1,2,1',2'}\frac{1}{(i\omega_i)}\\
\nonumber
&&+\beta\delta_{\omega_1,-\omega_2}\frac{U^2}{4}n_{F}\left(\frac{U}{2}\right)
\frac{\left[(i\omega_1)^2-\frac{U^2}{4}\right]\left[(i\omega_{1'})^2-\frac{U^2}{4}\right]}{(i\omega_1)^2 (i\omega_{1'})^2}\\
\nonumber
&&+\beta\delta_{\omega_2,\omega_{1'}}\frac{U^2}{4}\left[\frac{1}{2}-n_{F}\left(-\frac{U}{2}\right)\right]
\frac{\left[(i\omega_1)^2-\frac{U^2}{4}\right]\left[(i\omega_{2})^2-\frac{U^2}{4}\right]}{(i\omega_1)^2 (i\omega_{2})^2}\\
\label{eqlocalvertexmagnetic}
&&-\beta\delta_{\omega_1,\omega_{1'}}\frac{U^2}{4}n_{F}\left(-\frac{U}{2}\right)
\frac{\left[(i\omega_1)^2-\frac{U^2}{4}\right]\left[(i\omega_{2})^2-\frac{U^2}{4}\right]}{(i\omega_1)^2 (i\omega_{2})^2}
\end{eqnarray}
with $\omega_2'=\omega_1+\omega_2-\omega_1'$. 

With Eq. (\ref{eqdynamicchargesusceptibilitydef}) one gets the dynamic charge susceptibility
\begin{eqnarray}
\chi^{\text{charge},c}\left(i\nu\right)&=&-U \frac{n_{F}\left(\frac{U}{2}\right)-n_{F}\left(-\frac{U}{2}\right)}{(i\nu)^2-U^2}
+\beta\delta_{\nu,0}n_{F}\left(\frac{U}{2}\right)^2,\\
\chi^{\text{charge},dc}\left(i\nu\right)&=&U \frac{n_{F}\left(\frac{U}{2}\right)-n_{F}\left(-\frac{U}{2}\right)}{(i\nu)^2-U^2}
+\beta\delta_{\nu,0} n_F\left(\frac{U}{2}\right)n_F\left(-\frac{U}{2}\right),\\
\chi^{\text{charge}}\left(i\nu\right)&=&\beta\delta_{\nu,0}n_{F}\left(\frac{U}{2}\right).
\end{eqnarray}
With Eq. (\ref{eqdynamicspinsusceptibilitydef}) one gets the dynamic spin susceptibility
\begin{eqnarray}
\label{eqspinsusceptibilityclocalhubbardsite}
\chi^{\text{spin},c}\left(i\nu\right)&=&-\frac{U}{2}\frac{n_{F}\left(\frac{U}{2}\right)-n_{F}\left(-\frac{U}{2}\right)}{(i\nu)^2-U^2}
+\beta\delta_{\nu,0}\frac{1}{2}n_{F}\left(-\frac{U}{2}\right)^{2},\\
\label{eqspinsusceptibilitydclocalhubbardsite}
\chi^{\text{spin},dc}\left(i\nu\right)&=&\frac{U}{2}\frac{n_{F}\left(\frac{U}{2}\right)-n_{F}\left(-\frac{U}{2}\right)}{(i\nu)^2-U^2}
+\beta\delta_{\nu,0}\frac{1}{2}n_{F}\left(\frac{U}{2}\right)n_{F}\left(-\frac{U}{2}\right),\\
\label{eqspinsusceptibilitylocalhubbardsite}
\chi^{\text{spin}}\left(i\nu\right)&=&\beta\delta_{\nu,0}\frac{1}{2}n_{F}\left(-\frac{U}{2}\right).
\end{eqnarray}

\section{fRG flow equations}
\label{sec:appendixfrg}

A detailed derivation of the fRG flow equations that are used to solve the impurity problem, given in the form of a semi-infinite tight binding chain,
(\ref{eqhamiltonandersonmodel}) can be found in Ref. \onlinecite{Kin13}. In the following we show that this formalism can be generalized to impurity problems in the form of a semi infinite
$N$-chain ladder like, for example, Eq. \ref{eqhamiltontwositeandersonmodel}. The derivation is completely analog to the case before and
we just present the important steps.

The Hamiltonian of a $N$-chain ladder with an interaction term on the first rung is given by
\begin{eqnarray}\label{eqhamiltonnsiteandersonmodel}
\nonumber
\hat{H}_{\text{N-site-And}}&=&U\sum_{\sigma}\sum_{j=1}^{N}\hat{n}_{d,j,\uparrow}\hat{n}_{d,j,\downarrow}
-t^{\perp}_0\sum_{j=1}^{N-1}\sum_{\sigma}\left(d^{\dag}_{j,\sigma}d_{j+1,\sigma}+H.c.\right)
-t_{0}\sum_{\sigma}\sum_{j=1}^{N}\left(d^{\dag}_{j,\sigma}b_{1,j,\sigma}+H.c.\right)\\
&&-\sum_{i=1}^{\infty}\sum_{j=1}^{N}\sum_{\sigma}t_i\left(b^{\dag}_{i,j,\sigma}b_{i+1,j,\sigma}+H.c.\right)
-\sum_{i=1}^{\infty}\sum_{j=1}^{N-1}\sum_{\sigma}t^{\perp}_{i}\left(b^{\dag}_{i,j,\sigma}b_{i,j+1,\sigma}+H.c.\right),
\end{eqnarray}
where at we mulitiply the hopping between the core and the remaining bath rungs by a factor $\Lambda$, i.e. $t_L \rightarrow \Lambda t_L$.

The fRG flow is implemented in an effective theory on the bath rung $b_{L+1}$ which follows from the original theory (\ref{eqhamiltonnsiteandersonmodel})
by integrating out the core and all bath rungs with index $i>L+1$ in a functional integral representation. Up to the fourth order in the fields, the effective action
is given by
\begin{eqnarray}
\nonumber
S^{\text{eff}}\left(\bar{b}_{L+1},b_{L+1}\right)&=&-\frac{1}{\beta}\sum_{i\omega}\sum_{\sigma}\bar{b}_{L+1,\sigma}(i\omega)\hat{Q}^{\text{eff},\Lambda}_{\sigma}(i\omega)b_{L+1,\sigma}(i\omega)\\
&&-\frac{(\Lambda t_{L})^4}{4\beta^3}\sum_{i\omega_1,i\omega_2,\atop i\omega_{1'},i\omega_{2'}}\sum_{i_1,i_2,\atop i_{1'},i_{2'}}\sum_{\sigma_1,\sigma_2,\atop \sigma_{1'},\sigma_{2'}}\bar{b}_{L+1,i_1,\sigma_1}(i\omega_1)\bar{b}_{L+1,i_2,\sigma_2}(i\omega_2)\\
\nonumber
&&\times\mathcal{G}^{c,(2)}_{core}\left(i\omega_1,b_{L},i_1,\sigma_1;i\omega_2,b_{L},i_2,\sigma_2|i\omega_{1'},b_{L},i_{1'},\sigma_{1'};i\omega_{2'},b_{L},i_{2'},\sigma_{2'}\right)\\ 
&&\times b_{L+1,i_{1'},\sigma_{1'}}(i\omega_{1'})b_{L+1,i_{2'},\sigma_{2'}}(i\omega_{2'}) \delta_{\omega_1+\omega_2,\omega_{1'}+\omega_{2'}}\delta_{\sigma_1+\sigma_2,\sigma_{1'}+\sigma_{2'}}
\end{eqnarray}
with
\begin{equation}
\hat{Q}^{\text{eff},\Lambda}_{\sigma}(i\omega)=
i\omega\textbf{1}-\hat{t}^{\perp}_{L+1}-(\Lambda t_L)^2\hat{\mathcal{G}}^{c,(1)}_{\text{core},\sigma}(i\omega,b_L,b_L)-t^2_{L+1}\hat{g}_{b_{L+2},b_{L+3},...}(i\omega,b_{L+2},b_{L+2}).
\end{equation}
Here we used the abbreviation $\bar{b}_{L+1,\sigma}=\left(\bar{b}_{L+1,1,\sigma},\bar{b}_{L+1,2,\sigma},...,\bar{b}_{L+1,N,\sigma}\right)$ for vectors in the $b_{L+1}$ rung subspace. Matrices
in this space are denoted by a hat. $\mathcal{G}^{c,(n)}_{\text{core}}$ is the connected n-particle Green's function of the isolated core and $g_{b_{L+2},b_{L+3},...}$ the 
one-particle Green's function of the bath. $\hat{t}^{\perp}_{L+1}$ is the free hopping matrix on rung $L+1$. Note, that an additional frequency-dependent local term, as for example
the local self-energy that arise in a DMFT cycle can be easily included in $\hat{Q}^{\text{eff},\Lambda}_{\sigma}$.

The fRG flow equations form an infinite set of differential equations with respect to the parameter $\Lambda$ for the one-particle irreducible vertex
functions.\cite{Wet93,Sal01,Met12} We truncate them by neglecting the flow of the three-particle vertex and all higher vertex functions. Then we are left with a coupled set of
flow equations for the self-energy $\Sigma^{\Lambda}_{\text{eff}}$ and the two-particle vertex
$\Gamma^{\Lambda}_{\text{eff}}$. The latter can be separated into two different spin channels like in Eq. \ref{eqzerlegunggreensfunktionspinchannel}
and we denote the direct part as $V^{\Lambda}_{\text{eff}}$. The flow equations are given by
\begin{eqnarray}
\nonumber
\frac{d}{d\Lambda}\Sigma^{\Lambda}_{\text{eff}}\left(i\omega,i_1,i_{1'}\right)&=&-\frac{1}{\beta}\sum_{i\omega'}\sum_{i_2,i_{2'}}
S^{\Lambda}_{\text{eff}}\left(i\omega',i_{2'},i_2\right)
\big[2 V^{\Lambda}_{\text{eff}}\left(i_1,i\omega;i_2,i\omega'|i_{1'},i\omega;i_{2'},i\omega'\right)\\
&&-V^{\Lambda}_{\text{eff}}\left(i_1,i\omega;i_2,i\omega'|i_{2'},i\omega';i_{1'},i\omega\right)\big]\label{eqflussgleichungselbstenergie}\\
\nonumber
\frac{d}{d\Lambda}V^{\Lambda}_{\text{eff}}\left(i_1,i\omega_1;i_2,i\omega_2|i_{1'},i\omega_{1'};i_{2'},i\omega_{2'}\right)&=&\Phi^{\Lambda}_{\text{pp}}\left(i_1,i\omega_1;i_2,i\omega_2|i_{1'},i\omega_{1'};i_{2'},i\omega_{2'}\right)
+\Phi^{\Lambda}_{\text{dph}}\left(i_1,i\omega_1;i_2,i\omega_2|i_{1'},i\omega_{1'};i_{2'},i\omega_{2'}\right)\\
&&+\Phi^{\Lambda}_{\text{crph}}\left(i_1,i\omega_1;i_2,i\omega_2|i_{1'},i\omega_{1'};i_{2'},i\omega_{2'}\right)\label{eqflussgleichungzweiteilchenvertex}
\end{eqnarray}
with
\begin{eqnarray}
\label{eqppchannel}
\Phi^{\Lambda}_{\text{pp}}\left(i_1,i\omega_1;i_2,i\omega_2|i_{1'},i\omega_{1'};i_{2'},i\omega_{2'}\right)
&=&\frac{1}{\beta}\sum_{i\omega_3,i\omega_4}\sum_{i_3,i_4\atop i_{3'},i_{4'}}L^{\Lambda}\left(i\omega_3,i\omega_4,i_{3'},i_3,i_{4'},i_4\right)\\
\nonumber
&&\times V^{\Lambda}_{\text{eff}}\left(i_3,i\omega_3;i_4,i\omega_4|i_{1'},i\omega_{1'};i_{2'},i\omega_{2'}\right)
V^{\Lambda}_{\text{eff}}\left(i_1,i\omega_1;i_2,i\omega_2|i_{3'},i\omega_3;i_{4'},i\omega_4\right)\\
\label{eqdphchannel}
\Phi^{\Lambda}_{\text{dph}}\left(i_1,i\omega_1;i_2,i\omega_2|i_{1'},i\omega_{1'};i_{2'},i\omega_{2'}\right)
&=&-\frac{1}{\beta}\sum_{i\omega_3,i\omega_4}\sum_{i_3,i_4\atop i_{3'},i_{4'}}L^{\Lambda}\left(i\omega_3,i\omega_4,i_{3'},i_3,i_{4'},i_4\right)\\
&&\times\big[2 V^{\Lambda}_{\text{eff}}\left(i_1,i\omega_1;i_3,i\omega_3|i_{1'},i\omega_{1'};i_{4'},i\omega_4\right)
V^{\Lambda}_{\text{eff}}\left(i_2,i\omega_2;i_4,i\omega_4|i_{2'},i\omega_{2'};i_{3'},i\omega_3\right)\nonumber\\
&&- V^{\Lambda}_{\text{eff}}\left(i_1,i\omega_1;i_3,i\omega_3|i_{1'},i\omega_{1'};i_{4'},i\omega_4\right)
V^{\Lambda}_{\text{eff}}\left(i_2,i\omega_2;i_4,i\omega_4|i_{3'},i\omega_3;i_{2'},i\omega_{2'}\right)\nonumber\\
&&- V^{\Lambda}_{\text{eff}}\left(i_1,i\omega_1;i_3,i\omega_3|i_{4'},i\omega_4;i_{1'},i\omega_{1'}\right)
V^{\Lambda}_{\text{eff}}\left(i_2,i\omega_2;i_4,i\omega_4|i_{2'},i\omega_{2'};i_{3'},i\omega_3\right)\big]\nonumber\\
\label{eqcrphchannel}
\Phi^{\Lambda}_{\text{crph}}\left(i_1,i\omega_1;i_2,i\omega_2|i_{1'},i\omega_{1'};i_{2'},i\omega_{2'}\right)
&=&\frac{1}{\beta}\sum_{i\omega_3,i\omega_4}\sum_{i_3,i_4\atop i_{3'},i_{4'}}L^{\Lambda}\left(i\omega_3,i\omega_4,i_{3'},i_3,i_{4'},i_4\right)\\
\nonumber
&&\times V^{\Lambda}_{\text{eff}}\left(i_2,i\omega_2;i_3,i\omega_3|i_{4'},i\omega_4;i_{1'},i\omega_{1'}\right)
V^{\Lambda}_{\text{eff}}\left(i_1,i\omega_1;i_4,i\omega_4|i_{3'},i\omega_3;i_{2'},i\omega_{2'}\right)
\end{eqnarray}
and the single scale propagator
\begin{equation}
\hat{S}^{\Lambda}_{\text{eff}}(i\omega)=\hat{\mathcal{G}}^{\Lambda}_{\text{eff}}(i\omega)\frac{d}{d\Lambda}\left[\hat{Q}^{\text{eff},\Lambda}(i\omega)\right]\hat{\mathcal{G}}^{\Lambda}_{\text{eff}}(i\omega).
\end{equation}
The function $L^{\Lambda}$ is defined as
\begin{equation}
L^{\Lambda}\left(i\omega_1,i\omega_2,i_1,i_2,i_3,i_4\right)=\mathcal{G}^{\Lambda}_{\text{eff}}\left(i\omega_1,i_1,i_2\right)S^{\Lambda}_{\text{eff}}\left(i\omega_2,i_3,i_4\right)
+S^{\Lambda}_{\text{eff}}\left(i\omega_1,i_1,i_2\right)\mathcal{G}^{\Lambda}_{\text{eff}}\left(i\omega_2,i_3,i_4\right)
\end{equation}
with the Green's function
\begin{equation}
\left[\hat{\mathcal{G}}^{\Lambda}_{\text{eff}}(i\omega)\right]^{-1}=\hat{Q}^{\text{eff},\Lambda}(i\omega)-\hat{\Sigma}^{\Lambda}_{\text{eff}}(i\omega).
\end{equation}
We use the following replacement in the flow equation for the vertex function (\ref{eqflussgleichungzweiteilchenvertex}), which is motivated by the fulfillment of
Ward identities in the fRG flow\cite{Kat04}
\begin{equation}
\hat{S}^{\Lambda}_{\text{eff}}\rightarrow-\frac{d\hat{\mathcal{G}}^{\Lambda}_{\text{eff}}}{d\Lambda}=\hat{S}^{\Lambda}_{\text{eff}}-\hat{\mathcal{G}}^{\Lambda}_{\text{eff}}\frac{d\hat{\Sigma}^{\Lambda}_\text{eff}}{d\Lambda}.
\end{equation}
The initial conditions for $\Lambda=0$ are
\begin{eqnarray}
\Sigma^{\Lambda=0}_{\text{eff}}\left(i\omega,i_1,i_{1'}\right)&=&0,\\
V^{\Lambda=0}_{\text{eff}}\left(i_1,i\omega_1;i_2,i\omega_2|i_{1'},i\omega_{1'};i_{2'},i\omega_{2'}\right)&=&t_{L}^4
\mathcal{G}^{c,(2)}_{\text{core}}\left(b_L,i_1,\uparrow,i\omega_1;b_L,i_2,\downarrow,i\omega_2|b_L,i_{1'},\uparrow,i\omega_{1'};b_L,i_{2'},\downarrow,i\omega_{2'}\right).
\end{eqnarray}
In the most simple approximation the flow of $V^{\Lambda}_{\text{eff}}$ is neglected and only the flow equation for the self-energy (\ref{eqflussgleichungselbstenergie}) is integrated 
('approximation 1'). Integrating the full set of flow equations (\ref{eqflussgleichungselbstenergie}) and (\ref{eqflussgleichungzweiteilchenvertex}) is denoted as 'approximation 2'.

Finally one needs relations that connects the vertices of the effective theory with those of the original theory $\Sigma$ and $\Gamma$.
In the setup of the effective theory one can derive the local Green's function on rung $b_{L+1}$ from the effective self-energy 
$\hat{\Sigma}_{\text{eff}}\equiv\hat{\Sigma}^{\Lambda=1}_{\text{eff}}$
\begin{equation}
\hat{\mathcal{G}}_{\sigma}(i\omega,b_{l+1},b_{L+1})=\left[\hat{Q}^{\text{eff},\Lambda=1}_{\sigma}(i\omega)-\hat{\Sigma}_{\text{eff}}(i\omega)\right]^{-1}.
\end{equation}
The same Green's function can be derived in the setup of the original theory and one can use its functional dependence on the dot self-energy
$\hat{\Sigma}(i\omega)$ to derive a relation between $\hat{\Sigma}_{\text{eff}}$ and $\hat{\Sigma}$. For $L=0$ this relation is given by
\begin{equation}
\hat{\Sigma}(i\omega)=i\omega\textbf{1}-\hat{t}^{\perp}_{0}
-t_{0}^{2}\left[t_{0}^{2}\hat{\mathcal{G}}^{c,(1)}_{\text{core},\sigma}(i\omega,b_L,b_L)+\hat{\Sigma}_{\text{eff}}(i\omega)\right]^{-1}.
\end{equation}
In a similar way one gets the local 1PI vertex function on the dot site from the vertex function of the effective theory
$\Gamma^{\Lambda=1}_{\text{eff}}\equiv\Gamma_{\text{eff}}$. The connected two-particle
Green's function on rung $b_{L+1}$ is given by
\begin{eqnarray}
\nonumber
\mathcal{G}^{c,(2)}\left(b_{L+1},\alpha_1;b_{L+1},\alpha_2|b_{L+1},\alpha_{1'};b_{L+1},\alpha_{2'}\right)
&=&-\sum_{\beta_1,\beta_2 \atop \beta_{1'},\beta_{2'}}\hat{\mathcal{G}}(b_{L+1}\alpha_1;b_{L+1},\beta_1)
\hat{\mathcal{G}}(b_{L+1}\alpha_2;b_{L+1},\beta_2)\Gamma_{\text{eff}}\left(\beta_1,\beta_2|\beta_{1'},\beta_{2'}\right)\\
&&\times\hat{\mathcal{G}}(b_{L+1}\beta_{1'};b_{L+1},\alpha_{1'})\hat{\mathcal{G}}(b_{L+1}\beta_{2'};b_{L+1},\alpha_{2'}).
\end{eqnarray}
Here greek indices are super-indices that contain Matsubara frequency, spin and channel-index.
By amputing the Green's function that connect the dot rung with rung $b_{L+1}$ one gets the vertex function on the dot rung
\begin{eqnarray}
\nonumber
\Gamma(d,\alpha_1;d,\alpha_2|d,\alpha_{1'};d,\alpha_{2'})
&=&-\sum_{\beta_1,\beta_2 \atop \beta_{1'},\beta_{2'}}\left[\hat{\mathcal{G}}(b_{L+1},\beta_1;d,\alpha_1)\right]^{-1}
\left[\hat{\mathcal{G}}(b_{L+1},\beta_2;d,\alpha_2)\right]^{-1}\\
\nonumber
&&\times\mathcal{G}^{c,(2)}\left(b_{L+1},\beta_1;b_{L+1},\beta_2|b_{L+1},\beta_{1'};b_{L+1},\beta_{2'}\right)\\
&&\times\left[\hat{\mathcal{G}}(d,\alpha_{1'};b_{L+1},\beta_{1'})\right]^{-1}
\left[\hat{\mathcal{G}}(d,\alpha_{1'};b_{L+1},\beta_{1'})\right]^{-1}.
\end{eqnarray}
\end{widetext}
\bibliography{literatur}

\begin{thebibliography}{67}%
\makeatletter
\providecommand \@ifxundefined [1]{%
 \@ifx{#1\undefined}
}%
\providecommand \@ifnum [1]{%
 \ifnum #1\expandafter \@firstoftwo
 \else \expandafter \@secondoftwo
 \fi
}%
\providecommand \@ifx [1]{%
 \ifx #1\expandafter \@firstoftwo
 \else \expandafter \@secondoftwo
 \fi
}%
\providecommand \natexlab [1]{#1}%
\providecommand \enquote  [1]{``#1''}%
\providecommand \bibnamefont  [1]{#1}%
\providecommand \bibfnamefont [1]{#1}%
\providecommand \citenamefont [1]{#1}%
\providecommand \href@noop [0]{\@secondoftwo}%
\providecommand \href [0]{\begingroup \@sanitize@url \@href}%
\providecommand \@href[1]{\@@startlink{#1}\@@href}%
\providecommand \@@href[1]{\endgroup#1\@@endlink}%
\providecommand \@sanitize@url [0]{\catcode `\\12\catcode `\$12\catcode
  `\&12\catcode `\#12\catcode `\^12\catcode `\_12\catcode `\%12\relax}%
\providecommand \@@startlink[1]{}%
\providecommand \@@endlink[0]{}%
\providecommand \url  [0]{\begingroup\@sanitize@url \@url }%
\providecommand \@url [1]{\endgroup\@href {#1}{\urlprefix }}%
\providecommand \urlprefix  [0]{URL }%
\providecommand \Eprint [0]{\href }%
\providecommand \doibase [0]{http://dx.doi.org/}%
\providecommand \selectlanguage [0]{\@gobble}%
\providecommand \bibinfo  [0]{\@secondoftwo}%
\providecommand \bibfield  [0]{\@secondoftwo}%
\providecommand \translation [1]{[#1]}%
\providecommand \BibitemOpen [0]{}%
\providecommand \bibitemStop [0]{}%
\providecommand \bibitemNoStop [0]{.\EOS\space}%
\providecommand \EOS [0]{\spacefactor3000\relax}%
\providecommand \BibitemShut  [1]{\csname bibitem#1\endcsname}%
\let\auto@bib@innerbib\@empty
%</preamble>
\bibitem [{\citenamefont {Mott}(1968)}]{Mott68}%
  \BibitemOpen
  \bibfield  {author} {\bibinfo {author} {\bibfnamefont {N.~F.}\ \bibnamefont
  {Mott}},\ }\href {\doibase 10.1103/RevModPhys.40.677} {\bibfield  {journal}
  {\bibinfo  {journal} {Rev. Mod. Phys.}\ }\textbf {\bibinfo {volume} {40}},\
  \bibinfo {pages} {677} (\bibinfo {year} {1968})}\BibitemShut {NoStop}%
\bibitem [{\citenamefont {Gebhard}(1997)}]{Geb97}%
  \BibitemOpen
  \bibfield  {author} {\bibinfo {author} {\bibfnamefont {F.}~\bibnamefont
  {Gebhard}},\ }\href@noop {} {\emph {\bibinfo {title} {The Mott
  Metal-Insulator Transition}}}\ (\bibinfo  {publisher} {Springer},\ \bibinfo
  {year} {1997})\BibitemShut {NoStop}%
\bibitem [{\citenamefont {Hubbard}(1963)}]{Hub63}%
  \BibitemOpen
  \bibfield  {author} {\bibinfo {author} {\bibfnamefont {J.}~\bibnamefont
  {Hubbard}},\ }\href {\doibase 10.1098/rspa.1963.0204} {\bibfield  {journal}
  {\bibinfo  {journal} {Proceedings of the Royal Society of London. Series A.
  Mathematical and Physical Sciences}\ }\textbf {\bibinfo {volume} {276}},\
  \bibinfo {pages} {238} (\bibinfo {year} {1963})}\BibitemShut {NoStop}%
\bibitem [{\citenamefont {Gutzwiller}(1963)}]{Gut63}%
  \BibitemOpen
  \bibfield  {author} {\bibinfo {author} {\bibfnamefont {M.~C.}\ \bibnamefont
  {Gutzwiller}},\ }\href {\doibase 10.1103/PhysRevLett.10.159} {\bibfield
  {journal} {\bibinfo  {journal} {Phys. Rev. Lett.}\ }\textbf {\bibinfo
  {volume} {10}},\ \bibinfo {pages} {159} (\bibinfo {year} {1963})}\BibitemShut
  {NoStop}%
\bibitem [{\citenamefont {Kanamori}(1963)}]{Kan63}%
  \BibitemOpen
  \bibfield  {author} {\bibinfo {author} {\bibfnamefont {J.}~\bibnamefont
  {Kanamori}},\ }\href {\doibase 10.1143/PTP.30.275} {\bibfield  {journal}
  {\bibinfo  {journal} {Progress of Theoretical Physics}\ }\textbf {\bibinfo
  {volume} {30}},\ \bibinfo {pages} {275} (\bibinfo {year} {1963})}\BibitemShut
  {NoStop}%
\bibitem [{\citenamefont {McWhan}\ \emph {et~al.}(1973)\citenamefont {McWhan},
  \citenamefont {Menth}, \citenamefont {Remeika}, \citenamefont {Brinkman},\
  and\ \citenamefont {Rice}}]{McWhan73}%
  \BibitemOpen
  \bibfield  {author} {\bibinfo {author} {\bibfnamefont {D.~B.}\ \bibnamefont
  {McWhan}}, \bibinfo {author} {\bibfnamefont {A.}~\bibnamefont {Menth}},
  \bibinfo {author} {\bibfnamefont {J.~P.}\ \bibnamefont {Remeika}}, \bibinfo
  {author} {\bibfnamefont {W.~F.}\ \bibnamefont {Brinkman}}, \ and\ \bibinfo
  {author} {\bibfnamefont {T.~M.}\ \bibnamefont {Rice}},\ }\href {\doibase
  10.1103/PhysRevB.7.1920} {\bibfield  {journal} {\bibinfo  {journal} {Phys.
  Rev. B}\ }\textbf {\bibinfo {volume} {7}},\ \bibinfo {pages} {1920} (\bibinfo
  {year} {1973})}\BibitemShut {NoStop}%
\bibitem [{\citenamefont {Lee}\ \emph {et~al.}(2006)\citenamefont {Lee},
  \citenamefont {Nagaosa},\ and\ \citenamefont {Wen}}]{Lee06}%
  \BibitemOpen
  \bibfield  {author} {\bibinfo {author} {\bibfnamefont {P.~A.}\ \bibnamefont
  {Lee}}, \bibinfo {author} {\bibfnamefont {N.}~\bibnamefont {Nagaosa}}, \ and\
  \bibinfo {author} {\bibfnamefont {X.-G.}\ \bibnamefont {Wen}},\ }\href
  {\doibase 10.1103/RevModPhys.78.17} {\bibfield  {journal} {\bibinfo
  {journal} {Rev. Mod. Phys.}\ }\textbf {\bibinfo {volume} {78}},\ \bibinfo
  {pages} {17} (\bibinfo {year} {2006})}\BibitemShut {NoStop}%
\bibitem [{\citenamefont {Gutzwiller}(1965)}]{Gut65}%
  \BibitemOpen
  \bibfield  {author} {\bibinfo {author} {\bibfnamefont {M.~C.}\ \bibnamefont
  {Gutzwiller}},\ }\href {\doibase 10.1103/PhysRev.137.A1726} {\bibfield
  {journal} {\bibinfo  {journal} {Phys. Rev.}\ }\textbf {\bibinfo {volume}
  {137}},\ \bibinfo {pages} {A1726} (\bibinfo {year} {1965})}\BibitemShut
  {NoStop}%
\bibitem [{\citenamefont {Brinkman}\ and\ \citenamefont {Rice}(1970)}]{Bri70}%
  \BibitemOpen
  \bibfield  {author} {\bibinfo {author} {\bibfnamefont {W.~F.}\ \bibnamefont
  {Brinkman}}\ and\ \bibinfo {author} {\bibfnamefont {T.~M.}\ \bibnamefont
  {Rice}},\ }\href {\doibase 10.1103/PhysRevB.2.4302} {\bibfield  {journal}
  {\bibinfo  {journal} {Phys. Rev. B}\ }\textbf {\bibinfo {volume} {2}},\
  \bibinfo {pages} {4302} (\bibinfo {year} {1970})}\BibitemShut {NoStop}%
\bibitem [{\citenamefont {Georges}\ \emph {et~al.}(1996)\citenamefont
  {Georges}, \citenamefont {Kotliar}, \citenamefont {Krauth},\ and\
  \citenamefont {Rozenberg}}]{Geo96}%
  \BibitemOpen
  \bibfield  {author} {\bibinfo {author} {\bibfnamefont {A.}~\bibnamefont
  {Georges}}, \bibinfo {author} {\bibfnamefont {G.}~\bibnamefont {Kotliar}},
  \bibinfo {author} {\bibfnamefont {W.}~\bibnamefont {Krauth}}, \ and\ \bibinfo
  {author} {\bibfnamefont {M.~J.}\ \bibnamefont {Rozenberg}},\ }\href {\doibase
  10.1103/RevModPhys.68.13} {\bibfield  {journal} {\bibinfo  {journal} {Rev.
  Mod. Phys.}\ }\textbf {\bibinfo {volume} {68}},\ \bibinfo {pages} {13}
  (\bibinfo {year} {1996})}\BibitemShut {NoStop}%
\bibitem [{\citenamefont {Pruschke}\ \emph {et~al.}(1995)\citenamefont
  {Pruschke}, \citenamefont {Jarrell},\ and\ \citenamefont
  {Freericks}}]{Pru95}%
  \BibitemOpen
  \bibfield  {author} {\bibinfo {author} {\bibfnamefont {T.}~\bibnamefont
  {Pruschke}}, \bibinfo {author} {\bibfnamefont {M.}~\bibnamefont {Jarrell}}, \
  and\ \bibinfo {author} {\bibfnamefont {J.}~\bibnamefont {Freericks}},\ }\href
  {\doibase 10.1080/00018739500101526} {\bibfield  {journal} {\bibinfo
  {journal} {Advances in Physics}\ }\textbf {\bibinfo {volume} {44}},\ \bibinfo
  {pages} {187} (\bibinfo {year} {1995})}\BibitemShut {NoStop}%
\bibitem [{\citenamefont {Kotliar}\ and\ \citenamefont
  {Vollhardt}(2004)}]{Kot04}%
  \BibitemOpen
  \bibfield  {author} {\bibinfo {author} {\bibfnamefont {G.}~\bibnamefont
  {Kotliar}}\ and\ \bibinfo {author} {\bibfnamefont {D.}~\bibnamefont
  {Vollhardt}},\ }\href {\doibase 10.1063/1.1712502} {\bibfield  {journal}
  {\bibinfo  {journal} {Physics Today}\ }\textbf {\bibinfo {volume} {57}},\
  \bibinfo {pages} {53} (\bibinfo {year} {2004})}\BibitemShut {NoStop}%
\bibitem [{\citenamefont {Maier}\ \emph {et~al.}(2005)\citenamefont {Maier},
  \citenamefont {Jarrell}, \citenamefont {Pruschke},\ and\ \citenamefont
  {Hettler}}]{Mai05}%
  \BibitemOpen
  \bibfield  {author} {\bibinfo {author} {\bibfnamefont {T.}~\bibnamefont
  {Maier}}, \bibinfo {author} {\bibfnamefont {M.}~\bibnamefont {Jarrell}},
  \bibinfo {author} {\bibfnamefont {T.}~\bibnamefont {Pruschke}}, \ and\
  \bibinfo {author} {\bibfnamefont {M.~H.}\ \bibnamefont {Hettler}},\ }\href
  {\doibase 10.1103/RevModPhys.77.1027} {\bibfield  {journal} {\bibinfo
  {journal} {Rev. Mod. Phys.}\ }\textbf {\bibinfo {volume} {77}},\ \bibinfo
  {pages} {1027} (\bibinfo {year} {2005})}\BibitemShut {NoStop}%
\bibitem [{\citenamefont {Kotliar}\ \emph {et~al.}(2001)\citenamefont
  {Kotliar}, \citenamefont {Savrasov}, \citenamefont {P\'alsson},\ and\
  \citenamefont {Biroli}}]{Kot01}%
  \BibitemOpen
  \bibfield  {author} {\bibinfo {author} {\bibfnamefont {G.}~\bibnamefont
  {Kotliar}}, \bibinfo {author} {\bibfnamefont {S.~Y.}\ \bibnamefont
  {Savrasov}}, \bibinfo {author} {\bibfnamefont {G.}~\bibnamefont {P\'alsson}},
  \ and\ \bibinfo {author} {\bibfnamefont {G.}~\bibnamefont {Biroli}},\ }\href
  {\doibase 10.1103/PhysRevLett.87.186401} {\bibfield  {journal} {\bibinfo
  {journal} {Phys. Rev. Lett.}\ }\textbf {\bibinfo {volume} {87}},\ \bibinfo
  {pages} {186401} (\bibinfo {year} {2001})}\BibitemShut {NoStop}%
\bibitem [{\citenamefont {Lichtenstein}\ and\ \citenamefont
  {Katsnelson}(2000)}]{Lic00}%
  \BibitemOpen
  \bibfield  {author} {\bibinfo {author} {\bibfnamefont {A.~I.}\ \bibnamefont
  {Lichtenstein}}\ and\ \bibinfo {author} {\bibfnamefont {M.~I.}\ \bibnamefont
  {Katsnelson}},\ }\href {\doibase 10.1103/PhysRevB.62.R9283} {\bibfield
  {journal} {\bibinfo  {journal} {Phys. Rev. B}\ }\textbf {\bibinfo {volume}
  {62}},\ \bibinfo {pages} {R9283} (\bibinfo {year} {2000})}\BibitemShut
  {NoStop}%
\bibitem [{\citenamefont {Toschi}\ \emph {et~al.}(2007)\citenamefont {Toschi},
  \citenamefont {Katanin},\ and\ \citenamefont {Held}}]{Tos07}%
  \BibitemOpen
  \bibfield  {author} {\bibinfo {author} {\bibfnamefont {A.}~\bibnamefont
  {Toschi}}, \bibinfo {author} {\bibfnamefont {A.~A.}\ \bibnamefont {Katanin}},
  \ and\ \bibinfo {author} {\bibfnamefont {K.}~\bibnamefont {Held}},\ }\href
  {\doibase 10.1103/PhysRevB.75.045118} {\bibfield  {journal} {\bibinfo
  {journal} {Phys. Rev. B}\ }\textbf {\bibinfo {volume} {75}},\ \bibinfo
  {pages} {045118} (\bibinfo {year} {2007})}\BibitemShut {NoStop}%
\bibitem [{\citenamefont {Held}\ \emph {et~al.}(2008)\citenamefont {Held},
  \citenamefont {Katanin},\ and\ \citenamefont {Toschi}}]{Hel08}%
  \BibitemOpen
  \bibfield  {author} {\bibinfo {author} {\bibfnamefont {K.}~\bibnamefont
  {Held}}, \bibinfo {author} {\bibfnamefont {A.~A.}\ \bibnamefont {Katanin}}, \
  and\ \bibinfo {author} {\bibfnamefont {A.}~\bibnamefont {Toschi}},\ }\href
  {\doibase 10.1143/PTPS.176.117} {\bibfield  {journal} {\bibinfo  {journal}
  {Progress of Theoretical Physics Supplement}\ }\textbf {\bibinfo {volume}
  {176}},\ \bibinfo {pages} {117} (\bibinfo {year} {2008})}\BibitemShut
  {NoStop}%
\bibitem [{\citenamefont {Rohringer}\ \emph {et~al.}(2011)\citenamefont
  {Rohringer}, \citenamefont {Toschi}, \citenamefont {Katanin},\ and\
  \citenamefont {Held}}]{Roh11}%
  \BibitemOpen
  \bibfield  {author} {\bibinfo {author} {\bibfnamefont {G.}~\bibnamefont
  {Rohringer}}, \bibinfo {author} {\bibfnamefont {A.}~\bibnamefont {Toschi}},
  \bibinfo {author} {\bibfnamefont {A.}~\bibnamefont {Katanin}}, \ and\
  \bibinfo {author} {\bibfnamefont {K.}~\bibnamefont {Held}},\ }\href {\doibase
  10.1103/PhysRevLett.107.256402} {\bibfield  {journal} {\bibinfo  {journal}
  {Phys. Rev. Lett.}\ }\textbf {\bibinfo {volume} {107}},\ \bibinfo {pages}
  {256402} (\bibinfo {year} {2011})}\BibitemShut {NoStop}%
\bibitem [{\citenamefont {Rubtsov}\ \emph {et~al.}(2008)\citenamefont
  {Rubtsov}, \citenamefont {Katsnelson},\ and\ \citenamefont
  {Lichtenstein}}]{Rub08}%
  \BibitemOpen
  \bibfield  {author} {\bibinfo {author} {\bibfnamefont {A.~N.}\ \bibnamefont
  {Rubtsov}}, \bibinfo {author} {\bibfnamefont {M.~I.}\ \bibnamefont
  {Katsnelson}}, \ and\ \bibinfo {author} {\bibfnamefont {A.~I.}\ \bibnamefont
  {Lichtenstein}},\ }\href {\doibase 10.1103/PhysRevB.77.033101} {\bibfield
  {journal} {\bibinfo  {journal} {Phys. Rev. B}\ }\textbf {\bibinfo {volume}
  {77}},\ \bibinfo {pages} {033101} (\bibinfo {year} {2008})}\BibitemShut
  {NoStop}%
\bibitem [{\citenamefont {Rubtsov}\ \emph {et~al.}(2009)\citenamefont
  {Rubtsov}, \citenamefont {Katsnelson}, \citenamefont {Lichtenstein},\ and\
  \citenamefont {Georges}}]{Rub09}%
  \BibitemOpen
  \bibfield  {author} {\bibinfo {author} {\bibfnamefont {A.~N.}\ \bibnamefont
  {Rubtsov}}, \bibinfo {author} {\bibfnamefont {M.~I.}\ \bibnamefont
  {Katsnelson}}, \bibinfo {author} {\bibfnamefont {A.~I.}\ \bibnamefont
  {Lichtenstein}}, \ and\ \bibinfo {author} {\bibfnamefont {A.}~\bibnamefont
  {Georges}},\ }\href {\doibase 10.1103/PhysRevB.79.045133} {\bibfield
  {journal} {\bibinfo  {journal} {Phys. Rev. B}\ }\textbf {\bibinfo {volume}
  {79}},\ \bibinfo {pages} {045133} (\bibinfo {year} {2009})}\BibitemShut
  {NoStop}%
\bibitem [{\citenamefont {Rohringer}\ \emph {et~al.}(2013)\citenamefont
  {Rohringer}, \citenamefont {Toschi}, \citenamefont {Hafermann}, \citenamefont
  {Held}, \citenamefont {Anisimov},\ and\ \citenamefont {Katanin}}]{Roh13}%
  \BibitemOpen
  \bibfield  {author} {\bibinfo {author} {\bibfnamefont {G.}~\bibnamefont
  {Rohringer}}, \bibinfo {author} {\bibfnamefont {A.}~\bibnamefont {Toschi}},
  \bibinfo {author} {\bibfnamefont {H.}~\bibnamefont {Hafermann}}, \bibinfo
  {author} {\bibfnamefont {K.}~\bibnamefont {Held}}, \bibinfo {author}
  {\bibfnamefont {V.~I.}\ \bibnamefont {Anisimov}}, \ and\ \bibinfo {author}
  {\bibfnamefont {A.~A.}\ \bibnamefont {Katanin}},\ }\href {\doibase
  10.1103/PhysRevB.88.115112} {\bibfield  {journal} {\bibinfo  {journal} {Phys.
  Rev. B}\ }\textbf {\bibinfo {volume} {88}},\ \bibinfo {pages} {115112}
  (\bibinfo {year} {2013})}\BibitemShut {NoStop}%
\bibitem [{\citenamefont {Slezak}\ \emph {et~al.}(2009)\citenamefont {Slezak},
  \citenamefont {Jarrell}, \citenamefont {Maier},\ and\ \citenamefont
  {Deisz}}]{Sle09}%
  \BibitemOpen
  \bibfield  {author} {\bibinfo {author} {\bibfnamefont {C.}~\bibnamefont
  {Slezak}}, \bibinfo {author} {\bibfnamefont {M.}~\bibnamefont {Jarrell}},
  \bibinfo {author} {\bibfnamefont {T.}~\bibnamefont {Maier}}, \ and\ \bibinfo
  {author} {\bibfnamefont {J.}~\bibnamefont {Deisz}},\ }\href
  {http://stacks.iop.org/0953-8984/21/i=43/a=435604} {\bibfield  {journal}
  {\bibinfo  {journal} {Journal of Physics: Condensed Matter}\ }\textbf
  {\bibinfo {volume} {21}},\ \bibinfo {pages} {435604} (\bibinfo {year}
  {2009})}\BibitemShut {NoStop}%
\bibitem [{\citenamefont {Metzner}\ \emph {et~al.}(2012)\citenamefont
  {Metzner}, \citenamefont {Salmhofer}, \citenamefont {Honerkamp},
  \citenamefont {Meden},\ and\ \citenamefont {Sch\"onhammer}}]{Met12}%
  \BibitemOpen
  \bibfield  {author} {\bibinfo {author} {\bibfnamefont {W.}~\bibnamefont
  {Metzner}}, \bibinfo {author} {\bibfnamefont {M.}~\bibnamefont {Salmhofer}},
  \bibinfo {author} {\bibfnamefont {C.}~\bibnamefont {Honerkamp}}, \bibinfo
  {author} {\bibfnamefont {V.}~\bibnamefont {Meden}}, \ and\ \bibinfo {author}
  {\bibfnamefont {K.}~\bibnamefont {Sch\"onhammer}},\ }\href {\doibase
  10.1103/RevModPhys.84.299} {\bibfield  {journal} {\bibinfo  {journal} {Rev.
  Mod. Phys.}\ }\textbf {\bibinfo {volume} {84}},\ \bibinfo {pages} {299}
  (\bibinfo {year} {2012})}\BibitemShut {NoStop}%
\bibitem [{\citenamefont {Uebelacker}\ and\ \citenamefont
  {Honerkamp}(2012)}]{Ueb12}%
  \BibitemOpen
  \bibfield  {author} {\bibinfo {author} {\bibfnamefont {S.}~\bibnamefont
  {Uebelacker}}\ and\ \bibinfo {author} {\bibfnamefont {C.}~\bibnamefont
  {Honerkamp}},\ }\href {\doibase 10.1103/PhysRevB.86.235140} {\bibfield
  {journal} {\bibinfo  {journal} {Phys. Rev. B}\ }\textbf {\bibinfo {volume}
  {86}},\ \bibinfo {pages} {235140} (\bibinfo {year} {2012})}\BibitemShut
  {NoStop}%
\bibitem [{\citenamefont {Giering}\ and\ \citenamefont
  {Salmhofer}(2012)}]{Gie12}%
  \BibitemOpen
  \bibfield  {author} {\bibinfo {author} {\bibfnamefont {K.-U.}\ \bibnamefont
  {Giering}}\ and\ \bibinfo {author} {\bibfnamefont {M.}~\bibnamefont
  {Salmhofer}},\ }\href {\doibase 10.1103/PhysRevB.86.245122} {\bibfield
  {journal} {\bibinfo  {journal} {Phys. Rev. B}\ }\textbf {\bibinfo {volume}
  {86}},\ \bibinfo {pages} {245122} (\bibinfo {year} {2012})}\BibitemShut
  {NoStop}%
\bibitem [{\citenamefont {Kinza}\ \emph {et~al.}(2013)\citenamefont {Kinza},
  \citenamefont {Ortloff}, \citenamefont {Bauer},\ and\ \citenamefont
  {Honerkamp}}]{Kin13}%
  \BibitemOpen
  \bibfield  {author} {\bibinfo {author} {\bibfnamefont {M.}~\bibnamefont
  {Kinza}}, \bibinfo {author} {\bibfnamefont {J.}~\bibnamefont {Ortloff}},
  \bibinfo {author} {\bibfnamefont {J.}~\bibnamefont {Bauer}}, \ and\ \bibinfo
  {author} {\bibfnamefont {C.}~\bibnamefont {Honerkamp}},\ }\href {\doibase
  10.1103/PhysRevB.87.035111} {\bibfield  {journal} {\bibinfo  {journal} {Phys.
  Rev. B}\ }\textbf {\bibinfo {volume} {87}},\ \bibinfo {pages} {035111}
  (\bibinfo {year} {2013})}\BibitemShut {NoStop}%
\bibitem [{\citenamefont {Ran\ifmmode~\mbox{\c{c}}\else \c{c}\fi{}on}\ and\
  \citenamefont {Dupuis}(2011{\natexlab{a}})}]{Ran11}%
  \BibitemOpen
  \bibfield  {author} {\bibinfo {author} {\bibfnamefont {A.}~\bibnamefont
  {Ran\ifmmode~\mbox{\c{c}}\else \c{c}\fi{}on}}\ and\ \bibinfo {author}
  {\bibfnamefont {N.}~\bibnamefont {Dupuis}},\ }\href {\doibase
  10.1103/PhysRevB.84.174513} {\bibfield  {journal} {\bibinfo  {journal} {Phys.
  Rev. B}\ }\textbf {\bibinfo {volume} {84}},\ \bibinfo {pages} {174513}
  (\bibinfo {year} {2011}{\natexlab{a}})}\BibitemShut {NoStop}%
\bibitem [{\citenamefont {Ran\ifmmode~\mbox{\c{c}}\else \c{c}\fi{}on}\ and\
  \citenamefont {Dupuis}(2011{\natexlab{b}})}]{Ran11b}%
  \BibitemOpen
  \bibfield  {author} {\bibinfo {author} {\bibfnamefont {A.}~\bibnamefont
  {Ran\ifmmode~\mbox{\c{c}}\else \c{c}\fi{}on}}\ and\ \bibinfo {author}
  {\bibfnamefont {N.}~\bibnamefont {Dupuis}},\ }\href {\doibase
  10.1103/PhysRevB.83.172501} {\bibfield  {journal} {\bibinfo  {journal} {Phys.
  Rev. B}\ }\textbf {\bibinfo {volume} {83}},\ \bibinfo {pages} {172501}
  (\bibinfo {year} {2011}{\natexlab{b}})}\BibitemShut {NoStop}%
\bibitem [{\citenamefont {Reuther}\ and\ \citenamefont
  {Thomale}(2013)}]{Reu13}%
  \BibitemOpen
  \bibfield  {author} {\bibinfo {author} {\bibfnamefont {J.}~\bibnamefont
  {Reuther}}\ and\ \bibinfo {author} {\bibfnamefont {R.}~\bibnamefont
  {Thomale}},\ }\href@noop {} {\bibfield  {journal} {\bibinfo  {journal} {ArXiv
  e-prints}\ } (\bibinfo {year} {2013})},\ \Eprint
  {http://arxiv.org/abs/1309.3262} {arXiv:1309.3262 [cond-mat.str-el]}
  \BibitemShut {NoStop}%
\bibitem [{\citenamefont {Husemann}\ and\ \citenamefont
  {Salmhofer}(2009)}]{Hus09}%
  \BibitemOpen
  \bibfield  {author} {\bibinfo {author} {\bibfnamefont {C.}~\bibnamefont
  {Husemann}}\ and\ \bibinfo {author} {\bibfnamefont {M.}~\bibnamefont
  {Salmhofer}},\ }\href {\doibase 10.1103/PhysRevB.79.195125} {\bibfield
  {journal} {\bibinfo  {journal} {Phys. Rev. B}\ }\textbf {\bibinfo {volume}
  {79}},\ \bibinfo {pages} {195125} (\bibinfo {year} {2009})}\BibitemShut
  {NoStop}%
\bibitem [{\citenamefont {Xiang}\ \emph {et~al.}(2012)\citenamefont {Xiang},
  \citenamefont {Wang}, \citenamefont {Wang},\ and\ \citenamefont
  {Lee}}]{Xia12}%
  \BibitemOpen
  \bibfield  {author} {\bibinfo {author} {\bibfnamefont {Y.-Y.}\ \bibnamefont
  {Xiang}}, \bibinfo {author} {\bibfnamefont {W.-S.}\ \bibnamefont {Wang}},
  \bibinfo {author} {\bibfnamefont {Q.-H.}\ \bibnamefont {Wang}}, \ and\
  \bibinfo {author} {\bibfnamefont {D.-H.}\ \bibnamefont {Lee}},\ }\href
  {\doibase 10.1103/PhysRevB.86.024523} {\bibfield  {journal} {\bibinfo
  {journal} {Phys. Rev. B}\ }\textbf {\bibinfo {volume} {86}},\ \bibinfo
  {pages} {024523} (\bibinfo {year} {2012})}\BibitemShut {NoStop}%
\bibitem [{\citenamefont {Wang}\ \emph {et~al.}(2013)\citenamefont {Wang},
  \citenamefont {Li}, \citenamefont {Xiang},\ and\ \citenamefont
  {Wang}}]{Wan13}%
  \BibitemOpen
  \bibfield  {author} {\bibinfo {author} {\bibfnamefont {W.-S.}\ \bibnamefont
  {Wang}}, \bibinfo {author} {\bibfnamefont {Z.-Z.}\ \bibnamefont {Li}},
  \bibinfo {author} {\bibfnamefont {Y.-Y.}\ \bibnamefont {Xiang}}, \ and\
  \bibinfo {author} {\bibfnamefont {Q.-H.}\ \bibnamefont {Wang}},\ }\href
  {\doibase 10.1103/PhysRevB.87.115135} {\bibfield  {journal} {\bibinfo
  {journal} {Phys. Rev. B}\ }\textbf {\bibinfo {volume} {87}},\ \bibinfo
  {pages} {115135} (\bibinfo {year} {2013})}\BibitemShut {NoStop}%
\bibitem [{\citenamefont {Rohringer}\ \emph {et~al.}(2012)\citenamefont
  {Rohringer}, \citenamefont {Valli},\ and\ \citenamefont {Toschi}}]{Roh12}%
  \BibitemOpen
  \bibfield  {author} {\bibinfo {author} {\bibfnamefont {G.}~\bibnamefont
  {Rohringer}}, \bibinfo {author} {\bibfnamefont {A.}~\bibnamefont {Valli}}, \
  and\ \bibinfo {author} {\bibfnamefont {A.}~\bibnamefont {Toschi}},\ }\href
  {\doibase 10.1103/PhysRevB.86.125114} {\bibfield  {journal} {\bibinfo
  {journal} {Phys. Rev. B}\ }\textbf {\bibinfo {volume} {86}},\ \bibinfo
  {pages} {125114} (\bibinfo {year} {2012})}\BibitemShut {NoStop}%
\bibitem [{\citenamefont {Metzner}\ and\ \citenamefont
  {Vollhardt}(1989)}]{Met89}%
  \BibitemOpen
  \bibfield  {author} {\bibinfo {author} {\bibfnamefont {W.}~\bibnamefont
  {Metzner}}\ and\ \bibinfo {author} {\bibfnamefont {D.}~\bibnamefont
  {Vollhardt}},\ }\href {\doibase 10.1103/PhysRevLett.62.324} {\bibfield
  {journal} {\bibinfo  {journal} {Phys. Rev. Lett.}\ }\textbf {\bibinfo
  {volume} {62}},\ \bibinfo {pages} {324} (\bibinfo {year} {1989})}\BibitemShut
  {NoStop}%
\bibitem [{\citenamefont {Economou}(2006)}]{Eco06}%
  \BibitemOpen
  \bibfield  {author} {\bibinfo {author} {\bibfnamefont {E.~N.}\ \bibnamefont
  {Economou}},\ }\href@noop {} {\emph {\bibinfo {title} {Greens Functions in
  Quantum Physics}}}\ (\bibinfo  {publisher} {Springer},\ \bibinfo {year}
  {2006})\BibitemShut {NoStop}%
\bibitem [{Note1()}]{Note1}%
  \BibitemOpen
  \bibinfo {note} {Here and in the following we denote $t^{\ast }$ by
  $t$.}\BibitemShut {Stop}%
\bibitem [{\citenamefont {Georges}\ and\ \citenamefont
  {Kotliar}(1992)}]{Geo92}%
  \BibitemOpen
  \bibfield  {author} {\bibinfo {author} {\bibfnamefont {A.}~\bibnamefont
  {Georges}}\ and\ \bibinfo {author} {\bibfnamefont {G.}~\bibnamefont
  {Kotliar}},\ }\href {\doibase 10.1103/PhysRevB.45.6479} {\bibfield  {journal}
  {\bibinfo  {journal} {Phys. Rev. B}\ }\textbf {\bibinfo {volume} {45}},\
  \bibinfo {pages} {6479} (\bibinfo {year} {1992})}\BibitemShut {NoStop}%
\bibitem [{\citenamefont {Jarrell}(1992)}]{Jar92}%
  \BibitemOpen
  \bibfield  {author} {\bibinfo {author} {\bibfnamefont {M.}~\bibnamefont
  {Jarrell}},\ }\href {\doibase 10.1103/PhysRevLett.69.168} {\bibfield
  {journal} {\bibinfo  {journal} {Phys. Rev. Lett.}\ }\textbf {\bibinfo
  {volume} {69}},\ \bibinfo {pages} {168} (\bibinfo {year} {1992})}\BibitemShut
  {NoStop}%
\bibitem [{\citenamefont {Bulla}\ \emph {et~al.}(1998)\citenamefont {Bulla},
  \citenamefont {Hewson},\ and\ \citenamefont {Pruschke}}]{Bul98}%
  \BibitemOpen
  \bibfield  {author} {\bibinfo {author} {\bibfnamefont {R.}~\bibnamefont
  {Bulla}}, \bibinfo {author} {\bibfnamefont {A.~C.}\ \bibnamefont {Hewson}}, \
  and\ \bibinfo {author} {\bibfnamefont {T.}~\bibnamefont {Pruschke}},\ }\href
  {http://stacks.iop.org/0953-8984/10/i=37/a=021} {\bibfield  {journal}
  {\bibinfo  {journal} {Journal of Physics: Condensed Matter}\ }\textbf
  {\bibinfo {volume} {10}},\ \bibinfo {pages} {8365} (\bibinfo {year}
  {1998})}\BibitemShut {NoStop}%
\bibitem [{\citenamefont {Bulla}(1999)}]{Bul99}%
  \BibitemOpen
  \bibfield  {author} {\bibinfo {author} {\bibfnamefont {R.}~\bibnamefont
  {Bulla}},\ }\href {\doibase 10.1103/PhysRevLett.83.136} {\bibfield  {journal}
  {\bibinfo  {journal} {Phys. Rev. Lett.}\ }\textbf {\bibinfo {volume} {83}},\
  \bibinfo {pages} {136} (\bibinfo {year} {1999})}\BibitemShut {NoStop}%
\bibitem [{\citenamefont {Rozenberg}\ \emph {et~al.}(1992)\citenamefont
  {Rozenberg}, \citenamefont {Zhang},\ and\ \citenamefont {Kotliar}}]{Roz92}%
  \BibitemOpen
  \bibfield  {author} {\bibinfo {author} {\bibfnamefont {M.~J.}\ \bibnamefont
  {Rozenberg}}, \bibinfo {author} {\bibfnamefont {X.~Y.}\ \bibnamefont
  {Zhang}}, \ and\ \bibinfo {author} {\bibfnamefont {G.}~\bibnamefont
  {Kotliar}},\ }\href {\doibase 10.1103/PhysRevLett.69.1236} {\bibfield
  {journal} {\bibinfo  {journal} {Phys. Rev. Lett.}\ }\textbf {\bibinfo
  {volume} {69}},\ \bibinfo {pages} {1236} (\bibinfo {year}
  {1992})}\BibitemShut {NoStop}%
\bibitem [{\citenamefont {Caffarel}\ and\ \citenamefont
  {Krauth}(1994)}]{Caf94}%
  \BibitemOpen
  \bibfield  {author} {\bibinfo {author} {\bibfnamefont {M.}~\bibnamefont
  {Caffarel}}\ and\ \bibinfo {author} {\bibfnamefont {W.}~\bibnamefont
  {Krauth}},\ }\href {\doibase 10.1103/PhysRevLett.72.1545} {\bibfield
  {journal} {\bibinfo  {journal} {Phys. Rev. Lett.}\ }\textbf {\bibinfo
  {volume} {72}},\ \bibinfo {pages} {1545} (\bibinfo {year}
  {1994})}\BibitemShut {NoStop}%
\bibitem [{\citenamefont {Liebsch}\ and\ \citenamefont {Ishida}(2012)}]{Lie12}%
  \BibitemOpen
  \bibfield  {author} {\bibinfo {author} {\bibfnamefont {A.}~\bibnamefont
  {Liebsch}}\ and\ \bibinfo {author} {\bibfnamefont {H.}~\bibnamefont
  {Ishida}},\ }\href {http://stacks.iop.org/0953-8984/24/i=5/a=053201}
  {\bibfield  {journal} {\bibinfo  {journal} {Journal of Physics: Condensed
  Matter}\ }\textbf {\bibinfo {volume} {24}},\ \bibinfo {pages} {053201}
  (\bibinfo {year} {2012})}\BibitemShut {NoStop}%
\bibitem [{\citenamefont {Hewson}(1993)}]{Hew93}%
  \BibitemOpen
  \bibfield  {author} {\bibinfo {author} {\bibfnamefont {A.}~\bibnamefont
  {Hewson}},\ }\href@noop {} {\emph {\bibinfo {title} {The Kondo Problem to
  Heavy Fermions}}}\ (\bibinfo  {publisher} {Cambridge University Press},\
  \bibinfo {year} {1993})\BibitemShut {NoStop}%
\bibitem [{\citenamefont {Wetterich}(1993)}]{Wet93}%
  \BibitemOpen
  \bibfield  {author} {\bibinfo {author} {\bibfnamefont {C.}~\bibnamefont
  {Wetterich}},\ }\href {\doibase 10.1016/0370-2693(93)90726-X} {\bibfield
  {journal} {\bibinfo  {journal} {Physics Letters B}\ }\textbf {\bibinfo
  {volume} {301}},\ \bibinfo {pages} {90} (\bibinfo {year} {1993})}\BibitemShut
  {NoStop}%
\bibitem [{\citenamefont {Salmhofer}\ and\ \citenamefont
  {Honerkamp}(2001)}]{Sal01}%
  \BibitemOpen
  \bibfield  {author} {\bibinfo {author} {\bibfnamefont {M.}~\bibnamefont
  {Salmhofer}}\ and\ \bibinfo {author} {\bibfnamefont {C.}~\bibnamefont
  {Honerkamp}},\ }\href {\doibase 10.1143/PTP.105.1} {\bibfield  {journal}
  {\bibinfo  {journal} {Progress of Theoretical Physics}\ }\textbf {\bibinfo
  {volume} {105}},\ \bibinfo {pages} {1} (\bibinfo {year} {2001})}\BibitemShut
  {NoStop}%
\bibitem [{\citenamefont {Bl\"umer}(2003)}]{Blu03}%
  \BibitemOpen
  \bibfield  {author} {\bibinfo {author} {\bibfnamefont {N.}~\bibnamefont
  {Bl\"umer}},\ }\emph {\bibinfo {title} {Metal-Insulator Transition and
  Optical Conductivity in High Dimensions}},\ \href@noop {} {\bibinfo {type}
  {Phd thesis}},\ \bibinfo  {school} {RWTH Aachen University} (\bibinfo {year}
  {2003})\BibitemShut {NoStop}%
\bibitem [{\citenamefont {Potthoff}(2001)}]{Pot01}%
  \BibitemOpen
  \bibfield  {author} {\bibinfo {author} {\bibfnamefont {M.}~\bibnamefont
  {Potthoff}},\ }\href {\doibase 10.1103/PhysRevB.64.165114} {\bibfield
  {journal} {\bibinfo  {journal} {Phys. Rev. B}\ }\textbf {\bibinfo {volume}
  {64}},\ \bibinfo {pages} {165114} (\bibinfo {year} {2001})}\BibitemShut
  {NoStop}%
\bibitem [{Note2()}]{Note2}%
  \BibitemOpen
  \bibinfo {note} {Not to be confused with the two-site cluster DMFT
  scheme.}\BibitemShut {Stop}%
\bibitem [{\citenamefont {Stanescu}\ and\ \citenamefont
  {Kotliar}(2004)}]{Stan04}%
  \BibitemOpen
  \bibfield  {author} {\bibinfo {author} {\bibfnamefont {T.~D.}\ \bibnamefont
  {Stanescu}}\ and\ \bibinfo {author} {\bibfnamefont {G.}~\bibnamefont
  {Kotliar}},\ }\href {\doibase 10.1103/PhysRevB.70.205112} {\bibfield
  {journal} {\bibinfo  {journal} {Phys. Rev. B}\ }\textbf {\bibinfo {volume}
  {70}},\ \bibinfo {pages} {205112} (\bibinfo {year} {2004})}\BibitemShut
  {NoStop}%
\bibitem [{Note3()}]{Note3}%
  \BibitemOpen
  \bibinfo {note} {Here we follow the notation of Chapter 8 and 11 in Ref.
  \protect \rev@citealpnum {Ave12}.}\BibitemShut {Stop}%
\bibitem [{\citenamefont {Vidberg}\ and\ \citenamefont {Serene}(1977)}]{Vid77}%
  \BibitemOpen
  \bibfield  {author} {\bibinfo {author} {\bibfnamefont {H.}~\bibnamefont
  {Vidberg}}\ and\ \bibinfo {author} {\bibfnamefont {J.}~\bibnamefont
  {Serene}},\ }\href {\doibase 10.1007/BF00655090} {\bibfield  {journal}
  {\bibinfo  {journal} {Journal of Low Temperature Physics}\ }\textbf {\bibinfo
  {volume} {29}},\ \bibinfo {pages} {179} (\bibinfo {year} {1977})}\BibitemShut
  {NoStop}%
\bibitem [{\citenamefont {Raas}\ and\ \citenamefont {Uhrig}(2009)}]{Raa09}%
  \BibitemOpen
  \bibfield  {author} {\bibinfo {author} {\bibfnamefont {C.}~\bibnamefont
  {Raas}}\ and\ \bibinfo {author} {\bibfnamefont {G.~S.}\ \bibnamefont
  {Uhrig}},\ }\href {\doibase 10.1103/PhysRevB.79.115136} {\bibfield  {journal}
  {\bibinfo  {journal} {Phys. Rev. B}\ }\textbf {\bibinfo {volume} {79}},\
  \bibinfo {pages} {115136} (\bibinfo {year} {2009})}\BibitemShut {NoStop}%
\bibitem [{\citenamefont {M\"uller-Hartmann}(1989{\natexlab{a}})}]{Mue89}%
  \BibitemOpen
  \bibfield  {author} {\bibinfo {author} {\bibfnamefont {E.}~\bibnamefont
  {M\"uller-Hartmann}},\ }\href {\doibase 10.1007/BF01311397} {\bibfield
  {journal} {\bibinfo  {journal} {Zeitschrift f\"ur Physik B Condensed Matter}\
  }\textbf {\bibinfo {volume} {74}},\ \bibinfo {pages} {507} (\bibinfo {year}
  {1989}{\natexlab{a}})}\BibitemShut {NoStop}%
\bibitem [{\citenamefont {M\"uller-Hartmann}(1989{\natexlab{b}})}]{Mue89b}%
  \BibitemOpen
  \bibfield  {author} {\bibinfo {author} {\bibfnamefont {E.}~\bibnamefont
  {M\"uller-Hartmann}},\ }\href {\doibase 10.1007/BF01312686} {\bibfield
  {journal} {\bibinfo  {journal} {Zeitschrift f\"ur Physik B Condensed Matter}\
  }\textbf {\bibinfo {volume} {76}},\ \bibinfo {pages} {211} (\bibinfo {year}
  {1989}{\natexlab{b}})}\BibitemShut {NoStop}%
\bibitem [{\citenamefont {Karrasch}\ \emph {et~al.}(2008)\citenamefont
  {Karrasch}, \citenamefont {Hedden}, \citenamefont {Peters}, \citenamefont
  {Pruschke}, \citenamefont {Sch\"onhammer},\ and\ \citenamefont
  {Meden}}]{Kar08}%
  \BibitemOpen
  \bibfield  {author} {\bibinfo {author} {\bibfnamefont {C.}~\bibnamefont
  {Karrasch}}, \bibinfo {author} {\bibfnamefont {R.}~\bibnamefont {Hedden}},
  \bibinfo {author} {\bibfnamefont {R.}~\bibnamefont {Peters}}, \bibinfo
  {author} {\bibfnamefont {T.}~\bibnamefont {Pruschke}}, \bibinfo {author}
  {\bibfnamefont {K.}~\bibnamefont {Sch\"onhammer}}, \ and\ \bibinfo {author}
  {\bibfnamefont {V.}~\bibnamefont {Meden}},\ }\href
  {http://stacks.iop.org/0953-8984/20/i=34/a=345205} {\bibfield  {journal}
  {\bibinfo  {journal} {Journal of Physics: Condensed Matter}\ }\textbf
  {\bibinfo {volume} {20}},\ \bibinfo {pages} {345205} (\bibinfo {year}
  {2008})}\BibitemShut {NoStop}%
\bibitem [{\citenamefont {Sch\"afer}\ \emph {et~al.}(2013)\citenamefont
  {Sch\"afer}, \citenamefont {Rohringer}, \citenamefont {Gunnarsson},
  \citenamefont {Ciuchi}, \citenamefont {Sangiovanni},\ and\ \citenamefont
  {Toschi}}]{Schae13}%
  \BibitemOpen
  \bibfield  {author} {\bibinfo {author} {\bibfnamefont {T.}~\bibnamefont
  {Sch\"afer}}, \bibinfo {author} {\bibfnamefont {G.}~\bibnamefont
  {Rohringer}}, \bibinfo {author} {\bibfnamefont {O.}~\bibnamefont
  {Gunnarsson}}, \bibinfo {author} {\bibfnamefont {S.}~\bibnamefont {Ciuchi}},
  \bibinfo {author} {\bibfnamefont {G.}~\bibnamefont {Sangiovanni}}, \ and\
  \bibinfo {author} {\bibfnamefont {A.}~\bibnamefont {Toschi}},\ }\href
  {\doibase 10.1103/PhysRevLett.110.246405} {\bibfield  {journal} {\bibinfo
  {journal} {Phys. Rev. Lett.}\ }\textbf {\bibinfo {volume} {110}},\ \bibinfo
  {pages} {246405} (\bibinfo {year} {2013})}\BibitemShut {NoStop}%
\bibitem [{Note4()}]{Note4}%
  \BibitemOpen
  \bibinfo {note} {To be more precise, one should compare the ration $\protect
  \frac {U}{\sigma }$, where $\sigma $ is the standard deviation of the
  noninteracting density of states. Anyhow, one has $\sigma =t$ for the Bethe
  lattice and $\sigma =2t$ for the two-dimensional square lattice, so that both
  criteria are equivalent in our case.}\BibitemShut {Stop}%
\bibitem [{\citenamefont {Taranto}\ \emph {et~al.}(2013)\citenamefont
  {Taranto}, \citenamefont {Andergassen}, \citenamefont {Bauer}, \citenamefont
  {Held}, \citenamefont {Katanin}, \citenamefont {Metzner}, \citenamefont
  {Rohringer},\ and\ \citenamefont {Toschi}}]{Tar13}%
  \BibitemOpen
  \bibfield  {author} {\bibinfo {author} {\bibfnamefont {C.}~\bibnamefont
  {Taranto}}, \bibinfo {author} {\bibfnamefont {S.}~\bibnamefont
  {Andergassen}}, \bibinfo {author} {\bibfnamefont {J.}~\bibnamefont {Bauer}},
  \bibinfo {author} {\bibfnamefont {K.}~\bibnamefont {Held}}, \bibinfo {author}
  {\bibfnamefont {A.}~\bibnamefont {Katanin}}, \bibinfo {author} {\bibfnamefont
  {W.}~\bibnamefont {Metzner}}, \bibinfo {author} {\bibfnamefont
  {G.}~\bibnamefont {Rohringer}}, \ and\ \bibinfo {author} {\bibfnamefont
  {A.}~\bibnamefont {Toschi}},\ }\href@noop {} {\bibfield  {journal} {\bibinfo
  {journal} {ArXiv e-prints}\ } (\bibinfo {year} {2013})},\ \Eprint
  {http://arxiv.org/abs/1307.3475} {arXiv:1307.3475 [cond-mat.str-el]}
  \BibitemShut {NoStop}%
\bibitem [{\citenamefont {Ortloff}()}]{Ort}%
  \BibitemOpen
  \bibfield  {author} {\bibinfo {author} {\bibfnamefont {J.}~\bibnamefont
  {Ortloff}},\ }\href@noop {} {\bibinfo {type} {Phd thesis (unpublished)}},\
  \bibinfo  {school} {University of W\"urzburg}\BibitemShut {NoStop}%
\bibitem [{\citenamefont {Park}\ \emph {et~al.}(2011)\citenamefont {Park},
  \citenamefont {Haule},\ and\ \citenamefont {Kotliar}}]{Par11}%
  \BibitemOpen
  \bibfield  {author} {\bibinfo {author} {\bibfnamefont {H.}~\bibnamefont
  {Park}}, \bibinfo {author} {\bibfnamefont {K.}~\bibnamefont {Haule}}, \ and\
  \bibinfo {author} {\bibfnamefont {G.}~\bibnamefont {Kotliar}},\ }\href
  {\doibase 10.1103/PhysRevLett.107.137007} {\bibfield  {journal} {\bibinfo
  {journal} {Phys. Rev. Lett.}\ }\textbf {\bibinfo {volume} {107}},\ \bibinfo
  {pages} {137007} (\bibinfo {year} {2011})}\BibitemShut {NoStop}%
\bibitem [{\citenamefont {Toschi}\ \emph {et~al.}(2012)\citenamefont {Toschi},
  \citenamefont {Arita}, \citenamefont {Hansmann}, \citenamefont
  {Sangiovanni},\ and\ \citenamefont {Held}}]{Tos12}%
  \BibitemOpen
  \bibfield  {author} {\bibinfo {author} {\bibfnamefont {A.}~\bibnamefont
  {Toschi}}, \bibinfo {author} {\bibfnamefont {R.}~\bibnamefont {Arita}},
  \bibinfo {author} {\bibfnamefont {P.}~\bibnamefont {Hansmann}}, \bibinfo
  {author} {\bibfnamefont {G.}~\bibnamefont {Sangiovanni}}, \ and\ \bibinfo
  {author} {\bibfnamefont {K.}~\bibnamefont {Held}},\ }\href {\doibase
  10.1103/PhysRevB.86.064411} {\bibfield  {journal} {\bibinfo  {journal} {Phys.
  Rev. B}\ }\textbf {\bibinfo {volume} {86}},\ \bibinfo {pages} {064411}
  (\bibinfo {year} {2012})}\BibitemShut {NoStop}%
\bibitem [{\citenamefont {Liu}\ \emph {et~al.}(2012)\citenamefont {Liu},
  \citenamefont {Harriger}, \citenamefont {Luo}, \citenamefont {Wang},
  \citenamefont {Ewings}, \citenamefont {Guidi}, \citenamefont {Park},
  \citenamefont {Haule}, \citenamefont {Kotliar}, \citenamefont {Hayden},\ and\
  \citenamefont {Dai}}]{Liu12}%
  \BibitemOpen
  \bibfield  {author} {\bibinfo {author} {\bibfnamefont {M.}~\bibnamefont
  {Liu}}, \bibinfo {author} {\bibfnamefont {L.~W.}\ \bibnamefont {Harriger}},
  \bibinfo {author} {\bibfnamefont {H.}~\bibnamefont {Luo}}, \bibinfo {author}
  {\bibfnamefont {M.}~\bibnamefont {Wang}}, \bibinfo {author} {\bibfnamefont
  {R.~A.}\ \bibnamefont {Ewings}}, \bibinfo {author} {\bibfnamefont
  {T.}~\bibnamefont {Guidi}}, \bibinfo {author} {\bibfnamefont
  {H.}~\bibnamefont {Park}}, \bibinfo {author} {\bibfnamefont {K.}~\bibnamefont
  {Haule}}, \bibinfo {author} {\bibfnamefont {G.}~\bibnamefont {Kotliar}},
  \bibinfo {author} {\bibfnamefont {S.~M.}\ \bibnamefont {Hayden}}, \ and\
  \bibinfo {author} {\bibfnamefont {P.}~\bibnamefont {Dai}},\ }\href {\doibase
  10.1038/nphys2268} {\bibfield  {journal} {\bibinfo  {journal} {Nat. Phys.}\
  }\textbf {\bibinfo {volume} {8}},\ \bibinfo {pages} {376} (\bibinfo {year}
  {2012})}\BibitemShut {NoStop}%
\bibitem [{\citenamefont {Negele}\ and\ \citenamefont {Orland}(1988)}]{Neg88}%
  \BibitemOpen
  \bibfield  {author} {\bibinfo {author} {\bibfnamefont {J.~W.}\ \bibnamefont
  {Negele}}\ and\ \bibinfo {author} {\bibfnamefont {H.}~\bibnamefont
  {Orland}},\ }\href@noop {} {\emph {\bibinfo {title} {Quantum many-particle
  systems}}}\ (\bibinfo  {publisher} {Addison-Wesley},\ \bibinfo {year}
  {1988})\BibitemShut {NoStop}%
\bibitem [{\citenamefont {Hafermann}\ \emph {et~al.}(2009)\citenamefont
  {Hafermann}, \citenamefont {Jung}, \citenamefont {Brener}, \citenamefont
  {Katsnelson}, \citenamefont {Rubtsov},\ and\ \citenamefont
  {Lichtenstein}}]{Haf09}%
  \BibitemOpen
  \bibfield  {author} {\bibinfo {author} {\bibfnamefont {H.}~\bibnamefont
  {Hafermann}}, \bibinfo {author} {\bibfnamefont {C.}~\bibnamefont {Jung}},
  \bibinfo {author} {\bibfnamefont {S.}~\bibnamefont {Brener}}, \bibinfo
  {author} {\bibfnamefont {M.~I.}\ \bibnamefont {Katsnelson}}, \bibinfo
  {author} {\bibfnamefont {A.~N.}\ \bibnamefont {Rubtsov}}, \ and\ \bibinfo
  {author} {\bibfnamefont {A.~I.}\ \bibnamefont {Lichtenstein}},\ }\href
  {http://stacks.iop.org/0295-5075/85/i=2/a=27007} {\bibfield  {journal}
  {\bibinfo  {journal} {EPL (Europhysics letter)}\ }\textbf {\bibinfo {volume}
  {85}},\ \bibinfo {pages} {27007} (\bibinfo {year} {2009})}\BibitemShut
  {NoStop}%
\bibitem [{\citenamefont {Katanin}(2004)}]{Kat04}%
  \BibitemOpen
  \bibfield  {author} {\bibinfo {author} {\bibfnamefont {A.~A.}\ \bibnamefont
  {Katanin}},\ }\href {\doibase 10.1103/PhysRevB.70.115109} {\bibfield
  {journal} {\bibinfo  {journal} {Phys. Rev. B}\ }\textbf {\bibinfo {volume}
  {70}},\ \bibinfo {pages} {115109} (\bibinfo {year} {2004})}\BibitemShut
  {NoStop}%
\bibitem [{\citenamefont {Avella}\ and\ \citenamefont {Mancini}(2012)}]{Ave12}%
  \BibitemOpen
  \bibfield  {author} {\bibinfo {author} {\bibfnamefont {A.}~\bibnamefont
  {Avella}}\ and\ \bibinfo {author} {\bibfnamefont {F.}~\bibnamefont
  {Mancini}},\ }\href@noop {} {\emph {\bibinfo {title} {Strongly Correlated
  Systems: Theoretical Methods}}},\ edited by\ \bibinfo {editor} {\bibfnamefont
  {A.}~\bibnamefont {Avella}},\ Springer Series in Solid-State Sciences\
  (\bibinfo  {publisher} {Springer},\ \bibinfo {address} {Berlin},\ \bibinfo
  {year} {2012})\BibitemShut {NoStop}%
\end{thebibliography}%

\end{document}